\documentclass[12pt,a4paper]{report}
\usepackage{amssymb}
\usepackage{graphicx}
\usepackage{amsmath}

\newcommand{\bea}{\begin{eqnarray}}
\newcommand{\beq}{\begin{equation}}
\newcommand{\eea}{\end{eqnarray}}
\newcommand{\eeq}{\end{equation}}

\input{tcilatex}
\usepackage{graphicx}
\begin{document}

\setcounter{page}{1}\renewcommand{\thepage}{\Roman{page}}

\begin{center}
\textbf{IN THE NAME OF GOD}

\bigskip 

\bigskip 

\bigskip 

\textbf{\bigskip }

\textbf{CHARGED ROTATING BLACK BRANES IN VARIOUS}

\textbf{DIMENSIONS\bigskip }

\bigskip 

By:\medskip

\textbf{A. KHODAM-MOHAMMADI\bigskip }

\bigskip 

\bigskip 

THESIS\medskip

SUBMITTED TO THE SCHOOL OF GRADUATE STUDIES IN PARTIAL FULFILLMENT OF THE
REQUIREMENTS FOR THE DEGREE OF DOCTOR OF PHILOSOPHY (Ph.D.)\medskip

IN\bigskip

PHYSICS

SHIRAZ UNIVERSITY

SHIRAZ, IRAN\medskip 

\bigskip 

\bigskip 

\bigskip 

\bigskip 

All rights reserved.

\textbf{SEPTEMBER 2004}

\textbf{\newpage }
\end{center}

\bigskip\ \bigskip\ \vspace{5cm}

{\large To:\bigskip }

\hspace{3cm}\textit{my parents,}$\bigskip $

$\hspace{4.5cm}$\textit{my wife,\bigskip }

$\hspace{5cm}$\textit{and\bigskip ...}

$\hspace{6cm}$\textit{Farin}

\begin{center}
\newpage
\end{center}

\bigskip\ \bigskip

\begin{center}
{\Large \textbf{Acknowledgments\bigskip }}
\end{center}

I would like to express my deepest gratitude to the following people, who
have helped and accompanied me during my time as a Ph.D. student at Shiraz
University:

First, I am indebted to my senior supervisor, Dr. M. H. Dehghani, for making
me interested in the subject of my research. My progress was always boosted
by his relevant scientific advice, while his patience and sincere cordiality
made working under his hand a pleasure.

Secondly, my thanks go to the members of my supervisory committee,
professors N. Riazi, N. Ghahramani, Dr. M. M. Golshan and Dr. H. R. Sepangi,
for valuable criticisms of this work.

\newpage

\bigskip \baselineskip=5mm\ \bigskip

\begin{center}
\textbf{Abstract\bigskip \medskip }

\textbf{Charged Rotating Black Branes in Various Dimensions\medskip }

\textbf{BY:\medskip }

\textbf{Abdolhosein Khodam Mohammadi\bigskip }
\end{center}

{\small In this thesis, two different aspects of asymptotically charged
rotating black branes in various dimensions are studied. In the first part,
the thermodynamics of these spacetimes is investigated, while in the second
part the no hair theorem for these spacetimes in four dimensions is
considered. In part I, first, the Euclidean actions of a d-dimensional
charged rotating black brane are computed through the use of the
counterterms renormalization method both in the canonical and the
grand-canonical ensemble, and it is shown that the logarithmic divergencies
associated to the Weyl anomalies and matter field vanish. Second, a
Smarr-type formula for the mass as a function of the entropy, the angular
momenta and the electric charge is obtained, which shows that these
quantities satisfy the first law of thermodynamics. Third, by using the
conserved quantities and the Euclidean actions, the thermodynamics
potentials of the system in terms of the temperature, the angular velocities
and the electric potential are obtained both in the canonical and the
grand-canonical ensemble. Fourth, a stability analysis in these two
ensembles is performed, which shows that the system is thermally stable.
This is in commensurable with the fact that there is no Hawking-Page phase
transition for black object with zero curvature horizon. Finally, the
logarithmic correction of the entropy due to the thermal fluctuation around
the equilibrium is calculated.}

{\small In part II, the cosmological defects are studied, and it is shown
that the Abelian Higgs field equations in the background of a
four-dimensional rotating charged black string have vortex solutions. These
solutions, which have axial symmetry, show that the rotating black string
can support the Abelian Higgs field as hair. It is found that in the case of
rotating black string, there exists an electric field coupled to the Higgs
scalar field. This electric field is due to an electric charge per unit
length, which increases as the rotation parameter becomes larger. Also it is
found that the vortex thickness decreases as the rotation parameter grows
up. Finally the self-gravity of the Abelian Higgs field is investigated, and
it is shown that the effect of the vortex is to induce a deficit angle in
the metric under consideration which decreases as the rotation parameter
increases.}


\baselineskip=7.5mm

\tableofcontents
\listoffigures%

\newpage \setcounter{page}{1}\renewcommand{\thepage}{\arabic{page}} \newpage

\chapter{\protect\bigskip Introduction\protect\bigskip}

Black holes are truly unique objects: For theoretical physicists they pose
various interesting problems, which may offer a way for solving the
difficult problems of quantum gravity. Currently about 20 stellar binaries
are known in our galaxy which are believed to contain black holes of some
solar masses, whereas supermassive black holes provide the only explanation
for the processes observed in the centers of active galaxies \cite{Frolov}.
The gravitational wave detectors GEO 600 \cite{Geo}, VIRGO \cite{Virgo} and
LIGO \cite{Ligo}, are used to directly observe processes involving black
holes, including collisions of black holes, in our cosmic neighborhood of
about 25 Mpc.

The theoretical aspect of black hole physics is the essential aim of this
thesis. Many works have been done on spherical black holes which are
asymptotically flat, de Sitter or anti-de Sitter. In recent years the
thermodynamics and no-hair theorem of static solutions of Einstein equation
such as Schwarzschild, Reissner-Nordstrom and ... black holes have been
considered. But there exist only a few works on these aspects for stationary
black holes with less symmetries. This is due to the difficulties which will
appear in considering less symmetric black holes. In this thesis we want to
consider the thermodynamics (part \ref{part1}) and no-hair theorem (part \ref%
{part2}) of these kind of black holes. Rotating charged black brane is an
example of black holes with less symmetries which we consider here.

\section{Thermodynamics of Black Holes (Part \protect\ref{part1})}

The connection between heat and mechanical energy was one of the most
interesting discoveries of thermodynamics. This realization provided the
first clues about how phenomena at the macroscopic level must arise from the
statistical properties of the mechanics of microscopic objects. The failure
of the classical statistical mechanics, was the first indication of quantum
mechanical effects. The study of systems at the macroscopic scale yields
insight into the more fundamental theories of nature and has allowed
physicists to take the first steps to understand these theories.

Although quantum mechanics seems to be quite sufficient for most
applications, the theory suffers from serious problems in its foundation,
especially when one attempts to develop a quantum theory of gravitation. As
an aid in understanding quantum gravity, one needs a system in which both
the quantum and the classical behavior exist. The black hole is such a
system. One hopes to gain insight into the nature of quantum gravity by
studying the thermodynamics of black holes. The black hole is an object that
is considered of classical mechanics and quantum mechanics: of the
macroscopic and the microscopic point of view.

The quantities of particular interest in gravitational thermodynamics are
the physical entropy $S$ and the temperature $\beta ^{-1}$, where these
quantities are respectively proportional to the area and surface gravity of
the event horizon(s) \cite{Bekenstein1,Haw1,Hawking1}. Other black hole
properties, such as energy, angular momentum and conserved charges can also
be given a thermodynamic interpretation. In finding the thermodynamic
quantities, one should use the quasilocal definitions for the thermodynamic
variables. By quasilocal, we mean that the quantity is constructed from
information that exists on the boundary of a gravitating system alone. Just
as the Gauss' law, such quasilocal quantities will yield information about
the spacetime contained within the system boundary. Two advantages of using
such a quasilocal method are the following: first, one is able to
effectively separate the gravitating system from the rest of the universe;
second, the formalism does not depend on the particular asymptotic behavior
of the system, so one can accommodate a wide class of spacetimes with the
same formalism.

Brown and York \cite{BrownY} developed a method for calculating energy and
other charges contained within a specified surface surrounding a gravitating
system. Although their quasilocal energy depends only on quantities defined
on the boundary of the gravitating system, it yields information about the
total gravitational energy contained within the boundary. Also they showed
that this quasilocal energy is the thermodynamic internal energy \cite%
{BrownY2}. The quasilocal energy and momentum are obtained from the
gravitational action via a Hamilton-Jacobi analysis. Although the analysis
of Brown and York was restricted to general relativity in four dimensions,
the thermodynamic variables were obtained from the action rather than from
the field equations.

In general, the problem of calculating gravitational thermodynamic
quantities remains a lively subject of interest. Because they typically
diverge for both asymptotically flat, asymptotically anti-de Sitter (AdS)
and de Sitter (dS) spacetimes, a common approach toward evaluating them has
been to carry out all computations relative to some other spacetime that is
regarded as the ground state for the class of spacetimes of interest. This
is done by taking the original action $I_{G}=I_{M}+I_{\partial M}$ for
gravity coupled to matter fields and subtracting from it a reference action $%
I_{0}$, which is a functional of the induced metric on the boundary $%
\partial M$. Conserved and/or thermodynamic quantities are then computed
relative to reference spacetime, which can then be taken to (spatial)
infinity if desired. This approach has been widely successful in providing a
description of gravitational thermodynamics in regions of both finite and
infinite spatial extent \cite{BrownY,BCM}. Unfortunately it suffers from
several drawbacks. The choice of reference spacetime is not always unique 
\cite{Chan}, nor is it always possible to embed a boundary with a given
induced metric into the reference background. Indeed, for Kerr spacetimes,
this latter problem forms a serious obstruction towards calculating the
subtraction energy, and calculations have only been performed in the
slow-rotating regime \cite{Martinez}. Currently, one of the useful theories
which helps us obtain the thermodynamic quantities is the `AdS/conformal
field theory (CFT)' correspondence which can be used to compute the
conserved quantities of asymptotically anti-de Sitter (AAdS) spacetimes.

This principle posits a relationship between supergravity or string theory
in bulk AdS spacetimes and conformal field theories on the boundary. It
offers the possibility that a full quantum theory of gravity could be
described by a well understood CFT theory.

Since quantum field theories in general contain counterterms, it is natural
from the AdS/CFT viewpoint to append a boundary term $I_{ct}$ to the action
that depends on the intrinsic geometry of the (timelike) boundary at large
spatial distances. This requirement, along with general covariance, implies
that these terms be functional of curvature invariants of the induced metric
and have no dependence on the extrinsic curvature of the boundary. An
algorithmic procedure \cite{Awad1} exists for constructing $I_{ct}$ for
asymptotically AdS spacetimes, and so its determination is unique. The
addition of $I_{ct}$ will not affect the bulk equations of motion, thereby
eliminating the need to embed the given geometry in a reference spacetime.

Although, there is no proof of this conjecture, it furnishes a means for
calculating the action and thermodynamic quantities intrinsically without
reliance on any reference spacetime \cite{Hening-Sk,Balasub,Od1}. A
dictionary is emerging that translates between different quantities in the
bulk gravity theory and its counterparts on the boundary, including the
partition functions and correlation functions. Another interesting
application of the AdS/CFT correspondence is the interpretation of
Hawking-Page phase transition between thermal AdS and AAdS black hole as the
confinement-deconfinement phases of the Yang-Mills (dual gauge) theory
defined on the AdS boundary \cite{Wit}.

The AdS/CFT correspondence has been also extended to the case of
asymptotically de Sitter spacetimes \cite{Stro,Deh1}. Although the (A)dS/CFT
correspondence applies for the case of special infinite boundary, it was
also employed to the computation of the conserved and thermodynamic
quantities in the case of a finite boundary \cite{Deh2}. This conjecture has
also been applied for the case of black objects with constant negative or
zero curvature horizons \cite{Dehghani1,Deh4}.

Due to the AdS/CFT correspondence, AAdS black holes continue to attract a
great deal of attention. For AAdS spacetimes, the presence of a negative
cosmological constant makes it possible to have a large variety of black
holes/branes, whose event horizons are hypersurfaces with positive, negative
or zero scalar curvature \cite{Man1}. Static and rotating uncharged
solutions of Einstein Relativity with negative cosmological constant and
with planar symmetry (planar, cylindrical and toroidal topology) may be
found in Ref. \cite{Lem1}. Unlike the zero cosmological constant planar
case, these solutions include the presence of black strings (or cylindrical
black holes) in the cylindrical model, and of toroidal black holes and black
membranes in the toroidal and planar models, respectively. The extension to
include the Maxwell field has been done in Ref. \cite{Lemos}. This metric
was the AAdS rotating type solutions of Einstein's equation. The authors in
Ref. \cite{Lemos} found the static and rotating pure electrically charged
black holes that are the electric counterparts of the cylindrical, toroidal
and planar black holes which is found in \cite{Lem1}. The metric with
electric charge and zero angular momentum was also discussed in Ref. \cite%
{Huang}.

The generalization of this AAdS charged rotating solution of
Einstein-Maxwell equations to the higher dimensions has been done in Ref. 
\cite{Awad2}. Many authors have considered thermodynamics and stability
conditions of these black holes \cite{Dehghani1,Pec}.

\section{The No-Hair Theorem (Part \protect\ref{part2})}

The conjecture that after the gravitational collapse of the matter field,
the resultant black hole is characterized by at most its electromagnetic
charge, mass, and angular momentum is known as the classical `no-hair'
theorem and was first proposed by Ruffini and Wheeler \cite{Ruf}. Nowadays
we are faced with the discovery of black hole solutions in many theories in
which Einstein's equation is coupled with some self interacting matter
fields, and therefore this conjecture needs more investigation.

In certain special cases the conjecture has been verified. For example, a
scalar field minimally coupled to gravity in asymptotically flat spacetimes
cannot provide hair for the black hole \cite{Sud}. But this conjecture
cannot extend to all forms of matter fields. It is known that some long
range Yang-Mills quantum hair could be painted on the black holes \cite{Eli1}%
. Explicit calculations have also been carried out which verify the
existence of a long range Nielsen-Olesen vortex solution as stable hair for
a Schwarzschild black hole in four dimensions \cite{Achucarro}. For the
extreme black hole of Einstein-Maxwell gravity, it has been shown that flux
expulsion occurs for thick strings (thick with respect to the radius of
horizon), while flux penetration occurs for thin strings \cite%
{Cha1,Cha2,Bon1,Bon2}. Of course, one may note that these situations fall
outside the scope of classical no-hair theorem due to the non-trivial
topology of the string configuration.

Recently, some effort has been made to extend these ideas to the case of
(anti-)de Sitter spacetimes. While a scalar field minimally coupled to
gravity in asymptotically de Sitter spacetimes cannot provide hair for the
black hole \cite{Torii1}, it has been shown that in asymptotically AdS
spacetimes there is a hairy black hole solution \cite{Torii3}. Also, in Ref. 
\cite{Eli1}, it is shown that there exists a solution to the $SU(2)$
Einstein-Yang-Mills equations which describes a stable Yang-Mills hairy
black hole that is asymptotically AdS. In addition the idea of
Nielsen-Olesen vortices has been extended to the case of asymptotically
(anti-)de Sitter spacetimes \cite{DehHiggs2}. The investigation of
Nielsen-Olesen vortices in the background of charged black holes was done in
Refs. \cite{Cha1,Bon2}. More recently the stability of the Abelian Higgs
field in AdS-Schwarzschild and Kerr-AdS backgrounds has been investigated
and it has been shown that these asymptotically AdS black holes can support
an Abelian Higgs field as hair \cite{Ghez2,Dehjal}. In the case of
stationary black hole solutions, the explicit calculations which can
investigate the existence of a long range Nielson-Olesen vortex solution as
a hair is escorted with much more difficulties due to the rotation parameter 
\cite{Ghez2}. In this thesis we study the Abelian Higgs hair in a four
dimensional rotating charged black string that is a stationary model with
cylindrical symmetry. Various features of this kind of solutions of
Einstein-Maxwell equation have been considered \cite{Dehghani1}. Here, we
want to investigate the influence of rotation on the vortex solution of
Einstein-Maxwell-Higgs equation. Since an analytic solution to the Abelian
Higgs field equation appears to be intractable, we confirm by numerical
calculations that the rotating charged black string can be dressed by
Abelian Higgs field as hair.

\section{Overview}

In chapter \ref{Ads/CFT}, we present the counterterm renormalization method
in order to obtain the finite action, the divergence free stress
energy-momentum tensor, and the conserved quantities of asymptotically AdS
spacetimes. In chapter \ref{thermodynamics}, the classical law of mechanics
of black holes is posed. The metric of a four dimensional black string,
which was constructed by Lemos \cite{Lemos},\ is introduced in chapter \ref%
{Lemosch} %
.\textbf{\ }In chapter 5, we study the thermodynamics of an ($n+1$%
)-dimensional charged rotating black brane. The conserved charges are
obtained and also the stability of the black brane is discussed. The
discussion of no-hair theorem and the theory of Abelian Higgs hair for black
holes is presented in chapter \ref{Higgstheory}, while the existence of \
Abelian Higgs hair for charged rotating black string is studied in chapter %
\ref{Higgs}. Finally, in chapter \ref{conclusion}, we summarize the results.

\part{Thermodynamics of Black Holes and AdS/CFT Correspondence\label{part1}}

\chapter{The Counterterm Renormalization\label{Ads/CFT}\protect\bigskip}

The counterterm renormalization which is based on the AdS/CFT
correspondence, was first proposed by Maldacena, and is known as the
Maldacena's conjecture \cite{Mald}. In this chapter, we first discuss the
AdS/CFT correspondence briefly, and then introduce the couterterm
renormalization method. However, we take more attention on the gravitational
aspect of the theory while the conformal field theory aspect of the
correspondence would not be covered as much.

\ The starting point of some exciting developments over the last few years,
AdS/CFT correspondence, was a conjecture about the duality of classical
supergravity on $AdS_{5}\times S^{5}$\ and $\mathcal{N}=4,$\ $U(N)$\ super
Yang-Mills theory in the large $N$\ limit. Moreover, Witten \cite{Witten}
suggested that via this identification any field theory action on $(n+1)$%
-dimensional anti-de Sitter spacetime gives rise to an effective action of a
field theory on the $n$-dimensional horizon of AdS spacetime. Most
importantly, this field theory on the AdS horizon should be a conformal
field theory, because the AdS symmetries act as conformal symmetries on the
AdS horizon.

The general correspondence formula is \cite{Witten} 
\begin{equation}
\int_{\Psi _{0}}\mathcal{D}\Psi e^{-S_{AdS}[\Psi ]}=\langle \exp \int d^{n}x%
\mathcal{O}(x)\Psi _{0}(x)\rangle ,
\end{equation}
where the functional integral on the left hand side is over all the fields $%
\Psi $\ whose asymptotic boundary values are $\Psi _{0},$\ and $\mathcal{O}$%
\ denotes the conformal operators of the boundary conformal field theory. In
the classical limit, the correspondence formula can be written as \cite%
{Witten, Gubser} 
\begin{equation}
S_{AdS}[\Psi _{0}]=W_{CFT}[\Psi _{0}],
\end{equation}
where $S_{AdS}$\ is the classical on-shell action of an AdS field theory,
expressed in terms of the field boundary values $\Psi _{0}$, and $W_{CFT}$\
is the CFT effective action. However, one should expect $S_{AdS}$\ to be
divergent as it stands, because of the divergence of the AdS metric on the
AdS horizon (see Appendix \ref{Symmetry}). Thus, in order to extract the
physically relevant information, the on-shell action has to be renormalized
by adding counterterms, which cancel the infinities. After defining the
renormalized, finite action by 
\begin{equation}
S_{AdS,fin}=S_{AdS}-S_{div},
\end{equation}
where $S_{div}$\ stands for the local counterterms and $S_{AdS,fin}$\ is the
CFT effective action, then the meaningful correspondence formula is 
\begin{equation}
S_{AdS,fin}=W_{CFT}.
\end{equation}
Given a field theory action on AdS spacetime and a suitable regularization
method, it is straightforward to calculate the renormalized on-shell action $%
S_{AdS,fin}.$\ On the other hand, the CFT effective action $W_{CFT}[\Psi
_{0}]$\ contains all the information about the conformal field theory living
on the AdS horizon. Moreover, any field theory on AdS spacetime, which
includes gravity, has a corresponding counterpart CFT, whose action might
not even be known.

Before closing this section, an important aspects related to the
correspondence formula shall be discussed briefly. Why can one be sure that $%
W_{CFT}$\ is the generating functional of a conformal field theory? The AdS
symmetries act as conformal symmetries on the AdS horizon. Thus, by virtue
of the invariance of the AdS action under AdS symmetries, $W_{CFT}$\ is
invariant under conformal transformations, as long as the counterterms do
not break these symmetries. This case, which is the generic one, is given
when all the counterterms are covariant. Hence, the CFT correlation
functions will obey all the restrictions imposed by conformal invariance. A
brief discussion of AdS symmetries and conformal symmetries will be given in
Appendix \ref{Symmetry}.

The famous exception appears when $S_{div}$\ contains at least one
non-covariant term, which inevitably breaks some of the AdS symmetries.
Therefore, the conformal symmetry of the CFT effective action will be broken
too. The interesting and encouraging fact about this is that the breaking of
conformal invariance is a strong signature of the quantum character of the
CFT and tells about the anomalies in the algebra of quantum conformal
operators. The most notable example is the Weyl anomaly, which is discussed
in section(\ref{anomaly}).

\section{\ The AdS/CFT Correspondence\label{AdSGr}}

In a generally covariant theory, it is unnatural to assign a local
energy-momentum density to the gravitational field. For instance, candidate
expressions depending only on the metric and its first derivatives will
always vanish at a given point in locally coordinates. Instead, one can
consider the so-called `quasilocal stress tensor', defined locally on the
boundary of a given spacetime region. Consider the gravitational action as a
functional of the boundary metric $\gamma _{\mu \nu }$. The quasilocal
stress tensor associated with a spacetime region has been defined by Brown
and York to be \cite{BrownY}:

\begin{equation}
\mathcal{T}^{ij}=\frac{2}{\sqrt{-\gamma }}\frac{\delta S_{grav}}{\delta
\gamma _{ij}},  \label{stress1}
\end{equation}
where $S_{grav}$\ is the gravitational action. The resulting stress tensor
typically diverges as the boundary is taken to infinity. However, one is
always free to add a boundary term to the action without disturbing the bulk
equations of motion. To obtain a finite stress tensor, Brown and York
propose a subtraction derived by embedding a boundary with the same
intrinsic metric $\gamma _{\mu \nu }$\ in some reference spacetime.
Unfortunately this prescription suffers from several drawbacks. The choice
of reference spacetime is not always unique \cite{Chan}, nor is it always
possible to embed a boundary with a given induced metric into the reference
background. Indeed, for Kerr spacetimes this latter problem forms a serious
obstruction toward calculating the subtraction energy, and calculations have
only been performed in the slow-rotating regime \cite{Martinez}. Therefore,
the method of subtraction of Brown-York is generally not well defined.

For asymptotically anti-de Sitter (AdS) spacetimes, the AdS/CFT
correspondence was an attractive resolution to this difficulty. In the
gravitational aspects, it says there is an equivalence between a
gravitational theory in a ($n+1$)-dimensional anti-de Sitter spacetime and \
a conformal theory in a $n$-dimensional spacetime which can in some sense be
viewed as the boundary of the higher dimensional spacetime. According to
this correspondence, Eq. (\ref{stress1}) can be interpreted as the
expectation value of the stress tensor in the CFT: 
\begin{equation}
\langle \mathcal{T}^{ij}\rangle =\frac{2}{\sqrt{-\gamma }}\frac{\delta
S_{eff}}{\delta \gamma _{ij}}.
\end{equation}
The divergences which appear as the boundary goes to infinity, are simply
the standard ultraviolet divergences of quantum field theory, and may be
removed by adding local counterterms to the action. These counterterms
depend only on the intrinsic geometry of the boundary and are defined once
and for all. By using this method, the ambiguous prescription involving
embedding the boundary in a reference spacetime is removed.

\subsection{Counterterm Method}

We write the standard action for the gravitational field in vacuum, in a
unit system with $G=1$, as \footnote{%
This fixes our conventions for the Riemann curvature to be $R_{\mu \nu
\lambda }{}^{\sigma }=-2\partial _{\lbrack \mu }\Gamma _{\nu ]\lambda
}{}^{\sigma }+2\Gamma _{\lambda \lbrack \mu }{}^{\rho }\Gamma _{\nu ]\rho
}{}^{\sigma }$, where the antisymmetrization is defined with strength one,
i.e. $[\mu \nu ]=\frac{1}{2}(\mu \nu -\nu \mu )$. Also $R_{\mu \nu }=R_{\mu
\lambda \nu }{}^{\lambda }$. With these conventions spheres have a positive
scalar curvature. The cosmological constant is written as $\Lambda
=-n(n-1)/2\ell ^{2}$; in this notation pure $AdS_{n+1}$ has radius $\ell $.}:

\begin{equation}
S=-\frac{1}{16\pi }\int_{M}d^{n+1}x\sqrt{-g}\left( R+\frac{n(n-1)}{\ell ^{2}}%
\right) +\frac{1}{8\pi }\int_{^{3}B}d^{n}x\sqrt{-\gamma }\Theta -\frac{1}{%
8\pi }\int_{\Sigma }d^{n}x\sqrt{h}K,  \label{action}
\end{equation}
where $\Theta $\ and $K$\ are the trace of the extrinsic curvature of the
boundaries with induced metric $\gamma _{ij}$\ and $h_{ij}$\ (Appendix \ref%
{boundary}). The first integral is the Einstein-Hilbert volume term. The
second term represents an integral over the three-timelike-boundary $^{3}B$\
and the third term is an integral over the three-spacelike-boundary $\Sigma $%
. These two surface terms are called the Gibbons-Hawking boundary terms. The
necessity of the boundary terms can be seen as follows. The variation of the
volume term in action (\ref{action}) with respect to $g^{\mu \nu }$\ is
given by 
\begin{equation}
\delta S=\delta S_{M_{n+1}}+\delta S_{M_{n}},
\end{equation}
where 
\begin{equation}
\delta S_{M_{n+1}}=\frac{1}{8\pi }\int_{M_{n+1}}d^{n+1}x\sqrt{-g}\delta
g^{\varsigma \zeta }\left[ -\frac{1}{2}g_{\varsigma \zeta }\left( R+\frac{%
n(n-1)}{\ell ^{2}}\right) +R_{\varsigma \zeta }\right]
\end{equation}
and 
\begin{equation}
\delta S_{M_{n}}=\frac{1}{8\pi }\int_{M_{n}}d^{n}x\sqrt{-\hat{g}}n_{\mu }%
\left[ \partial ^{\mu }(g_{\xi \nu }\delta g^{\xi \nu })-D_{\nu }(\delta
g^{\mu \nu })\right] .  \label{surfaction}
\end{equation}
Here $\hat{g}_{ij}$\ (i.e. $\gamma _{ij}$\ or $h_{ij})$ is the metric on $%
M_{n}$\ (i.e. $\Sigma $ or $^{3}B$) induced from $g_{\mu \nu }$\ and $n_{\mu
}$\ is the unit vector normal to $M_{n}$\ (Appendix \ref{boundary}). Now we
choose the metric in the following form: 
\begin{equation}
ds^{2}=g_{\mu \nu }dx^{\mu }dx^{\nu }=\frac{l^{2}}{4\rho ^{2}}d\rho ^{2}+%
\frac{1}{\rho }\hat{g}_{ij}dx^{i}dx^{j}.  \label{cordchoice}
\end{equation}
Then $n_{\mu }$\ and its covariant derivatives are given by, 
\begin{equation}
n^{\mu }=\left( \frac{2\rho }{l},0,...,0\right) ,\hspace{1.0941pc}\nabla
_{\rho }n^{\rho }=\nabla _{\rho }n^{i}=\nabla _{i}n^{\rho }=0,\hspace{%
1.0941pc}\nabla _{i}n^{j}=\frac{\rho }{l}\hat{g}_{ik}\hat{g}^{\prime jk},
\end{equation}
where the prime denotes the derivative with respect to $\rho $.\ In the
coordinate choice (\ref{cordchoice}), the surface terms $\delta S_{M_{n}}$\
in (\ref{surfaction}) have the following form 
\begin{equation}
\delta S_{M_{n}}=\underset{\rho \rightarrow 0}{\lim \frac{1}{8\pi }}%
\int_{M_{n}}d^{n}x\sqrt{-\hat{g}}\frac{2\rho }{l}\left[ \partial ^{\rho }(%
\hat{g}_{ij}\delta \hat{g}^{ij})\right] .  \label{bulkEinstein}
\end{equation}
Note that the terms containing $\delta \hat{g}^{\rho \rho }$\ or $\delta 
\hat{g}^{\rho i}$\ vanish. The variant $\delta S_{M_{n}}$\ contains the
derivative of $\delta \hat{g}^{ij}$\ with respect to $\rho $, which makes
the variational principle ill-defined. In order to have a well defined
variation principle on the boundary, the variation of the action, after
using the partial integration, should be written as 
\begin{equation}
\delta S_{M_{n}}=\underset{\rho \rightarrow 0}{\lim \frac{1}{8\pi }}%
\int_{M_{n}}d^{n}x\sqrt{-\hat{g}}\delta \hat{g}^{ij}\{...\}.  \label{deltaS}
\end{equation}
If the variation of the action on the boundary contains $(\delta \hat{g}%
^{ij})^{\prime }$, however, we can not partially integrate it with respect
to $\rho $\ on the boundary to rewrite the variation in the form of (\ref%
{deltaS}) since $\rho $\ is the coordinate expressing the direction
perpendicular to the boundary. Therefore the `extremization' of the action
is ambiguous. Such a problem was well studied in \cite{Gibones} for the
Einstein gravity. It is easy to show that the boundary term which can remove
the terms containing $(\delta \hat{g}^{ij})^{\prime }$ is \footnote{%
In the coordinate choice (\ref{cordchoice}), the action (\ref{Gib-How}) has
the form 
\begin{equation}
S_{b}^{GH}=-\frac{1}{4\pi }\int_{M_{n}}d^{n}x\sqrt{-\hat{g}}\frac{\rho }{l}%
\hat{g}_{ij}(\hat{g}^{ij})^{\prime }.  \label{17Ein}
\end{equation}
then the variation over the metric $\hat{g}^{ij}$\ gives 
\begin{equation}
\delta S_{b}^{GH}=-\frac{1}{4\pi }\int_{M_{n}}d^{n}x\sqrt{-\hat{g}}\frac{%
\rho }{l}\left[ \delta \hat{g}^{ij}\left\{ -\hat{g}_{ik}\hat{g}_{jl}(\hat{g}%
^{kl})^{\prime }-\frac{1}{2}\hat{g}_{ij}\hat{g}_{kl}(\hat{g}^{kl})^{\prime
}\right\} +\hat{g}_{ij}(\delta \hat{g}^{ij})^{\prime }\right] .  \label{18GH}
\end{equation}
On the other side, the surface terms in the variation of the bulk\ Einstein
action in (\ref{bulkEinstein}), have the form 
\begin{equation}
\delta S_{M_{n}}^{\text{Einstein}}=\underset{\rho \rightarrow 0}{\lim }\frac{%
1}{8\pi }\int_{M_{n}}d^{n}x\sqrt{-\hat{g}}\frac{2\rho }{l}\left[ \hat{g}%
_{ij}^{\prime }(\delta \hat{g}^{ij})+\hat{g}_{ij}(\delta \hat{g}%
^{ij})^{\prime }\right] .
\end{equation}
Then the terms containing $(\delta \hat{g}^{ij})^{\prime }$\ in (\ref{17Ein}%
) and (\ref{18GH}) are canceled with each other \cite{Od9911152}.}

\begin{equation}
S_{b}^{GH}=-\frac{1}{4\pi }\int_{M_{n}}d^{n}x\sqrt{-\hat{g}}\nabla _{\mu
}n^{\mu }.  \label{Gib-How}
\end{equation}

Variation of the action (\ref{action}) with respect to the metrics $\gamma
_{\mu \nu }$\ and $h_{\mu \nu }$\ of the boundary $^{3}B$\ \ and $\Sigma $\
gives \cite{BrownY}: 
\begin{equation}
\int_{^{3}B}d^{3}x\Pi ^{ij}\delta \gamma
_{ij}+\int_{t_{i}}^{t_{f}}d^{3}xP^{ij}\delta h_{ij}.
\end{equation}
Here, $\Pi ^{ij}$\ denotes the gravitational momentum conjugate to $\gamma
_{ij},$\ as defined with respect to the three boundary $^{3}B,$\ while $%
P^{ij}$\ denotes the gravitational momentum conjugate to $h_{ij},$\ as
defined with respect to the spacelike hypersurfaces $t_{i}$\ and $t_{f}$\
given as 
\begin{eqnarray}
\Pi ^{ij} &=&-\frac{1}{16\pi }\sqrt{-\gamma }(\Theta \gamma ^{ij}-\Theta
^{ij}),  \label{momentum1} \\
P^{ij} &=&\frac{1}{16\pi }\sqrt{h}(Kh^{ij}-K^{ij}),  \label{momentum2}
\end{eqnarray}
where $\Theta $ and $K\ $are trace of the extrinsic curvatures$\ \Theta
^{ij} $ and $K^{ij}$ of three-boundaries $^{3}B$ and $\Sigma $ (see Appendix %
\ref{boundary}).\ In the AdS frame work, it is natural to contribute only
the time like boundary term $^{3}B$\ \ in action (\ref{action}) and $\Pi
^{ij}$\ in Eq. (\ref{momentum1}).

The energy-momentum tensor for boundary $^{3}B$ can obtain as: 
\begin{equation}
\mathcal{T}^{ij}=-\frac{1}{8\pi }(\Theta \gamma ^{ij}-\Theta ^{ij}).
\label{E-Mtensor}
\end{equation}

Concrete computations show that in most spacetimes both the action integral (%
\ref{action}) and the energy-momentum tensor (\ref{E-Mtensor}) diverge as
the boundary $^{3}B\ $goes to infinity. We therefore think of these as the
unrenormalized quantities.

The divergences must be cancelled in order to achieve physically meaningful
expressions. Thus, one may add a counterterm

\begin{equation}
\tilde{S}=\frac{1}{8\pi }\int d^{n}x\sqrt{-\gamma }\mathcal{\tilde{L}},
\end{equation}
to the action, along with the corresponding counterterm energy-momentum
tensor: 
\begin{equation}
\mathcal{\tilde{T}}^{ab}=\frac{2}{\sqrt{-\gamma }}\frac{\delta \tilde{S}}{%
\delta \gamma _{ab}}.
\end{equation}
The counterterms, by definition, contain the divergent part of the
corresponding unrenormalized quantities, but finite terms may depend on the
details of the renormalization.

The counterterm should be a function of the boundary geometry only.
Moreover, suppose the counterterm is an analytical function of the boundary
geometry, and expand it as a power series in the metric and its derivatives.
Dimensional analysis shows that in $AdS_{n+1}$\ only terms of order $m<n/2$\
contribute to the divergent part of the action. (By terms of order $m,$ we
mean terms containing $2m$\ derivatives.) Therefore one may truncate the
series at this order and obtain a finite polynomial \cite{Balasub}. This
agrees with the expectations from the interpretation of the divergences in
terms of a dual boundary theory that obeys the usual axioms of quantum field
theory, including locality.

\subsection{The Counterterm Generating Algorithm}

The structure of divergences is tightly constrained by the Gauss-Codacci
equations (\ref{Gauss1})-(\ref{Gauss3}). Using Eq. (\ref{E-Mtensor}), the\
Gauss-Codacci equations for timelike hypersurface boundary $^{3}B$\ can be
written as \cite{Waldbook1}: 
\begin{align}
& G_{ab}=G_{ab}(\gamma )+n^{\mu }\nabla _{\mu }\mathcal{T}_{ab}-\frac{1}{2}%
\gamma _{ab}\left( \frac{\mathcal{T}^{2}}{n-1}-\mathcal{T}_{cd}\mathcal{T}%
^{cd}\right) +\frac{1}{n-1}\mathcal{T}_{ab}\mathcal{T},  \label{Codacci1} \\
& G_{a\mu }n^{\mu }=-\nabla ^{b}\mathcal{T}_{ba},  \label{Codacci2} \\
& G_{\mu \nu }n^{\mu }n^{\nu }=\frac{1}{2}\left( \frac{\mathcal{T}^{2}}{n-1}-%
\mathcal{T}_{cd}\mathcal{T}^{cd}-R(\gamma )\right) ,  \label{Codacci3}
\end{align}
where $n^{\mu }$\ is an outward pointing unit vector normal to the boundary $%
^{3}B$. We will always consider solutions of the bulk equations of motion.
So the following equations 
\begin{align}
& G_{ab}=\frac{1}{2}\frac{n(n-1)}{\ell ^{2}}\gamma _{ab},  \notag \\
& G_{a\mu }n^{\mu }=0,  \notag \\
& G_{\mu \nu }n^{\mu }n^{\nu }=\frac{1}{2}\frac{n(n-1)}{\ell ^{2}},
\label{Codacci4}
\end{align}
determine the left hand side of the Gauss-Codacci equations.

In principle, one could solve the Gauss-Codacci equations (\ref{Codacci1})-(%
\ref{Codacci3}) for the unrenormalized energy-momentum tensor $\mathcal{T}%
_{ab}$, and then identify its divergent part with $-\mathcal{\tilde{T}}_{ab}$%
. However, this strategy is rather complicated due to the presence of the
normal derivatives in (\ref{Codacci1}). The appearance of these normal
derivatives expresses the intuitive fact that, to determine the solution
throughout, both the boundary values and their derivatives are needed.
However, the counterterm should be determined independently of data that is
extrinsic to the boundary, such as the normal derivative.

Explicit computations show that one can find a coordinate system so that the
divergent part of the normal derivatives expressed in terms of the intrinsic
boundary data . We implement this observation covariantly, as follows. We
impose the constraint equation (\ref{Codacci3}): 
\begin{equation}
\frac{1}{n-1}\mathcal{\tilde{T}}^{2}-\mathcal{\tilde{T}}_{ab}\mathcal{\tilde{%
T}}^{ab}=\frac{n(n-1)}{\ell ^{2}}+R,  \label{Cond1}
\end{equation}
and further insist that the counterterm energy-momentum tensor must derive
from a counterterm action, which is itself intrinsic to the boundary: 
\begin{equation}
\mathcal{\tilde{T}}^{ab}=\frac{2}{\sqrt{-\gamma }}\frac{\delta }{\delta
\gamma _{ab}}\int d^{n}x\sqrt{-\gamma }\mathcal{\tilde{L}}.  \label{Cond2}
\end{equation}
As we will show, the conditions (\ref{Cond1}) and (\ref{Cond2}) fully
determine the counterterm. The form of (\ref{Cond2}) ensures that the
counterterm energy-momentum is conserved, which in turn implies (\ref%
{Codacci2}). It is important to stress that the remaining Gauss-Codacci
equations (\ref{Codacci1}) are also satisfied: they can be viewed as
expressions for the normal derivatives specified implicitly in our
construction. We note that the normal derivatives which are determined, do
not in general vanish.

We are now prepared to describe an algorithm that determines the counterterm
as an expansion in the parameter $\ell $. The leading order term scale as $%
\ell ^{-1}$\ and terms at a given order $\ell ^{2m-1}$\ with $m\geqslant 0$\
are denoted by $\mathcal{\tilde{T}}_{ab}^{(m)}$ and $\mathcal{\tilde{L}}%
^{(m)}$. The starting point is to note that the curvature term in (\ref%
{Cond1}) can be neglected to the leading order in $\ell $, so that the
metric is the only tensor characterizing the boundary geometry to the
leading order, and therefore $\mathcal{\tilde{T}}_{ab}^{(0)}$\ is
proportional to the metric. Explicit computation will be given in Subsec. (%
\ref{explicit}).

Higher order counterterms are now given by induction. Assuming that $%
\mathcal{\tilde{T}}_{ab}$ is known up to and including order $m-1,$\ the
following three steps determine $\mathcal{\tilde{T}}_{ab}^{(m)}$:

step 1:\ Insert the known terms in (\ref{Cond1}); the resulting expression
is a linear equation with the trace $\mathcal{\tilde{T}}^{(m)}$\ as the only
unknown.

step 2:\ With the trace $\mathcal{\tilde{T}}^{(m)}$ in hand, integrate (\ref%
{Cond2}) and find $\mathcal{\tilde{L}}^{(m)}$. This step is purely
algebraic, as discussed in the following subsection.

step 3:\ Finally, take the functional derivative of $\mathcal{\tilde{L}}%
^{(m)}$ with respect to $\gamma _{ab}$, and so find the full tensor $%
\mathcal{\tilde{T}}_{ab}^{(m)}$\ from (\ref{Cond2}).

The fact that $\mathcal{\tilde{T}}_{ab}^{(0)}$ is proportional to the metric 
$\gamma _{ab}$\ is crucial to make step 1 possible. We stress that higher
orders of $\mathcal{\tilde{T}}_{ab}$ in general will depend also on other
tensor structures.

\subsection{Some Comments on Weyl Rescaling}

Under Weyl rescaling we have 
\begin{equation}
g_{ij}^{\prime }=[\lambda (x)]^{2}g_{ij}(x)  \label{Weylres}
\end{equation}
where $\lambda (x)$\ is an arbitrary function. If the action $S$\ is
invariant under any such Weyl rescalings, the definition of energy momentum
tensor 
\begin{equation}
\delta S=-\frac{1}{2}\int_{\Omega }d^{n}x\sqrt{g(x)}T_{ij}(x)\delta
g^{ij}(x),
\end{equation}
implies the tracelessness of the energy momentum tensor, 
\begin{equation}
T_{i}^{i}=0.
\end{equation}
The inverse statement is true as well. Moreover, Eq. (\ref{2.29}) implies
that, if an action is invariant under Weyl rescaling, it is also conformally
invariant. However the inverse is not true, because, while $\lambda $\ is
arbitrary in Eq. (\ref{Weylres}), it is not so in Eq. (\ref{2.29}).

The integration in step 2 in previous section is interesting and deserves
comment. It is related to the behavior of the various terms under the local
Weyl variations which transform the metric as: 
\begin{equation}
\delta _{W}\gamma _{ab}=\sigma \gamma _{ab},
\end{equation}
where $\sigma $ is an arbitrary function. Consider the counterterm action at
the $m$th order and note that dimensional analysis gives the behavior under
a global Weyl rescaling. The result of a local Weyl variation can therefore
be written in the form \cite{Kraus9906}: 
\begin{equation}
\delta _{W}\int d^{n}x\sqrt{-\gamma }\mathcal{\tilde{L}}^{(m)}=\int d^{n}x%
\sqrt{-\gamma }\sigma \left( \frac{n-2m}{2}\mathcal{\tilde{L}}^{(m)}+\nabla
_{a}X^{a(m)}\right) ,
\end{equation}
where $X^{a(m)}$\ is some unspecified expression (involving $2m+1$\
derivatives). However, it follows from (\ref{Cond2}) that:

\begin{equation}
\delta _{W}\int d^{n}x\sqrt{-\gamma }\mathcal{\tilde{L}}^{(m)}=\frac{1}{2}%
\int d^{n}x\sqrt{-\gamma }\sigma \mathcal{\tilde{T}}^{(m)},
\end{equation}
and therefore 
\begin{equation}
(n-2m)\mathcal{\tilde{L}}^{(m)}=\mathcal{\tilde{T}}^{(m)},  \label{17}
\end{equation}
The practical significance of this identity is that it renders the
integration in step 2 almost trivial. We also note that \cite{Kraus9906}: 
\begin{equation}
\delta _{W}\int d^{n}x\sqrt{-\gamma }\mathcal{\tilde{T}}^{(m)}=\frac{n-2m}{2}%
\int d^{n}x\sqrt{-\gamma }\sigma {}\mathcal{\tilde{T}}^{(m)},
\end{equation}
and therefore $\sqrt{-\gamma }\mathcal{\tilde{T}}^{(m)}$ transforms as a
conformal density with Weyl weight $\frac{1}{2}(n-2m),$\ up to a total
derivative \cite{Bonora1986cq,Deser1993yx}. This constrains the form of the
counterterms.

In even dimensions it is clear that (\ref{17}) prevents $\mathcal{\tilde{T}}%
^{(n/2)}$\ from being obtained as the variation of any local action. This is
the origin of trace anomalies. For an even $n,$\ the trace $\mathcal{\tilde{T%
}}^{(n/2)}$\ is, therefore, identified with the trace anomaly of the dual
boundary theory. This result for the anomaly agrees with that of \cite%
{Hening-Sk}, as may be verified by looking at the explicit expressions given
below.

\subsection{Explicit Computations of Counterterms\label{explicit}}

At this point, we evaluate the first few orders of counterterms explicitly.

Expanding $\mathcal{\tilde{T}}={\gamma }^{ab}${$\mathcal{\tilde{T}}$}$_{ab}$%
\ as a power series of $\ell $: 
\begin{eqnarray}
{\mathcal{\tilde{T}}} &=&{\mathcal{\tilde{T}}}^{(0)}{+{\mathcal{\tilde{T}}}%
^{(1)}+{\mathcal{\tilde{T}}}^{(2)}+...}  \notag \\
&=&\frac{a_{0}}{\ell }+a_{1}\ell +a_{2}\ell ^{3}+a_{3}\ell ^{5}+...,
\label{Expan}
\end{eqnarray}
In leading order one may neglect $R$ in Eq. (\ref{Cond1}) and uses the fact
that $\gamma ^{ab}\gamma _{ab}=n$ to obtain 
\begin{gather}
\frac{1}{n-1}({\gamma }^{ab}{\mathcal{\tilde{T}}}_{ab}^{(0)})({\gamma }_{ab}{%
\mathcal{\tilde{T}}}^{(0)ab})-{\mathcal{\tilde{T}}}_{ab}^{(0)}{\mathcal{%
\tilde{T}}}^{(0)ab}=\frac{n(n-1)}{\ell ^{2}}  \notag \\
\rightarrow {\mathcal{\tilde{T}}}_{ab}^{(0)}{\mathcal{\tilde{T}}}^{(0)ab}=%
\frac{n(n-1)^{2}}{\ell ^{2}}\rightarrow {\mathcal{\tilde{T}}}_{ab}^{(0)}=-%
\frac{(n-1)}{\ell }{\gamma }_{ab}.
\end{gather}
Thus, 
\begin{equation}
{\mathcal{\tilde{T}}}^{(0)}=-\frac{n(n-1)}{\ell }~,  \label{leading}
\end{equation}
and by using Eq. (\ref{17}) one obtains: 
\begin{equation}
{\tilde{\mathcal{L}}}^{(0)}=-\frac{n-1}{\ell }~.
\end{equation}
Up to the first order by inserting Eq. (\ref{leading}) in Eq. (\ref{Expan})
and using Eq. (\ref{Cond1}) one obtains: 
\begin{gather}
{\mathcal{\tilde{T}}}^{2}={\mathcal{\tilde{T}}}^{(0)^{2}}+2a_{1}\ell {%
\mathcal{\tilde{T}}}^{(0)}=\frac{n^{2}(n-1)^{2}}{\ell ^{2}}+Rn(n-1)  \notag
\\
\rightarrow a_{1}=-\frac{R}{2}\rightarrow {\mathcal{\tilde{T}}}^{(1)}=-{%
\frac{\ell }{2}}R~.
\end{gather}
Now Eq. (\ref{17}) gives: 
\begin{equation}
{\tilde{\mathcal{L}}}^{(1)}=-{\frac{\ell }{2(n-2)}}R~,
\end{equation}
and the variation (\ref{Cond2}) yields: 
\begin{equation}
{\mathcal{\tilde{T}}}_{ab}^{(1)}={\frac{\ell }{n-2}}\left( R_{ab}-{\frac{1}{2%
}}\gamma _{ab}R\right) ~.
\end{equation}
By using this algorithm, one can generate the next few order in expansion
as: 
\begin{align}
\mathcal{\tilde{T}}& =-\frac{n(n-1)}{\ell }-\frac{\ell }{2}R-\frac{\ell ^{3}%
}{2(n-2)^{2}}\left( R_{ab}R^{ab}-\frac{n}{4(n-1)}R^{2}\right)  \notag \\
& +\frac{\ell ^{5}}{(n-2)^{3}(n-4)}\left\{ \frac{3n+2}{4(n-1)}RR_{ab}R^{ab}-%
\frac{n(n+2)}{16(n-1)^{2}}R^{3}\right.  \notag \\
& \left. -2R^{ab}R_{acbd}R^{cd}+\frac{n-2}{2(n-1)}R^{ab}\nabla _{a}\nabla
_{b}R-R^{ab}\square R_{ab}+\frac{1}{2(n-1)}R\square R\right\} +\cdots ~,
\label{trace}
\end{align}
\begin{align}
{\tilde{\mathcal{L}}}& =-\frac{n-1}{\ell }-\frac{\ell }{2(n-2)}R-\frac{\ell
^{3}}{2(n-2)^{2}(n-4)}\left( R_{ab}R^{ab}-\frac{n}{4(n-1)}R^{2}\right) 
\notag \\
& +\frac{\ell ^{5}}{(n-2)^{3}(n-4)(n-6)}\left\{ \frac{3n+2}{4(n-1)}%
RR_{ab}R^{ab}-\frac{n(n+2)}{16(n-1)^{2}}R^{3}\right.  \notag \\
& \left. -2R^{ab}R_{acbd}R^{cd}+\frac{n-2}{2(n-1)}R^{ab}\nabla _{a}\nabla
_{b}R-R^{ab}\square R_{ab}+\frac{1}{2(n-1)}R\square R\right\} +\cdots .
\label{counter}
\end{align}
The energy-momentum tensor can be obtained through the variation of Eq. (\ref%
{counter}): 
\begin{align}
\mathcal{\tilde{T}}_{ab}& =-\frac{n-1}{\ell }\gamma _{ab}+\frac{\ell }{n-2}%
\left( R_{ab}-\frac{1}{2}\gamma _{ab}R\right)  \notag \\
& +\frac{\ell ^{3}}{(n-2)^{2}(n-4)}\left\{ -\frac{1}{2}\gamma _{ab}\left(
R_{cd}R^{cd}-\frac{n}{4(n-1)}R^{2}\right) -\frac{n}{2(n-1)}RR_{ab}\right. 
\notag \\
& \left. +2R^{cd}R_{cadb}-\frac{n-2}{2(n-1)}\nabla _{a}\nabla _{b}R+\square
R_{ab}-\frac{1}{2(n-1)}\gamma _{ab}\square R\right\} +\cdots ~,  \label{full}
\end{align}
where $\square =\nabla _{a}\nabla ^{a}$ and $R$, $R_{abcd}$, and $R_{ab}$
are the Ricci scalar, Riemann and Ricci tensors of the boundary metric $%
\gamma _{ab}$.\ The most laborious step is to find the full energy
momentum-tensor from the counterterm. Accordingly, we have resisted carrying
out this computation to the fourth order \cite{Kraus9906}.\medskip\ At last
the divergent free or finite energy-momentum tensor can be obtained from $%
\mathcal{T}_{(\mathrm{finite})ab}=\mathcal{T}_{ab}+\mathcal{\tilde{T}}_{ab},$
therefore 
\begin{align}
\mathcal{T}_{(\mathrm{finite})}^{ab}& =\frac{1}{8\pi }\{(\Theta ^{ab}-\Theta
\gamma ^{ab})-\frac{n-1}{\ell }\gamma ^{ab}+\frac{\ell }{n-2}(R^{ab}-\frac{1%
}{2}R\gamma ^{ab})  \notag \\
& \ +\frac{\ell ^{3}\Upsilon (n-5)}{(n-4)(n-2)^{2}}[-\frac{1}{2}\gamma
^{ab}(R^{cd}R_{cd}-\frac{n}{4(n-1)}R^{2})-\frac{n}{(2n-2)}RR^{ab}  \notag \\
& \ +2R_{cd}R^{acbd}-\frac{n-2}{2(n-1)}\nabla ^{a}\nabla ^{b}R+\nabla
^{2}R^{ab}-\frac{1}{2(n-1)}\gamma ^{ab}\nabla ^{2}R]+...\},
\label{finitestress}
\end{align}
where $\Upsilon (x)$ is the step function which is equal to one for $x\geq 0$
and zero otherwise.

\section{The Other Divergent Terms and It's Counterterms}

Additional divergent terms may be created due to a matter field on manifold $%
M$, and Weyl anomalies. First, we will discuss briefly the Weyl anomaly in
the dual conformal field theory and then reproduce the logarithmic
divergence for the matter field which is related to anomalies in the dual
conformal field theories and the Weyl anomalies.

\subsection{Weyl Anomalies\label{anomaly}}

As mentioned before the breaking of AdS symmetry and consequently breaking
of conformal invariance is a signature of anomalies in the algebra of
quantum conformal operators. The conformally invariance is a consequence of
the action invariability under the Weyl rescalings.

Previous section dealt with the counterterms which were required to
regularize the gravity action. These were calculated for an arbitrary
boundary metric without any need to linearize around a given background. The
calculation of the Weyl anomalies can be found in Refs. \cite{Hening-Sk} and 
\cite{Duff}. Let $W$\ denote the effective action of a quantum field theory,
defined by 
\begin{equation}
W=-\ln \int \mathcal{D}\phi e^{-S[\phi ]}.
\end{equation}
Defining an infinitesimal Weyl rescaling by $\delta g_{ij}(x)=2\lambda
(x)g_{ij}(x),$\ the Weyl anomaly is given by 
\begin{equation}
\langle T_{i}^{i}(x)\rangle =\mathcal{A}[g_{ij}(x)]=\frac{1}{\sqrt{g(x)}}%
\frac{\delta W}{\delta \lambda (x)}.  \label{5.41}
\end{equation}

\subsubsection{Scale Invariance and Its Breaking by Non-Covariant
Counterterms:}

It is natural to extend the AdS/CFT correspondence on a pure AdS spacetime
to the arbitrary Einstein spaces $\Omega $\ with negative cosmological
constants \cite{Witten}. Such a space $\Omega $\ possesses a horizon
manifold $\partial \Omega $, on which it determines a conformal structure.
For simplicity, $\partial \Omega $\ is assumed to have the topology of a $n$%
-sphere in the sequel. Then, in generalization of Eq. (\ref{2.5}), there is
a set of coordinates on $\Omega $\ for which the metric takes the form \cite%
{Graham} 
\begin{equation}
ds^{2}=\frac{l^{2}}{x_{0}^{2}}[\left( dx_{0}\right) ^{2}+\hat{g}_{ij}(%
\mathbf{x},x_{0})dx^{i}dx^{j}],
\end{equation}
where $\hat{g}_{ij}(x,0)=\hat{g}_{ij}(x)$\ is the horizon metric. For the
following consideration it is useful to use the dimensionless variable $\rho
=x_{0}^{2}/l^{2}$ and write the metric as 
\begin{equation}
ds^{2}=\frac{l^{2}}{4\rho ^{2}}\left( d\rho \right) ^{2}+\frac{1}{\rho }\hat{%
g}_{ij}(\mathbf{x},\rho )dx^{i}dx^{j}.  \label{5.45}
\end{equation}
Besides any coordinate symmetries, the metric (\ref{5.45}) is invariant
under 
\begin{equation}
\rho \rightarrow \sigma \rho \hspace{1.0304pc}\text{and\hspace{1.0304pc}}%
\hat{g}_{ij}\rightarrow \sigma \hat{g}_{ij},\hspace{0.412pc}  \label{5.46}
\end{equation}
which constitutes a global rescaling of the horizon $\partial \Omega $.

Obviously, the gravity action on the manifold $\Omega $\ is invariant under
the rescaling (\ref{5.46}). Hence, the AdS/CFT correspondence implies the
scale invariance of the CFT effective action, if it were not for
non-covariant divergent terms, which have to be cancelled by non-covariant
counterterms. Such divergent terms have the form 
\begin{equation}
S_{div}=\ln \epsilon \int_{\partial \Omega }d^{n}x\sqrt{\hat{g}}\mathcal{L}%
_{c},  \label{log-div}
\end{equation}
which is known as `logarithmic divergence'. Hear, $\int_{\partial \Omega
}d^{n}x\sqrt{\hat{g}}\mathcal{L}_{c}$\ is itself scale invariant, and the
cut-off boundary is characterized by $\rho =\epsilon $. Hence, for the scale
transformation (\ref{5.46}) with $\sigma =1+2\lambda $\ one obtains 
\begin{equation}
\delta S_{div}=2\lambda \int_{\partial \Omega }d^{n}x\sqrt{\hat{g}}\mathcal{L%
}_{c},
\end{equation}
which, because of $S=S_{fin}+S_{div}$\ and $\delta S=0$, implies 
\begin{equation}
\frac{\partial S_{fin}}{\partial \lambda }=-2\int_{\partial \Omega }d^{n}x%
\sqrt{\hat{g}}\mathcal{L}_{c}.  \label{5.48}
\end{equation}
Equation (\ref{5.48}) shows that the scale invariance of the CFT effective
action is broken due to the necessity to renormalized with a non-covariant
counterterm. The right hand side of equation (\ref{5.48}) is the integrated
Weyl anomaly. Within the AdS/CFT correspondence, $S_{fin}$\ is identified
with the effective action $W$\ of the boundary CFT. Hence, comparing Eqs. (%
\ref{5.48}) and (\ref{5.41}) yields the anomaly 
\begin{equation}
\mathcal{A}=-2\mathcal{L}_{c}+D_{i}J^{i},
\end{equation}
where $J^{i}$\ is continuous, but otherwise arbitrary.

\subsection{The Logarithmic Divergencies}

The action of gravity in the presence of an electromagnetic field can be
written as 
\begin{equation}
S_{\mathrm{bulk}}=-\frac{1}{16\pi }\int_{M}d^{n+1}x\sqrt{-g}\left( R+\frac{%
n(n-1)}{\ell ^{2}}-\frac{1}{4}F^{2}\right) .  \label{pform}
\end{equation}
In the neighborhood of a boundary $\partial M$, we will assume that the
metric can be expressed in the form 
\begin{equation}
ds^{2}=\frac{\ell ^{2}}{x^{2}}dx^{2}+\frac{1}{x^{2}}\tilde{\gamma}%
_{ij}dx^{i}dx^{j},  \label{asym}
\end{equation}
with the induced hypersurface metric $\gamma _{ij}=\tilde{\gamma}%
_{ij}/\epsilon ^{2}$ where $\epsilon \ll 1$. The nondegenerate\ metric $%
\tilde{\gamma}_{ij}$ \ admitting the expansion 
\begin{equation}
\tilde{\gamma}_{ij}=\tilde{\gamma}_{ij}^{0}+x^{2}\tilde{\gamma}%
_{ij}^{2}+x^{4}\tilde{\gamma}_{ij}^{4}+x^{6}\tilde{\gamma}_{ij}^{6}....\text{%
.}  \label{exp}
\end{equation}
Here $\tilde{\gamma}_{ij}^{0}$\ is the leading order of the metric $\tilde{%
\gamma}_{ij}.$\ If $M$\ is an Einstein manifold with negative cosmological
constant, then according to \cite{Graham, FG} such an expansion always
exists. For solutions of gauged supergravity theories with matter fields,
demanding that this expansion is well-defined as $x\rightarrow 0,$\ will
impose conditions on the matter fields induced on $\partial M$. The
equations of motion derived from the action (\ref{pform}) are \cite{Rob} 
\begin{eqnarray}
R_{mn} &=&-\frac{n}{\ell ^{2}}g_{mn}+\frac{1}{2}%
F_{mq_{1}..q_{p-1}}F_{n}^{q_{1}..q_{p-1}}-\frac{1}{4(n-1)}F^{2}g_{mn}; 
\notag \\
R &=&-\frac{n(n+1)}{\ell ^{2}}+\frac{(n-3)}{4(n-1)}F^{2}.  \label{eqms}
\end{eqnarray}
Using Eq. (\ref{eqms}), it is easy to show that the on-shell bulk action can
be written as: 
\begin{equation}
S_{\mathrm{bulk}}=\frac{1}{8\pi }\int_{M}d^{n+1}x\sqrt{-g}(\frac{n}{\ell ^{2}%
}+\frac{1}{4(n-1)}F^{2}).  \label{action3}
\end{equation}
Let us assume that in the vicinity of the conformal boundary the field can
be expanded as a power series in $x$ as 
\begin{equation}
F_{\mu \nu }=F_{\mu \nu }^{0}+xdx_{[\mu }A_{\nu ]}^{1}+x^{2}F_{\mu \nu
}^{2}+x^{2}dx_{[\mu }A_{\nu ]}^{2}+x^{3}F_{\mu \nu }^{3}...,  \label{exp1}
\end{equation}
where $G_{\mu }^{i}\ $is a tensor dependent only on $x^{i}$ and $F_{\mu \nu
}^{0}\ $is the leading order of $F_{\mu \nu }$ or, in other word, is the
induced field on the boundary. In order to write down the explicit form of
the action, we have to behave as follow \cite{Rob}:$\ $

Step 1. Expand out the $\sqrt{-g}$ and $F^{2}$ as leading order of $\tilde{%
\gamma}_{ij}^{0}$\ and $F^{0}$.

Step 2. Classify the expansion of $\sqrt{-g}$ in two part, one without field
($F=0$) depending only on $\tilde{\gamma}_{ij}^{0}$, $\sqrt{-g}^{(1)}$, and
the other, $\sqrt{-g}^{(2)}$, which depend only on $F^{0}$.

By substituting the explicit form for the metric and the field strength as
functions of $\tilde{\gamma}^{0}$\ and $F_{\mu \nu }^{0}$ in Eq. (\ref%
{action3}), we find that one encounters with two kinds of divergent terms.
The first type of these divergent terms depend only on $\tilde{\gamma}^{0}$\
and its curvature invariants, given as: 
\begin{equation}
S^{(1)}=-\frac{(n-1)}{8\pi \ell \epsilon ^{n}}\int d^{n}x\sqrt{-\tilde{\gamma%
}^{0}}-\frac{(n-4)(n-1)\ell }{16\pi (n^{2}-4)\epsilon ^{n-2}}\int d^{n}x%
\sqrt{-\tilde{\gamma}^{0}}R^{0}+...  \label{usu}
\end{equation}
which can be removed (at least for $n$\ odd) by the counterterms given in
Sec.(\ref{explicit}). The second type of divergent terms in the bulk action
is 
\begin{equation}
S_{\mathrm{bulk}}^{(2)}=\frac{\ell }{8\pi }\int \frac{d^{n+1}x}{x^{n+1}}%
\sqrt{-\tilde{\gamma}^{0}}\{x^{4}\frac{(n+8)}{32(n-1)}(F_{\mu \nu
}^{0})^{2}+...\},
\end{equation}
and in the surface action is 
\begin{equation}
S_{\mathrm{surf}}=-\frac{1}{8\pi }\int d^{n}x\sqrt{-\tilde{\gamma}^{0}}\{%
\frac{(n-4)\ell }{32(n-1)\epsilon ^{n-4}}(F_{\mu \nu }^{0})^{2}+...\}.
\end{equation}
Hear $R^{0}$, $R_{ij}^{0}$\ and $R_{ijkl}^{0}$\ are the Ricci scalar, Ricci
tensor and Riemann tensor respectively of the metric $\tilde{\gamma}%
_{ij}^{0} $. There are no divergences for $n<4$. In $n=4$\ there is a
logarithmic divergence due to the Weyl anomaly term \cite{Hening-Sk} 
\begin{equation}
S_{\mathrm{log}}=-\frac{\ell ^{3}}{64\pi }\ln \epsilon \int d^{4}x\sqrt{-%
\tilde{\gamma}^{0}}\left[ (R_{ij}^{0})^{2}-\frac{1}{3}(R^{0})^{2}\right] ,
\label{Ano4}
\end{equation}
and an additional logarithmic divergence in the action given by 
\begin{equation}
S_{\mathrm{log}}^{\mathrm{em}}=\frac{\ell }{64\pi }\ln \epsilon \int d^{4}x%
\sqrt{-\tilde{\gamma}^{0}}(F_{\mu \nu }^{0})^{2}.  \label{logem4}
\end{equation}
In $n>4$\ the $F_{\mu \nu }^{0}$\ will cause a power law divergence in the
action \cite{Rob} 
\begin{equation}
S_{\mathrm{div}}=-\frac{\ell }{256\pi \epsilon ^{n-4}}\frac{(n-8)}{(n-4)}%
\int d^{n}x\sqrt{-\tilde{\gamma}^{0}}(F_{\mu \nu }^{0})^{2},  \label{logemd}
\end{equation}
which can be removed by a counterterm of the form 
\begin{equation}
S_{\mathrm{ct}}=\frac{\ell }{256\pi }\int d^{n}x\sqrt{-\gamma }\frac{(n-8)}{%
(n-4)}(F_{\mu \nu })^{2}.  \label{Actct2}
\end{equation}

In $n=6$\ as well as the logarithmic divergence associated with the Weyl
anomaly of the dual theory, which is given by \cite{Rob} 
\begin{eqnarray}
S_{\mathrm{log}} &=&\frac{\ell ^{3}}{8^{4}\pi }\ln \epsilon \int d^{6}x\sqrt{%
-\tilde{\gamma}^{0}}\{\frac{3}{50}(R^{0})^{3}+R^{(0)ij}R^{(0)kl}R_{ijkl}^{0}-%
\frac{1}{2}R^{0}R^{(0)ij}R_{ij}^{0}  \notag \\
&&+\frac{1}{5}R^{(0)ij}D_{i}^{0}D_{j}^{0}R^{0}-\frac{1}{2}R^{(0)ij}\Box
^{0}R_{ij}^{0}+\frac{1}{20}R^{0}\Box ^{0}R^{0}\},  \label{Ano6}
\end{eqnarray}
as was found in \cite{Hening-Sk}, there is an anomaly of the form 
\begin{eqnarray}
S_{\mathrm{log}}^{\mathrm{em}} &=&\frac{\ell ^{3}}{8\pi }\ln \epsilon \int
d^{6}x\sqrt{-\tilde{\gamma}^{0}}\{\frac{1}{16}R^{0}(F_{2}^{0})^{2}-\frac{1}{8%
}R^{(0)ij}(F_{2}^{0})_{i}^{\;l}(F_{2}^{0})_{jl}  \notag \\
&&+\frac{1}{64}(F_{2}^{0})^{ij}\left[
D_{j}^{(0)}D^{(0)k}F_{ki}^{0}-D_{i}^{(0)}D^{(0)k}F_{kj}^{0}\right] \},
\label{logem6}
\end{eqnarray}
due to field anomaly.

\section{Conserved Charges and the AdS/CFT Correspondence}

One of the important applications of AdS/CFT correspondence is the
computation of conserved charges for the spacetimes. In this section, we
give the formalism of calculating the conserved charges of a spacetime
through the use of Brown and York formalism and AdS/CFT correspondence.

The start point for this subject is to write the spacetime metric as the
usual ADM decomposition \cite{ADM}: 
\begin{equation}
ds^{2}=g_{\mu \nu }dx^{\mu }dx^{\nu
}=-N^{2}dt^{2}+h_{ij}(dx^{i}+V^{i}dt)(dx^{j}+V^{j}dt),
\end{equation}
where $N$ is the laps function and $V^{i}$ is the shift vector (see Appendix %
\ref{boundary}). Throughout this analysis, it is assumed that the
hypersurface foliation $\Sigma $ is orthogonal to $^{3}B,$ meaning that on
the boundary $^{3}B$ the hyper-surface normal $u^{\mu }$ and the
three-boundary normal $n^{\mu }$ satisfy $(u.n)\mid _{^{3}B}=0$ (see Appendix%
\ref{boundary})$.$ In the canonical formalism, the boundary $B$ is specified
as a fixed surface in $\Sigma $. The Hamiltonian must evolve the system in a
manner consistent with the presence of this boundary, and cannot generate
transformations that map the canonical variables across $B$. This means that
the component of the shift vector normal to the boundary must be restricted
to vanish, $V^{i}n_{i}\mid _{B}=0$. From a spacetime point of view, this is
the condition that the two--boundary evolves into a three--surface that
contains the unit normal $u^{\mu }$ to the hypersurfaces $\Sigma $.
Therefore, $u^{\mu }$ and $n^{\mu }$ are orthogonal on $^{3}B$. The metric
on $^{3}B$ can be decomposed as 
\begin{equation}
\gamma _{ij}dx^{i}dx^{j}=-N^{2}dt^{2}+\sigma
_{ab}(dx^{a}+V^{a}dt)(dx^{b}+V^{b}dt)\ ,
\end{equation}
where $x^{a}$ ($a=1,2$) and $\sigma _{ab}$ are coordinates and induced
metric on $B$ respectively. The extrinsic curvature of $B$ as a surface
embedded in $\Sigma $ is denoted by $k_{ab}$. These tensors can be viewed as
spacetime tensors $\sigma _{\mu \nu }$ and $k_{\mu \nu }$, or as tensors on $%
\Sigma $ or $^{3}B$ by using indices $i$, $j$, $k$, $l$. Also, $\sigma _{\nu
}^{\mu }$ is the projection tensor onto $B$.

Now we define proper energy surface density $\varepsilon $\ which is the
projection of stress tensor normal to the space like surface $B,$ while
proper momentum density $j_{a}$ and spatial stress $s^{ab}$\ are the
normal--tangential and tangential--tangential projections of the stress
tensor respectively. On base of these definitions the variation of
gravitational action on the three-boundary $^{3}B$ can be written as 
\begin{equation}
\delta S_{\mathrm{cl}}=\int d^{3}x\sqrt{\sigma }(-\varepsilon \delta
N+j_{a}\delta V^{a}+\frac{1}{2}Ns^{ab}\delta \sigma _{ab})\delta \gamma
_{ij}.
\end{equation}
Then on the two-surface $B,$ we have 
\begin{eqnarray}
\varepsilon &\equiv &-\frac{1}{\sqrt{\sigma }}\frac{\delta S_{\mathrm{cl}}}{%
\delta N}=u_{i}u_{j}\mathcal{T}^{ij},  \label{surfen} \\
\text{\ }j_{a} &\equiv &\frac{1}{\sqrt{\sigma }}\frac{\delta S_{\mathrm{cl}}%
}{\delta V^{a}}=-\sigma _{ai}u_{j}\mathcal{T}^{ij},  \label{surfang} \\
s^{ab} &\equiv &\frac{2}{\sqrt{-\gamma }}\frac{\delta S_{\mathrm{cl}}}{%
\delta \sigma _{ab}}=\sigma _{i}^{a}\sigma _{j}^{a}\mathcal{T}^{ij},
\label{spatstres}
\end{eqnarray}
where the second equality in Eqs. (\ref{surfen})-(\ref{spatstres}) follow
from definition (\ref{stress1}) for $\mathcal{T}^{ij}$ and the relationships 
\begin{eqnarray}
\frac{\partial \gamma _{ij}}{\partial N} &=&-2\frac{u_{i}u_{j}}{N}, \\
\frac{\partial \gamma _{ij}}{\partial V^{a}} &=&-2\frac{\sigma a_{a(i}u_{j)}%
}{N}, \\
\frac{\partial \gamma _{ij}}{\partial \sigma _{ab}} &=&\sigma _{i}^{a}\sigma
_{j}^{a}.
\end{eqnarray}
The total quasilocal energy of the system is 
\begin{equation}
E=\int_{B}d^{2}x\sqrt{\sigma }\varepsilon ,
\end{equation}
which can be meaningfully associated with the thermodynamic energy of the
system \cite{Creighton}.

When there is a Killing vector field $\xi $ on the boundary $^{3}B,$ an
associated conserved charge is defined by \cite{BrownY} 
\begin{equation}
\mathcal{Q}(\xi )=-\int_{B}d^{2}x\sqrt{\sigma }(u_{i}\mathcal{T}^{ij}\xi
_{j}).  \label{conscharg}
\end{equation}
From Eqs. (\ref{surfen})-(\ref{spatstres}) one can see that $-u_{i}\mathcal{T%
}^{ij}=(\varepsilon u^{i}+j^{i}).$ Hence we have 
\begin{equation}
\mathcal{Q}(\xi )=-\int_{B}d^{2}x\sqrt{\sigma }(\varepsilon u^{i}+j^{i})\xi
_{i}.  \label{conscharge}
\end{equation}
For boundaries with time like Killing vector ($\xi =\partial /\partial t$)
and rotational Killing vector fields ($\varsigma =\partial /\partial \phi $)
we obtain 
\begin{eqnarray}
M &=&\int_{B}d^{2}x\sqrt{\sigma }(\varepsilon u^{i}+j^{i})\xi _{i},
\label{consmass} \\
J &=&\int_{B}d^{2}x\sqrt{\sigma }j^{i}\varsigma _{i},  \label{consang}
\end{eqnarray}
provided the surface $B$ contains the orbits of $\varsigma .$ These
quantities are respectively the conserved mass and angular momentum of the
system enclosed by the boundary.

By computing $\varepsilon $ and $j_{a}$, one can find the conserved mass and
angular momentum. But for infinite spacetimes, the surface stress
energy-momentum tensor diverges. To solve this problem, one may use the
AdS/CFT correspondence conjecture and obtain the finite surface stress
energy-momentum tensor as obtained in the Eq. (\ref{finitestress}). Examples
of this kind of computations will be given in chapters \ref{Lemosch}\ and %
\ref{Khodam}.

\chapter{Thermodynamics of Black Holes\label{thermodynamics}\protect\bigskip}

The defining feature of a black hole is its future event horizon or\ shortly
its event horizon. This one way membrane separates events which are inside
the horizon from those that are outside the horizon; the events inside the
event horizon are never within the causal past of the ones outside. This
feature can be seen in figure (\ref{fig3.1}), which depicts the formation of
a black hole by the spherical collapse of a star. The conformal
transformation that brings points at infinity to a finite distance preserves
the light-cone structure so that light rays travelling radially inwards or
outwards travel on lines inclined by $45^{\circ }$; these rays are called
null. In Fig. (\ref{fig3.1}), the points at infinity include timelike and
spacelike infinities (i$^{+}$, i$^{-}$\ and i$^{0}$), and null infinities ($%
I^{+}$\ and \ $I^{0}$). Each point represents a sphere\textsl{.} Notice that
the event horizon in figure (\ref{fig3.1}) is a null surface, so the events
within it are never in the past light-cone of events outside of the horizon.
However, charged black holes have two inner and outer horizons. In this
case, one may encounter with an extreme black hole with a degenerate
horizon, if one choose the black hole's parameters properly. The presence of
an event horizon causes some unusual effects on quantum fields existing in
the black hole spacetime. By acting as a one-way membrane, the event horizon
can trap one of the `virtual' particle-pairs produced by quantum processes.
The escaping particle (which is no longer virtual) appears to have been
radiated from the black hole. This process is shown in figure (\ref{fig3.2}).%
\FRAME{ftbpFU}{2.6948in}{3.1531in}{0pt}{\Qcb{\textsl{Conformal diagram of a
black hole formed from the spherical collapse of a star.}}}{\Qlb{fig3.1}}{%
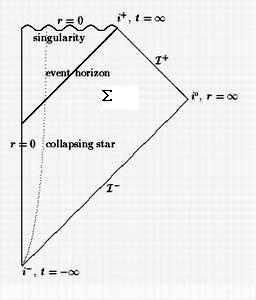}{\raisebox{-3.1531in}{\includegraphics[height=3.1531in]{fig31.jpg}}}\FRAME{ftbpFU}{3.1704in}{1.8922in}{0pt%
}{\Qcb{\textsl{Pair production (right) and (left) in `Hawking radiation'.}}}{%
\Qlb{fig3.2}}{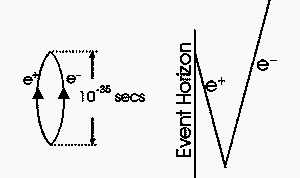}{\raisebox{-1.8922in}{\includegraphics[height=1.8922in]{fig32.jpg}}}

The radiation, known as Hawking radiation, has a thermal spectrum (if one
neglects scattering of the gravitational field) with a temperature
proportional to the surface gravity of the event horizon. Although this
picture is drawn from semi-classical quantum field theory, it should be
qualitatively correct since the gravitational field at the event horizon of
a black hole need not be very strong.

Another strange property of black holes was realized by Wheeler: one can
forever hide information from the outside world by dropping it into a black
hole \footnote{%
Recently, S. W. Hawking in `17th International Conference on General
Relativity and Gravitation in Dublin' said that: black holes, the mysterious
massive vortexes formed from collapsed stars, do not destroy everything they
consume but instead eventually fire out matter and energy ``in a mangled
form.'' Also he told ``There is no baby universe branching off, as I once
thought. The information remains firmly in our universe.'' Then he said
``I'm sorry to disappoint science fiction fans, but if information is
preserved, there is no possibility of using black holes to travel to other
universes,'' he said. ``If you jump into a black hole, your mass energy will
be returned to our universe, but in a mangled form, which contains the
information about what you were like, but in an unrecognizable state.'' He
added: `` information is not lost in the formation and evaporation of black
holes. The way the information gets out seems to be that a true event
horizon never forms, just an apparent horizon.'' By these statements, in
spite of the old idea about black holes, the information does not hide in
black holes and black holes never fully evaporate. `\emph{The information is
preserved}'.}. Indeed, this property seems to pose a problem with the second
law of thermodynamics since a black hole will quickly return to a very
simple state even if an object with a large amount of entropy is dropped
into it. Bekenstein \cite{Beken1, Beken2} speculating that the black hole
itself is a thermodynamic system with the area of the event horizon
representing the entropy. Since the area always increases when matter of
positive energy is added to a black hole, the problem with the second law of
thermodynamics can be resolved. The fact that the black hole is also
surrounded by quantum fields with a thermal spectrum seems to support
Bekenstein's speculation. Various attempts for understanding the entropy of
black holes in terms of the number of quantum states contained within or
near the event horizon have been made (two recent reviews are given in
references \cite{Beken3, Frolov95}). Therefore, black holes are interesting
systems to study, if only theoretically, as their classical and
semi-classical properties may hint at the nature of a quantum theory of
gravity. Black holes can be treated as a thermodynamic system whose
properties must be reproduced in the statistics of the quantum fields.
However, the thermodynamic properties of black holes must first be
understood.

\section{Classical Black Hole Thermodynamics}

In this section, using reference \cite{Townsend}, we will give a brief
review of the null hypersurfaces and Killing horizons. Also the laws of
classical black hole mechanics will be described.

\subsection{The Null Hypersurfaces and Killing Horizons}

Let $S(x)$\ be a smooth function of the spacetime coordinates $x^{\mu }$\
and consider a family of hypersurfaces $S=\mathrm{const}$. The vector fields
normal to the hypersurface are 
\begin{equation}
l=\tilde{f}(x)(\partial ^{\mu }S)\frac{\partial }{\partial x^{\mu }},
\end{equation}
where $\tilde{f}$\ \ is an arbitrary non-zero function. If $l^{2}=0$\ for a
particular hypersurface $\mathcal{N}$ in the family, then $\mathcal{N}$\ is
said to be a null hypersurface. For example the surface $r=2M$\ is a null
hypersurface for Schwarzschild spacetime. In the null hypersurfaces, $l$\ is
itself a tangent vector.

\begin{definition}
A null hypersurface $\mathcal{N}$ is a Killing horizon of a Killing vector
field $\xi $ if, on $\mathcal{N}$, $\xi $ is normal to $\mathcal{N}$.
\end{definition}

Let $l$\ be normal to $\mathcal{N}$\ such that $l\cdot Dl^{\mu }=0$\ (affine
parameterization). Then, since, on $\mathcal{N}$, 
\begin{equation}
\xi =fl
\end{equation}
for some function $f$, it follows that 
\begin{eqnarray}
\xi \cdot D\xi ^{\mu } &=&\xi \cdot D(fl^{\mu })  \notag \\
&=&\frac{\xi \cdot Df}{f}fl^{\mu }=\kappa \xi ^{\mu },\hspace{1.0228pc}\text{%
on }\mathcal{N}\text{ }
\end{eqnarray}
where $\kappa =\xi \cdot \partial \ln \left| f\right| $\ is called the
surface gravity. In this way, the surface gravity can be obtained from 
\begin{equation}
\kappa ^{2}=-\frac{1}{2}\left( D^{\mu }\xi ^{\nu }\right) \left( D_{\mu }\xi
_{\nu }\right) \mid _{\mathcal{N}}.  \label{kapa2}
\end{equation}

\begin{proposition}
$\kappa $\ is constant on orbits of $\xi $.{\large \ \newline
}
\end{proposition}

\begin{proof}
Let $t$\ be tangent to $\mathcal{N}$. Then, since (\ref{kapa2}) is valid everywhere
on $\mathcal{N}$\  
\begin{equation}
t\cdot \partial \kappa ^{2}=-\left( D^{\mu }\xi ^{\nu }\right) t^{\rho
}D_{\rho }D_{\mu }\xi _{\nu }\mid _{\mathcal{N}}  \label{Propos1}
\end{equation}
but for Killing vector $\xi ^{\nu }$\ we have  
\begin{equation}
D_{\rho }D_{\mu }\xi ^{\nu }=R_{\hspace{0.2046pc}\mu \rho \sigma }^{\nu 
\hspace{0.2046pc}}\xi ^{\sigma }  \label{Killing}
\end{equation}
where $R_{\hspace{0.2054pc}\mu \rho \sigma }^{\nu \hspace{0.2054pc}}$\ is
the Riemann tensor. Hence Eq. (\ref{Propos1}) can be written as  
\begin{equation}
t\cdot \partial \kappa ^{2}=-\left( D^{\mu }\xi ^{\nu }\right) t^{\rho
}R_{\nu \mu \rho }^{\hspace{1.0228pc}\sigma }\xi _{\sigma }.
\end{equation}
Now, $\xi $\ is tangent to $\mathcal{N}$\ (in addition to being normal to it).
Choosing $t=\xi $\ we have  
\begin{eqnarray}
\xi \cdot \partial \kappa ^{2} &=&-\left( D^{\mu }\xi ^{\nu }\right) R_{\nu
\mu \rho \sigma }^{\hspace{1.0266pc}}\xi ^{\rho }\xi ^{\sigma }  \notag \\
&=&0\hspace{1.0228pc}\text{(since }R_{\nu \mu \rho \sigma }^{\hspace{1.0304pc%
}}=-R_{\nu \mu \sigma \rho }^{\hspace{1.0343pc}}\text{)}
\end{eqnarray}
so $\kappa $\ is constant on orbits of $\xi .$
\end{proof}

For surface gravity, also there is an interpretation as follow:

\begin{quote}
Surface gravity of the event horizon can be thought as the force required to
hold a unit test mass on the event horizon in place by an observer who is
far from the black hole .
\end{quote}

A bifurcate Killing horizon is a pair of null surfaces, $\mathcal{N}_{A}$\
and $\mathcal{N}_{B}$, which intersect on a spacelike 2-surface, $C$\
(called the `bifurcation surface'), such that $\mathcal{N}_{A}$\ and $%
\mathcal{N}_{B}$\ are each Killing horizons with respect to the same Killing
field $\xi ^{a}$. It follows that $\xi ^{a}$\ must vanish on $C$;
conversely, if a Killing field, $\xi ^{a}$, vanishes on a two-dimensional
spacelike surface, $C$, then $C$\ will be the bifurcation surface of a
bifurcate Killing horizon associated with $\xi ^{a}$\ (see \cite{Waldbook1}
for further discussion).

\begin{proposition}
If $\mathcal{N}$\ is a bifurcate Killing horizon of $\xi $, with bifurcation
2-surface, $C$, then $\kappa ^{2}$\ is constant on $\mathcal{N}$.
\end{proposition}

\begin{proof}
$\kappa ^{2}$\ is constant on each orbit of $\xi $. The value of this
constant is the value of $\kappa ^{2}$\ at the limit point of the orbit on $C
$, so $\kappa ^{2}$\ is constant on $\mathcal{N}$\ if it is constant on $C$. But we
saw previously that  
\begin{eqnarray}
t\cdot \partial \kappa ^{2} &=&-\left( D^{\mu }\xi ^{\nu }\right) t^{\rho
}R_{\nu \mu \rho }^{\hspace{1.019pc}\sigma }\xi _{\sigma }\mid _{\mathcal{N}}
\notag \\
&=&0\text{ \ \ \ on }\mathcal{C}\text{\ (since }\xi _{\sigma }\mid _{%
\mathcal{C}}=0\text{).}
\end{eqnarray}
Since $t$\ can be any tangent to $C$, $\kappa ^{2}$\ is constant on $C$, and
hence on $\mathcal{N}.$
\end{proof}

\subsection{Conserved Charges}

Let $V$\ be a volume of spacetime on a spacelike hypersurface $\Sigma $,
with boundary $B$. To every Killing vector field $\xi ,$ we can associate
the Komar integral \cite{DeWitt} 
\begin{equation}
Q_{\xi }(V)=\frac{c}{16\pi }\oint_{B}dS_{\mu \nu }D^{\mu }\xi ^{\nu }
\end{equation}
for some constant $c.$\ Using Gauss' law 
\begin{equation}
Q_{\xi }(V)=\frac{c}{8\pi }\int_{V}dS_{\mu }D_{\nu }D^{\mu }\xi ^{\nu },
\end{equation}
and the fact that 
\begin{equation}
D_{\nu }D_{\mu }\xi ^{\nu }=R_{\mu \nu }\xi ^{\nu },
\end{equation}
we obtain 
\begin{eqnarray}
Q_{\xi }(V) &=&\frac{c}{8\pi }\int_{V}dS_{\mu }R_{\hspace{0.4247pc}\nu
}^{\mu }\xi ^{\nu }  \notag \\
&=&\frac{c}{8\pi }\int dS_{\mu }(T_{\hspace{0.4247pc}\nu }^{\mu }\xi ^{\nu }-%
\frac{1}{2}T\xi ^{\mu })\text{ \ \ \ \ \ \ \ \ \ \ (by Einstein's equation)}
\notag \\
&=&\frac{1}{8\pi }\int dS_{\mu }J^{\mu }(\xi ),
\end{eqnarray}
where $J^{\mu }(\xi )$ defined as: 
\begin{equation}
J^{\mu }(\xi )=c(T_{\hspace{0.4263pc}\nu }^{\mu }\xi ^{\nu }-\frac{1}{2}T\xi
^{\mu }).
\end{equation}
One can verify that the current $J^{\mu }(\xi )$ is conserved. To prove
this, using the fact that $D_{\mu }T^{\mu \nu }=0.$ We obtain 
\begin{equation}
D_{\mu }J^{\mu }(\xi )=c\left( T^{\mu \nu }D_{\mu }\xi ^{\nu }-\frac{1}{2}%
TD_{\mu }\xi ^{\mu }\right) -\frac{c}{2}\xi \cdot \partial T.
\end{equation}
Since for Killing vector $D_{\mu }\xi ^{\nu }=D_{\mu }\xi ^{\mu }=0,$ then 
\begin{equation}
D_{\mu }J^{\mu }(\xi )=\frac{c}{2}\xi \cdot \partial T.\text{\ \ }
\end{equation}
Now using Einstein's equation, one can show that $D_{\mu }J^{\mu }(\xi
)\varpropto \xi ^{\mu }\partial _{\mu }R$ which is zero.

Since $J^{\mu }(\xi )$\ is a `conserved current', the charge $Q_{\xi }(V)$\
is time-independent provided $J^{\mu }(\xi )$\ vanishes on $B$.

If one chooses time-translation Killing vector field $\xi =k,$\ then one can
obtain quasi-local energy in the volume $V$\ by 
\begin{equation}
E(V)=-\frac{1}{8\pi }\oint_{B}dS_{\mu \nu }D^{\mu }k^{\nu }.
\end{equation}
For Schwarzschild spacetime one can obtain $E(V)=M$\ for any $V$\ with $B$\
outside the horizon.

In axisymmetric spacetimes, by choosing $\xi =m=\partial /\partial \phi $\
and $c=1$, we can calculate the angular momentum as 
\begin{equation}
J(V)=\frac{1}{16\pi }\oint_{B}dSD^{\mu }m^{\nu }.
\end{equation}

It is worthwhile to mention that the Komar integral formalism is valid only
for asymptotically flat spacetimes. For asymptotically\ AdS or dS
spacetimes, one should use the formalism which was given in chapter \ref%
{Ads/CFT}.

\subsection{ The Laws of Black Hole Mechanics}

Previously, we showed that $\kappa ^{2}$\ is constant on a bifurcate Killing
horizon. The proof fails if we have only part of a Killing horizon, without
the bifurcation 2-sphere, as happens in gravitational collapse. However in
this case, if one accepts some conditions which will be discussed, then we
have the following laws.

\subsubsection{Zeroth Law:}

The surface gravity $\kappa $\ is constant on the event horizon $H$ if $%
T_{\mu \nu }$ obeys the dominant energy condition. This resembles the zeroth
law of thermodynamics, which says that the temperature is constant in
thermodynamic equilibrium.

\subsubsection{Smarr's Formula:}

Let $\Sigma $\ be a spacelike hypersurface in a stationary exterior black
hole spacetime with an inner boundary, $H$, on the event horizon and another
boundary at $i_{0}$\ (see figure(\ref{fig3.1})). The surface $H$\ is a
2-sphere that can be considered as the `boundary' of the black hole.

Applying Gauss' law to the Komar integral for $J$\ we have 
\begin{eqnarray}
J &=&\frac{1}{8\pi }\int_{\Sigma }ds_{\mu }D_{\nu }D^{\mu }m^{\nu }+\frac{1}{%
16\pi }\oint_{H}ds_{\mu \nu }D^{\mu }m^{\nu }  \notag \\
&=&\frac{1}{8\pi }\int_{\Sigma }ds_{\mu }R_{\hspace{0.4135pc}\nu }^{\mu
}m^{\nu }+J_{H}.
\end{eqnarray}
Using Einstein equation, one obtains: 
\begin{equation}
J=\int_{\Sigma }ds_{\mu }(T_{\hspace{0.4135pc}\nu }^{\mu }m^{\nu }-\frac{1}{2%
}Tm^{\mu })+J_{H}.
\end{equation}
In the absence of matter other than electromagnetic field, we have 
\begin{equation}
T_{\mu \nu }=T_{\mu \nu }(F)=\frac{1}{4\pi }(F_{\mu \lambda }F_{\hspace{%
0.3055pc}\nu }^{\lambda }-\frac{1}{4}F^{\rho \lambda }F_{\rho \lambda
}g_{\mu \nu }),
\end{equation}
where 
\begin{equation}
F_{\mu \nu }=\partial _{\mu }A_{\nu }-\partial _{\nu }A_{\mu }.
\end{equation}
\ Since $g^{\mu \nu }T_{\mu \nu }(F)=T(F)=0,$\ we have 
\begin{equation}
J=\int_{\Sigma }ds_{\mu }T_{\hspace{0.4151pc}\nu }^{\mu }(F)m^{\nu }+J_{H}
\label{6.64}
\end{equation}
for an isolated black hole (i.e. $T_{\mu \nu }=$\ $T_{\mu \nu }(F)$).

Now apply Gauss' law to the Komar integral for the total energy (= mass). 
\begin{equation}
M=-\frac{1}{4\pi }\int_{\Sigma }ds_{\mu }R_{\hspace{0.4295pc}\nu }^{\mu
}k^{\nu }-\frac{1}{8\pi }\oint_{H}ds_{\mu \nu }D^{\mu }k^{\nu }.
\end{equation}
By inserting 
\begin{equation}
\xi =k+\Omega _{H}m,
\end{equation}
we have 
\begin{equation}
M=\int_{\Sigma }ds_{\mu }(-2T_{\hspace{0.4311pc}\nu }^{\mu }k^{\nu }+Tk^{\mu
})-\frac{1}{8\pi }\oint_{H}ds_{\mu \nu }(D^{\mu }\xi ^{\nu }-\Omega
_{H}D^{\mu }m^{\nu })
\end{equation}
since $\Omega _{H}$\ is constant on $H,$\ for $T_{\mu \nu }=T_{\mu \nu }(F)$%
\ we have 
\begin{equation}
M=-2\int_{\Sigma }ds_{\mu }T_{\hspace{0.4327pc}\nu }^{\mu }(F)k^{\nu
}+2\Omega _{H}J_{H}-\frac{1}{8\pi }\oint_{H}ds_{\mu \nu }D^{\mu }\xi ^{\nu }.
\end{equation}
Using (\ref{6.64}) for an isolated black hole, 
\begin{equation}
M=-2\int_{\Sigma }ds_{\mu }T_{\hspace{0.4343pc}\nu }^{\mu }(F)\xi ^{\nu
}+2\Omega _{H}J-\frac{1}{8\pi }\oint_{H}ds_{\mu \nu }D^{\mu }\xi ^{\nu }.
\end{equation}
The first term can be written as: 
\begin{equation}
-2\int_{\Sigma }ds_{\mu }T_{\hspace{0.4359pc}\nu }^{\mu }(F)\xi ^{\nu }=\Phi
_{H}Q,
\end{equation}
and for the third term we have \cite{Townsend} 
\begin{equation}
-\frac{1}{8\pi }\oint_{H}ds_{\mu \nu }D^{\mu }\xi ^{\nu }=\frac{\kappa A}{%
4\pi },
\end{equation}
where $A$\ is the `area of the horizon'. Also $\Phi _{H}$ and $Q$\ are the
co-rotating electric potential on the horizon and electrical charge, which
are defined by 
\begin{equation}
\Phi =\xi ^{\nu }A_{\nu },
\end{equation}
\begin{equation}
Q=\int_{\Sigma }ds_{\mu }D_{\nu }F^{\mu \nu }=\oint_{\partial \Sigma
}F^{0i}ds_{i}.
\end{equation}
\ Hence 
\begin{equation}
M=\frac{\kappa A}{4\pi }+2\Omega _{H}J+\Phi _{H}Q.  \label{Smarr}
\end{equation}
This is Smarr's formula for the mass of charged rotating black holes.

\subsubsection{First Law:}

If a stationary black hole of mass $M$, charge $Q$\ and angular momentum $J$%
, with event horizon of area $A$, surface gravity $\kappa $, electric
surface potential $\Phi _{H}$\ and angular velocity $\Omega _{H}$, is
perturbed such that it settles down to another black hole with mass $%
M+\delta M$\ charge $Q+\delta Q$\ and angular momentum $J+\delta J$, then
the conservation of energy requires that 
\begin{equation}
\delta M=\frac{\kappa \delta A}{8\pi }+\Omega _{H}\delta J+\Phi _{H}\delta Q.
\label{firstlaw}
\end{equation}
This has the same form as the first law of thermodynamics, and since $\kappa 
$\ is the analog of temperature, the area plays the role of entropy. This
statement of the first law uses the fact that the event horizon of a
stationary black hole must be a Killing horizon.

To prove Eq. (\ref{firstlaw}), for $Q=0,$ we use the Smarr's formula for
mass, Eq. (\ref{Smarr}), then by Euler's theorem for homogeneous function $%
M(A,J)$, we have 
\begin{eqnarray}
A\frac{\partial M}{\partial A}+J\frac{\partial M}{\partial J} &=&\frac{1}{2}M
\notag \\
&=&\frac{\kappa }{8\pi }A+\Omega _{H}J
\end{eqnarray}

At the first step of this calculation, we use the fact that $A$\ and $J$\
both have dimension of $M^{2}$\ so the function $M(A,J)$\ must be
homogeneous of degree $1/2,$ and at the second step we use Eq. (\ref{Smarr})
Therefore 
\begin{equation}
A\left( \frac{\partial M}{\partial A}-\frac{\kappa }{8\pi }\right) +J\left( 
\frac{\partial M}{\partial J}-\Omega _{H}\right) =0.
\end{equation}
But $A$\ and $J$\ are free parameters so, 
\begin{equation}
\frac{\partial M}{\partial A}=\frac{\kappa }{8\pi },\text{ \ \ \ \ \ \ }%
\frac{\partial M}{\partial J}=\Omega _{H}.\text{\ \ }
\end{equation}
In the case $Q\neq 0,$ we can generalize this equation and write 
\begin{equation}
\frac{\partial M}{\partial Q}=\Phi _{H}.
\end{equation}
\medskip

\subsubsection{The Second Law (Hawking's Area Theorem):}

The analogy of area and entropy is confirmed by the second law of black hole
mechanics \cite{Hawking71}. This is a statement about non stationary
processes in a spacetime containing black hole, including collisions and
fusions of black holes. Two assumptions can be made:

\begin{enumerate}
\item The time evolution of the system must be under sufficient control.
This is implemented by requiring that the spacetime is strongly
asymptotically predictable (cosmic censorship hypothesis). In physical
formulation this explains that the complete gravitational collapse of a body
always results in a black hole rather than a naked singularity; i.e., all
singularities of gravitational collapse are `hidden' within horizon, and can
not be `seen' by distant observers.

\item The matter, represented by the stress energy tensor should behave
`reasonable'. This is done by imposing the `week energy condition' on the
stress energy tensor. For more detailed about energy conditions , we refer
to \cite{Waldbook1}.
\end{enumerate}

Under these assumptions the second law states that the total area of all
event horizons is non-decreasing, 
\begin{equation}
\delta A\geq 0.  \label{secondlaw}
\end{equation}
This is striking analogue of the entropy law of thermodynamics. For more
detail we refer to \cite{Townsend}.

\subsubsection{ The Third Law :}

Here several versions exist, and the status of this law does not seem to be
fully understood. We only touch upon this and refer to \cite{Waldbook1} for
a more detailed account. One version of the law states that the extreme
limit cannot be reached in finite time in any physical process. Here the
obvious problem is to define what a physical process is and to bring such
non-stationary processes under sufficient control. Another version, which
does not refer to non-stationary properties, states that black holes of
vanishing temperature (surface gravity) have vanishing entropy. This is in
obvious contradiction to the fact that the area of an extreme black hole can
be non-vanishing. There are however subtleties at the quantum level, and
these have been used as arguments in favor of the second version of the
third law. We will return to this when discussing quantum aspects of black
holes.

\section{Quantum Aspects of Black Holes and Black Hole Thermodynamics}

The laws of black hole mechanics have been known for quite some time, but
were mostly considered as a curious formal analogy. The most obvious reason
for not believing in a thermodynamic content is that a classical black hole
is just black: It cannot radiate and therefore one should assign temperature
zero to it, so that the interpretation of the surface gravity as temperature
has no physical content.

This changes dramatically when taking into account quantum effects. One can
analyze black holes in the context of quantum field theory in curved
backgrounds, where matter is described by quantum field theory while gravity
enters as a classical background, see for example \cite{Birrel}. In this
framework it was discovered that black holes can emit Hawking radiation \cite%
{Hawking1}. The spectrum is (almost) Planckian with a temperature, the
so-called Hawking temperature, which is indeed proportional to the surface
gravity, 
\begin{equation}
T_{H}=\frac{\kappa }{2\pi }.
\end{equation}
This motivates to take the analogy of area and entropy seriously. Since the
Hawking temperature fixes the factor of proportionality between temperature
and surface gravity, one finds the Bekenstein-Hawking area law, 
\begin{equation}
S=\frac{A}{4}.  \label{entropy}
\end{equation}
Before the discovery of Hawking radiation, Bekenstein had already given an
independent argument in favour of assigning entropy on black holes \cite%
{Bekenstein1, Bekenstein2}. He pointed out that in a spacetime containing a
black hole one could adiabatically transport matter into it. This reduces
the entropy in the observable world and thus violates the second law of
thermodynamics. He, therefore, proposed to assign entropy to black holes,
such that a generalized second law is valid, which states that the sum of
thermodynamic entropy and black hole entropy is non-decreasing. With the
discovery of Hawking radiation one can give an additional argument in favour
of this generalization: By Hawking radiation a black hole looses mass and
shrinks. This is not in contradiction with the second law of black hole
mechanics, because one can show that the weak energy condition is violated
in the near horizon region if the effect of quantum fields is taken into
account. Bekenstein's generalized second law claims that the loss in black
hole entropy is always (at least) compensated by the thermodynamic entropy
of the Hawking radiation, so that the total entropy is non-decreasing.

One example of unusual thermodynamic behavior of black holes is provided by
the mass dependence of the temperature of uncharged black holes. For the
Schwarzschild black hole one finds $\kappa =(4M)^{-1}$, which shows that the
specific heat is negative: The black hole heats up while loosing mass. This
behavior is unusual, but nevertheless not unexpected because gravity is a
purely attractive force. The fact that uncharged black holes seem to fully
decay into Hawking radiation leads to the information or unitarity problem
of quantum gravity, see for example \cite{Waldbook2}. Charged black holes
behave differently, since the Hawking temperature vanishes in the extreme
limit, and therefore extreme black holes are stable against decay by thermic
radiation.

We already mentioned that one version of the third law states that extreme
black holes have vanishing entropy. This statement depends on subtleties of
the quantum mechanical treatment of such objects \cite{Ghosh, Hooft}: The
entropy can be computed in semiclassical quantum gravity, i.e. by quantizing
gravity around a black hole configuration. One can use either the Euclidean
path integral formulation or a Minkowskian canonical framework. The result
for the entropy depends on whether the extreme limit is taken before or
after quantization: If one quantizes around extreme black holes the entropy
vanishes. But if one quantizes around general charged black hole
configurations, one finds an entropy that is non-vanishing when taking the
extreme limit. The second option seems to be more natural and it is the one
supported by string theory.

The identification of the area with entropy leads to several questions.
Standard thermodynamics provides a macroscopic effective description of
systems in terms of macroscopic observables like temperature and entropy. At
the fundamental microscopic level, systems are described by statistical
mechanics in terms of microstates which encode, say, the positions and
momenta of all particles that constitute the system. At this level one can
define the microscopic or statistical entropy as the quantity which
characterizes the degeneracy of microstates in a given macrostate, where the
macrostate is characterized by specifying the macroscopic observables.
Assuming ergodic behavior the macroscopic and microscopic entropy agree. One
should therefore address the question whether there exists a fundamental,
microscopic level of description of black holes, where one can identify
microstates and count how many of them lead to the same macrostate. The
macrostate of a black hole is characterized by its mass, charge and angular
momentum. Denoting the number of microstates leading to the same mass $M$,
charge $Q$ and angular momentum $J$ by $N(M,Q,J)$, the statistical or
microscopic black hole entropy is defined by

\begin{equation}
S_{micro}=\log N(M,Q,J).
\end{equation}
If the Bekenstein-Hawking entropy is the analogue of thermodynamic entropy
and if stationary black holes are the analogue of thermodynamic equilibrium
states, then the Bekenstein-Hawking entropy must coincide with the
microscopic entropy, 
\begin{equation}
S=S_{micro}.
\end{equation}

One of the astonishing properties of the Bekenstein-Hawking entropy is its
simple and universal behavior: the entropy is just proportional to the area.
The fact that the entropy is proportional to the area and not to the volume
has led to the speculation that quantum gravity \ is in some sense non-local
and admits a holographic representation on boundaries of spacetime \cite%
{Hooft, Susskind}.

\subsection{The Hawking Temperature}

As we mentioned in the previous sections, we can assign temperature to a
black hole. In order to calculate this temperature we start with the example
of the Schwarzschild metric. 
\begin{equation}
ds^{2}=-(1-\frac{2M}{r})dt^{2}+(1-\frac{2M}{r})^{-1}dr^{2}+r^{2}(d\theta
^{2}+\sin ^{2}\theta d\phi ^{2})
\end{equation}
This represents the gravitational field that a black hole would settle down
to, if it were non rotating. In the usual $r$\ and $t$\ coordinates there is
an apparent singularity at the Schwarzschild radius $r=2M$. However, this is
just caused by a bad choice of coordinates. One can choose other coordinates
in which the metric is regular there. If one puts $t=i\tau $\ one gets a
positive definite metric. We shall refer to such positive definite metrics
as Euclidean even though they may be curved. In the Euclidean-Schwarzschild
metric there is again an apparent singularity at $r=2M$. However, one can
define a new radial coordinate $x$\ to be $4M(1-2Mr^{-1})^{\frac{1}{2}}.$\
Then we have 
\begin{equation}
ds^{2}=x^{2}(\frac{d\tau }{4M})^{2}+(\frac{r^{2}}{4M^{2}})^{2}dx^{2}+r^{2}(d%
\theta ^{2}+\sin ^{2}\theta d\phi ^{2})
\end{equation}
\FRAME{ftbpFU}{4.7392in}{2.1534in}{0pt}{\Qcb{\textsl{Regularization of the
metric.}}}{\Qlb{fig3.4}}{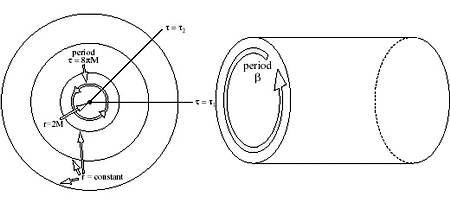}{\raisebox{-2.1534in}{\includegraphics[height=2.1534in]{fig34.jpg}}}where $\kappa =(4M)^{-1}$. The metric
in the $x-\tau $\ plane then becomes like the origin of polar coordinates if
one identifies the coordinate with period $8\pi M$. Similarly other
Euclidean black hole metrics will have apparent singularities on their
horizons which can be removed by identifying the imaginary time coordinate
with period $\beta =2\pi /\kappa .$

What is the significance of having imaginary time identified with some
period $\beta $? To see this, consider the amplitude to go from some field
configuration $\phi _{1}$\ on the surface $t_{1}$\ to a configuration $\phi
_{2}$\ on the surface $t_{2}$. This will be given by the matrix element of $%
\exp [iH(t_{2}-t_{1})].$\ However, one can also represent this amplitude as
a path integral over all fields between $t_{1}$\ and $t_{2}$\ which agree
with the given fields $\phi _{1}$\ and $\phi _{2}$\ on the two surfaces as
Fig.(\ref{fig3.3}).

\FRAME{ftbpFU}{3.2742in}{1.8602in}{0pt}{\Qcb{Surfaces $\protect\phi _{1}$
and $\protect\phi _{2}$ at constant times $t_{1}$ and $t_{2}.$}}{\Qlb{fig3.3}%
}{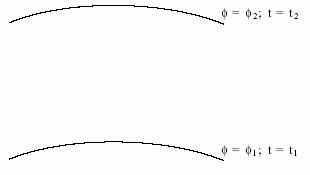}{\raisebox{-1.8602in}{\includegraphics[height=1.8602in]{fig33.jpg}}} 
\begin{eqnarray}
&<&\phi _{2},t_{2}\mid \phi _{1},t_{1}>=<\phi _{2}\mid \exp
(-iH(t_{2}-t_{1}))\mid \phi _{1}>  \notag \\
&=&\int D[\phi ]\exp (iS[\phi ])  \label{trans}
\end{eqnarray}
One now chooses the time separation $(t_{2}-t_{1})$\ to be pure imaginary
and equal to $\beta .$\ One also puts the initial field $\phi _{1}$\ equal
to the final field $\phi _{2}$\ and sums over a complete basis of states $%
\phi _{n}$. On the left, one has the expectation value of $e^{-\beta H}$\
summed over all states. This is just the thermodynamic partition function $Z$%
\ at the temperature $T=\beta ^{-1}.$\ On the right hand of the equation (%
\ref{trans}) one has a path integral. One puts $\phi _{1}=\phi _{2}$\ and 
\begin{equation*}
t_{2}-t_{1}=-i\beta ,\hspace{1.0738pc}\phi _{1}=\phi _{2}
\end{equation*}
\begin{eqnarray}
Z &=&\sum <\phi _{n}\mid \exp (-\beta H)\mid \phi _{n}>  \notag \\
&=&\int D[\phi ]\exp (-i\hat{I}[\phi ])
\end{eqnarray}
sums over all field configurations $\phi _{n}$. This means that effectively
one is doing the path integral over all fields $\phi $\ on a spacetime that
is identified periodically in the imaginary time direction with period $%
\beta $. Thus the partition function for the field $\phi $\ at temperature $%
T $\ \ is given by a path integral over all fields on a Euclidean spacetime.
This spacetime is periodic in the imaginary time direction with period $%
\beta =T^{-1}$. If one does the path integral in flat spacetime identified
with period in the imaginary time direction, one gets the usual result for
the partition function of black body radiation. However, as we have just
seen, the Euclidean- Schwarzschild solution is also periodic in imaginary
time with period $2\pi /\kappa $. This means that fields on the
Schwarzschild background will behave as if they were in a thermal state with
temperature $2\pi /\kappa $\ \cite{Hawking2}.

\chapter{Rotating Charged Black Strings in 4-Dimensions\protect\medskip 
\label{Lemosch}}

The theory of gravitational collapse and the theory of black holes are two
distinct but linked subjects. From the work of Oppenheimer and Snyder \cite%
{oppenheimer} and Penrose's theorem \cite{penrose1} we know that if general
relativity is correct, then realistic, slightly non-spherical, complete
collapse leads to the formation of a black hole and a singularity. There are
also studies hinting that the introduction of a cosmological constant ($%
\Lambda $) does not alter this picture \cite{hiscock}. Collapse of
cylindrical systems and other idealized models was used by Thorne to mimic
prolate collapse \cite{thorne1}. This study led to the formulation of the
hoop conjecture which states that horizons form when and only when a mass
gets compacted into a region whose circumference in every direction is less
than its Schwarzschild circumference, $4\pi M$. Thus, cylindrical collapsing
matter will not form a black hole. However, the hoop conjecture was given
for spacetimes with zero cosmological constant. In the presence of a
negative cosmological constant one can expect the occurrence of major
changes. Indeed, we show in this chapter that there are black hole solutions
with cylindrical symmetry if a negative cosmological constant is present (a
fact that does not happen for zero cosmological constant). These cylindrical
black holes are also called black strings. We study the charged rotating
black string and show that apart from spacetime being asymptotically anti-de
Sitter in the radial direction (and not asymptotically flat) the black
string solution has many similarities with the Kerr-Newman black hole. A 4-D
metric, $g_{\mu \nu }$\ ($\mu \nu =0,1,2,3$), with one Killing vector can be
written (in a particular instance) as, 
\begin{equation}
ds^{2}=g_{\mu \nu }dx^{\mu }dx^{\nu }=g_{mn}dx^{m}dx^{n}+\frac{r^{2}}{\ell
^{2}}dz^{2}\,,  \label{eq:12}
\end{equation}
where $g_{mn}$\ and $\phi $\ are metric functions, $m,n=0,1,2$\ and $z$\ is
the Killing coordinate. Equation (\ref{eq:12}) is invariant under $%
z\rightarrow -z$. A cylindrical symmetric metric can then be taken from (\ref%
{eq:12}) by imposing that the azimuthal coordinate, $\varphi $, also yields
a Killing direction.

We show that the theory has black holes similar to the Kerr-Newman black
holes, with a polynomial timelike singularity hidden behind the event and
Cauchy horizons. When the charge is zero, the rotating solution does not
resemble so much the Kerr solution, the singularity is spacelike hidden
behind a single event horizon. In addition, in the non-rotating uncharged
case, apart from the topology and asymptotics, the solution is identical to
the Schwarzschild solution.

Cylindrical symmetry, as emphasized by Thorne \cite{thorne1}, is an
idealized situation. It is possible that the Universe we live in, contains
an infinite cosmic string. It is also possible, however less likely, that
the Universe is crossed by an infinite black string. Yet, one can always
argue that close enough to a loop string, spacetime resembles the spacetime
of an infinite cosmic string. In the same way, one could argue that close
enough to a toroidal finite black hole, spacetime resembles the spacetime of
the infinite black string.

\section{Equations and Solutions}

We consider Einstein-Hilbert action in four dimensions with a cosmological
term in the presence of an electromagnetic field. The total action is 
\begin{equation}
S_{G}=\frac{1}{16\pi }\int_{\mathcal{M}}{d^{4}x\sqrt{-g}(R-2\Lambda -F^{\mu
\nu }F_{\mu \nu })}\,,  \label{eq:21}
\end{equation}
where $F_{\mu \nu }=\partial _{\mu }A_{\nu }-\partial _{\nu }A_{\mu }$ and $%
A_{\mu }$\ is the vector potential. We study solutions of the
Einstein-Maxwell equations with cylindrical symmetry. By this, we mean
spacetimes admitting three kinds of topology as \cite{chrusciel}

\begin{quote}
(i) $R\times S^{1}$, the standard cylindrically symmetric model.

(ii) $S^{1}\times S^{1}$\ the flat torus $T^{2}$\ model.

(iii) $R^{2}$\ .
\end{quote}

We will focus upon (i) and (ii). We then choose a cylindrical coordinate
system $(x^{0},x^{1},x^{2},x^{3})=(t,r,\varphi ,z)$\ with $-\infty
<t<+\infty $, $0\leq r<+\infty $, $-\infty <z<+\infty $ and $0\leq \varphi
<2\pi $. In the toroidal model (ii) the range of the coordinate $z$\ is $%
0\leq \alpha z<2\pi $. The electromagnetic four potential is given by $%
A_{\mu }=-h(r)\delta _{\mu }^{0}$, where $h(r)$\ is yet unknown function of
the radial coordinate $r$. Solving the Einstein-Maxwell equations yielded by
(\ref{eq:21}) for a static cylindrically symmetric spacetime we find, 
\begin{equation}
ds^{2}=-f(r)d{t}^{2}-\frac{dr^{2}}{f(r)}+r^{2}d{\varphi }^{2}+\frac{r^{2}}{%
\ell ^{2}}dz^{2},\   \label{eq:23}
\end{equation}
where 
\begin{equation}
f(r)=\frac{r^{2}}{\ell ^{2}}-\frac{b\ell }{r}+\frac{\lambda ^{2}\ell ^{2}}{%
r^{2}},  \label{eq:4102}
\end{equation}
and 
\begin{equation}
h(r)=\frac{\lambda \ell }{r}+\mathrm{const.}  \label{eq:24}
\end{equation}
where $b$\ and $\lambda $\ are integration constants. It is easy to show
that $\lambda /2$\ is the linear charge density of the $z$-line, and $b/4$
is the mass per unit length of the $z$-line as we will see in the next
section. Depending on the relative values of $b$\ and $\lambda $, the metric
(\ref{eq:23}) can represent a static black string. In this case there is a
black string that is not simply connected, i.e., closed curves encircling
the horizon cannot be shrunk to a point.

There is also a stationary solution that follows from equations (\ref{eq:21}%
) given by 
\begin{eqnarray}
ds^{2} &=&-\Xi ^{2}\left[ f(r)-\frac{a^{2}r^{2}}{\Xi ^{2}\ell ^{4}}\right]
dt^{2}-\frac{\Xi a\ell }{r}(b-\frac{\lambda ^{2}\ell }{r})2d\varphi dt 
\notag \\
&&+\left[ \Xi ^{2}r^{2}-a^{2}f(r)\right] d\varphi ^{2}+\frac{dr^{2}}{f(r)}+%
\frac{r^{2}}{\ell ^{2}}dz^{2},  \label{eq:25}
\end{eqnarray}
where 
\begin{equation}
A_{\mu }=-h(r)(\Xi \delta _{\mu }^{0}+a\delta _{\mu }^{2})\text{ \ \ \ \ \ \
\ and \ \ \ \ \ \ \ \ }\Xi ^{2}=1+\frac{a^{2}}{\ell ^{2}},  \label{Xi-A}
\end{equation}
where $a$\ is constant, $h(r)=\lambda \ell /r$, and the coordinates have the
same range as in the static case. Solution (\ref{eq:25}) can represent a
stationary black string. If one compactifies the $z$\ coordinate ($0\leq
\alpha z<2\pi $) one has a closed black string. In this case, one can also
put the coordinate $z$\ to rotate. However, this simply represents a bad
choice of coordinates. One can always find principal directions in which
spacetime rotates only along one of these coordinates ($\varphi $, say) as
in (\ref{eq:25}).

For an observer at radial infinity, the standard cylindrical spacetime model
(with $R\times S^{1}$\ topology) given by the metric (\ref{eq:25}) extends
uniformly over the infinite $z$-line. Thus one expects that, as $%
r\rightarrow \infty $, the total energy as well as the total charge is
infinite. The quantities that can be interpreted physically are the mass and
charge densities, i.e., mass and charge per unit length of the string. In
fact we have already found above the finite and well defined line charge
density (of the $z$-line) as an integration constant in Einstein-Maxwell
equations. For the close black string (the flat torus model with $%
S^{1}\times S^{1}$\ topology) the total energy and total charge are well
defined quantities. In order to properly define such quantities we use the
Hamiltonian formalism and the prescription of Brown and York \cite{BrownY}%
.\medskip

There is a suitable canonical form for the metric (\ref{eq:25}) as follow: 
\begin{equation}
ds^{2}=-{N^{0}}^{2}dt^{2}+R^{2}(N^{\varphi }dt+d\varphi )^{2}+\frac{dr^{2}}{%
f(r)}+\frac{r^{2}}{\ell ^{2}}dz^{2},  \label{eq:31}
\end{equation}
where 
\begin{equation}
{N^{0}}^{2}=f(r)\frac{r^{2}}{R^{2}}\,,\hspace{1.0421pc}\hspace*{0.1cm}%
N^{\varphi }=-\frac{\Xi a}{R^{2}}\left( \frac{b\ell }{r}-\frac{\lambda
^{2}\ell ^{2}}{r^{2}}\right) \,,\text{ \ \ \ \ \ }R^{2}=\Xi
^{2}r^{2}-a^{2}f(r)\,.\hspace{1.0421pc}  \label{eq:32}
\end{equation}
In metric (\ref{eq:31}) $N^{0}$\ and $N^{\varphi }$\ are respectively the
lapse and shift functions. Now we see that the metric given in equation (\ref%
{eq:31}) admits the two Killing vectors needed in order to define mass and
angular momentum: a timelike Killing vector $\xi =(\partial /\partial t)$\
and a spacelike (axial) Killing vector $\varsigma =(\partial /\partial
\varphi )$. The total mass as well as the total angular momentum per unit
length of this metric can be obtained from Eqs. (\ref{consmass}) and (\ref%
{consang}), that was expressed in chapter (\ref{Ads/CFT}), as follow \cite%
{Lemos, Dehghani1}: 
\begin{equation}
M=\frac{1}{8}b(3\Xi ^{2}-1),\hspace{1.0658pc}J=\frac{3}{8}ba\Xi .
\label{M-J-Q}
\end{equation}
Also the electric charge per unit length $Q$, are found by calculating the
flux of the electromagnetic field at infinity. So 
\begin{equation}
\text{\ }Q=\Xi \lambda /2.
\end{equation}

Let us recall that spacetime of (\ref{eq:25}) is pure anti-de Sitter metric,
if we choose $b=\lambda =0.$\ In this case 
\begin{equation}
ds^{2}=-\frac{r^{2}}{\ell ^{2}}dt^{2}+\frac{\ell ^{2}dr^{2}}{r^{2}}%
+r^{2}d\varphi ^{2}+\frac{r^{2}}{\ell ^{2}}dz^{2}.  \label{eq:27}
\end{equation}
This is also the background reference spacetime, since metric (\ref{eq:25})
reduces to (\ref{eq:27}) if the black hole is not present.

One can derive $b$\ and $\lambda $\ from equations (\ref{M-J-Q}), and
rewrite the metric (\ref{eq:25}) as 
\begin{eqnarray}
ds^{2} &=&-\left( \frac{r^{2}}{\ell ^{2}}-\frac{2(M+\Omega )\ell }{r}+\frac{%
4Q^{2}\ell ^{2}}{r^{2}}\right) dt^{2}-\frac{16J\ell }{3r}\left( 1-\frac{%
2Q^{2}\ell }{(M+\Omega )r}\right) dtd\varphi  \notag \\
&&+\left[ r^{2}+{\frac{4(M-\Omega )\ell ^{3}}{{r}}}\left( 1-{\frac{2}{{%
(M+\Omega )}}}\frac{Q^{2}\ell }{r}\right) \right] d\varphi ^{2}  \notag \\
&&+{\frac{dr^{2}}{\frac{r^{2}}{\ell ^{2}}{-\frac{2(3\Omega -M)\ell }{r}+%
\frac{3\Omega -M}{\Omega +M}\frac{4Q^{2}\ell ^{2}}{r^{2}}}}}+\frac{r^{2}}{%
\ell ^{2}}dz^{2}\,,  \label{eq:310}
\end{eqnarray}
where 
\begin{equation}
\Omega =\sqrt{M^{2}-\frac{8J^{2}}{9\ell ^{2\alpha }}}.  \label{eq:311}
\end{equation}
\medskip

\section{Causal Structure of the Charged Rotating Black String Spacetime}

In order to study the metric and its causal structure it is useful to define
the parameter $\alpha $\ (with units of angular momentum per unit mass),
that can show the effect of rotation ( called \emph{rotation parameter}) as: 
\begin{equation}
\frac{\alpha ^{2}}{\ell ^{2}}\equiv 1-\frac{\Omega }{M}  \label{eq:4001}
\end{equation}
such that 
\begin{equation}
1+\frac{\Omega }{M}=2(1-\frac{\alpha ^{2}}{2\ell ^{2}})\quad ,\quad 3\frac{%
\Omega }{M}-1=2(1-\frac{3}{2}\frac{\alpha ^{2}}{\ell ^{2}}).  \label{eq:4002}
\end{equation}
The relation between $J$\ and $\alpha $\ is given by 
\begin{equation}
J=\frac{3}{2}\alpha M\sqrt{1-\frac{\alpha ^{2}}{2\ell ^{2}}}.
\label{eq:4003}
\end{equation}
The range of $\alpha $\ is $0\leq \alpha /\ell \leq 1$. With these
definitions the metric (\ref{eq:310}) assumes the form 
\begin{eqnarray}
ds^{2} &=&-\left( \frac{r^{2}}{\ell ^{2}}-\frac{4M(1-\frac{\alpha ^{2}}{%
2\ell ^{2}})\ell }{r}+\frac{4Q^{2}\ell ^{2}}{r^{2}}\right) dt^{2}  \notag \\
&&-\frac{4\alpha M\ell \sqrt{1-\frac{\alpha ^{2}}{2\ell ^{2}}}}{r}\left( 1-%
\frac{Q^{2}\ell }{M(1-\frac{\alpha ^{2}}{2\ell ^{2}})r}\right) 2dtd\varphi 
\notag \\
&&+\left( \frac{r^{2}}{\ell ^{2}}-\frac{4M(1-\frac{3}{2}\frac{\alpha ^{2}}{%
\ell ^{2}})\ell }{r}+\frac{4Q^{2}\ell ^{2}}{r^{2}}\frac{(1-\frac{3}{2}\frac{%
\alpha ^{2}}{\ell ^{2}})}{(1-\frac{\alpha ^{2}}{2\ell ^{2}})}\right)
^{-1}dr^{2}  \notag \\
&&+\left[ r^{2}+\frac{4M\alpha ^{2}\ell }{r}\left( 1-\frac{Q^{2}\ell }{(1-%
\frac{\alpha ^{2}}{2\ell ^{2}})Mr}\right) \right] d\varphi ^{2}+\frac{r^{2}}{%
\ell ^{2}}dz^{2}.  \label{eq:4004}
\end{eqnarray}
It is worthwhile to mention that, there is a relation between the new
parameter $\alpha $ and the old parameter $a$ as: 
\begin{equation}
\alpha ^{2}=a^{2}(1+\frac{3}{2}\frac{a^{2}}{\ell ^{2}})^{-1}.
\end{equation}
Hence the old parameter $a,$ also may be called `rotation parameter'. In
order to compare metric (\ref{eq:4004}) with the well-known Kerr-Newman
metric, we write explicitly here the Kerr-Newman metric on the equatorial
plane 
\begin{eqnarray}
ds^{2} &=&-(1-\frac{2m}{r}+\frac{e^{2}}{r^{2}})dt^{2}-\frac{2ma}{r}(1-\frac{%
e^{2}}{2mr})2dtd\varphi  \notag \\
&&+\left[ r^{2}+a(1+\frac{2m}{r}-\frac{e^{2}}{r^{2}})\right] d\varphi
^{2}+r^{2}d\theta ^{2}  \notag \\
&&+(1-\frac{2m}{r}+\frac{a^{2}+e^{2}}{r^{2}})^{-1}dr^{2},  \label{eq:4005}
\end{eqnarray}
where $(m,a,e)$\ are the mass, specific angular momentum and charge
parameters of Kerr-Newman spacetimes, respectively. We can now see that the
metric for a rotating cylindrical symmetric asymptotically anti-de Sitter
spacetime, given in (\ref{eq:4004}), has many similarities with the metric
on the equatorial plane for the Kerr-Newman metric in (\ref{eq:4005}).

Metric (\ref{eq:4004}) has a singularity at $r=0$. The Kretschmann scalar or
scalar of Riemann tensor, $K=R^{\mu \nu \rho \sigma }R_{\mu \nu \rho \sigma
} $\ is 
\begin{equation}
K=\frac{{24}}{\ell ^{4}}{\left( 1+\frac{b^{2}\ell ^{6}}{2r^{6}}\right) -%
\frac{48\lambda ^{2}\ell ^{3}}{r^{7}}\left( b-\frac{7\lambda ^{2}\ell }{6r}%
\right) ,}  \label{eq:4006}
\end{equation}
where $b$\ and $\lambda $\ can be picked up from (\ref{eq:25}) and (\ref%
{eq:4004}). Thus $K$\ diverges at $r=0$. The solution has totally different
character depending on whether $r>0$\ or $r<0$. The important black hole
solution exists for $r>0$ or ($M>0$) which we consider.

To analyze the causal structure and follow the procedure of Boyer and
Lindquist \cite{boyer} and Carter \cite{carter1} we put metric (\ref{eq:4004}%
) in the form, 
\begin{equation}
ds^{2}=-f(r)\left( \Xi dt-ad\varphi \right) ^{2}+r^{2}\left( \Xi d\varphi -%
\frac{a}{\ell ^{2}}dt\right) ^{2}+\frac{dr^{2}}{f(r)}+\frac{r^{2}}{\ell ^{2}}%
dz^{2}.  \label{eq:4101}
\end{equation}
There are horizons whenever 
\begin{equation}
f(r)=0,  \label{eq:4107}
\end{equation}
i.e., at the roots of $f(r)$. One knows that the non-extremal situations in
the Kerr-Newman metric are given by $0\leq a^{2}/m^{2}\leq 1-e^{2}/m^{2}$.
Here, to have horizons one needs either one of the two conditions: 
\begin{equation}
0\leq \frac{\alpha ^{2}}{\ell ^{2}}\leq \frac{2}{3}-\frac{128}{81}\frac{Q^{6}%
}{M^{4}(1-\frac{1}{2}\frac{\alpha ^{2}}{\ell ^{2}})^{3}},  \label{eq:4108}
\end{equation}
or 
\begin{equation}
\frac{2}{3}<\frac{\alpha ^{2}}{\ell ^{2}}\leq 1.  \label{eq:4109}
\end{equation}
Thus there are five distinct cases depending on the value of the charge and
angular momentum:

\begin{quote}
(I) $0\leq \alpha ^{2}/\ell ^{2}\leq 2/3-(128/81)Q^{6}[M^{4}(1-\alpha
^{2}/2\ell ^{2})^{3}]^{-1}$, which yields the black hole solution with event
and Cauchy horizons.

(II) $\alpha ^{2}/\ell ^{2}=2/3-(128/81)Q^{6}[M^{4}(1-\alpha ^{2}/2\ell
^{2})^{3}]^{-1}$, which corresponds to the extreme case, where the two
horizons merge.

(III) $2/3-(128/81)Q^{6}[M^{4}(1-\alpha ^{2}/2\ell ^{2})^{3}]^{-1}<\alpha
^{2}/\ell ^{2}<\frac{2}{3}$, corresponding to naked singularities solutions.

(IV) $\alpha ^{2}/\ell ^{2}=2/3$, which gives a null singularity.

(V) $2/3<\alpha ^{2}/\ell ^{2}<1$, which gives a black hole solution with
one horizon.
\end{quote}

The most interesting solutions are given in items (I) and (II). Solutions
(IV) and (V) do not have partners in the Kerr-Newman family. In figure (\ref%
{fig4.1}), we show the black hole and naked singularity regions, and the
extremal black hole line dividing those two regions.\FRAME{ftbpFU}{6.6843cm}{%
6.6843cm}{0pt}{\Qcb{\textsl{The five regions and lines which yield solutions
of different natures are shown.}}}{\Qlb{fig4.1}}{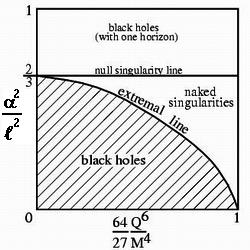}{\raisebox{-6.6843cm}
{\includegraphics[height=6.6843cm]{fig41.jpg}}} Now we analyze each item in turn.

\subsection{\ (I) Black Hole With Two Horizons\ }

This is the charged-rotating black string spacetime. As we will see this is
indeed very similar to the Kerr-Newman black hole. The structure has event
and Cauchy horizons, timelike singularities, and closed timelike curves.

Now, following Boyer and Lindquist, we choose a new angular coordinate which
straightens out the helicoidal null geodesics that pile up around the event
horizon. A good choice is 
\begin{equation}
\overline{\varphi }=\Xi \varphi -\frac{a}{\ell ^{2}}t.  \label{eq:4110}
\end{equation}
In this case the metric reads, 
\begin{equation}
ds^{2}=-f(r)\left( \Xi dt-\frac{a}{\Xi }d\overline{\varphi }\right)
^{2}+r^{2}d\overline{\varphi }^{2}+\frac{dr^{2}}{f(r)}+\frac{r^{2}}{\ell ^{2}%
}dz^{2}.  \label{eq:4111}
\end{equation}
The horizons are given at the zeros of the lapse function, i.e. when $f(r)=0$%
. We find that $f(r)$\ has two roots $r_{+}$\ and $r_{-}$\ given as 
\begin{equation}
r_{\pm }=b^{\frac{1}{3}}\frac{\sqrt{s}\pm \sqrt{2\sqrt{s^{2}-4q^{2}}-s}}{%
2\alpha }  \label{eq:4112}
\end{equation}
where, 
\begin{equation}
s=\left( \frac{1}{2}+\frac{1}{2}\sqrt{1-4\left( \frac{4q^{2}}{3}\right) ^{3}}%
\right) ^{\frac{1}{3}}+\left( \frac{1}{2}-\frac{1}{2}\sqrt{1-4\left( \frac{%
4q^{2}}{3}\right) ^{3}}\right) ^{\frac{1}{3}},  \label{eq:4114}
\end{equation}
\begin{equation}
q^{2}=\frac{\lambda ^{2}}{b^{\frac{4}{3}}}.  \label{eq:4115}
\end{equation}

Now we introduce a Kruskal coordinate patch around $r_{+}$\ and $r_{-}$. The
first patch constructed around $r_{+}$\ is valid for $r_{-}<r<\infty $.

In the region $r_{-}<r\leq r_{+}$\ the null Kruskal coordinates $U$\ and $V$%
\ are given by, 
\begin{align}
& U=\left( \frac{r_{+}-r}{\ell b^{\frac{1}{3}}}\right) ^{\frac{1}{2}}\left( 
\frac{r-r_{-}}{\ell b^{\frac{1}{3}}}\right) ^{-\frac{C}{B}\frac{{r_{-}}^{2}}{%
2{r_{+}}^{2}}}F\left( r\right) \exp \left( -\frac{A}{B}\frac{r_{+}-r_{-}}{2{%
r_{+}}^{2}}\frac{t}{\Xi \ell }\right)  \notag \\
& V=\left( \frac{r_{+}-r}{\ell b^{\frac{1}{3}}}\right) ^{\frac{1}{2}}\left( 
\frac{r-r_{-}}{\ell b^{\frac{1}{3}}}\right) ^{-\frac{C}{B}\frac{{r_{-}}^{2}}{%
2{r_{+}}^{2}}}F\left( r\right) \exp \left( \frac{A}{B}\frac{r_{+}-r_{-}}{2{%
r_{+}}^{2}}\frac{t}{\Xi \ell }\right) .,  \label{eq:4116}
\end{align}
For $r_{+}\leq r<\infty $ we put, 
\begin{align}
& U=-\left( \frac{r-r_{+}}{\ell b^{\frac{1}{3}}}\right) ^{\frac{1}{2}}\left( 
\frac{r-r_{-}}{\ell b^{\frac{1}{3}}}\right) ^{-\frac{C}{B}\frac{{r_{-}}^{2}}{%
2{r_{+}}^{2}}}F\left( r\right) \exp \left( -\frac{A}{B}\frac{r_{+}-r_{-}}{2{%
r_{+}}^{2}}\frac{t}{\Xi \ell }\right)  \notag \\
& V=\left( \frac{r-r_{+}}{\ell b^{\frac{1}{3}}}\right) ^{\frac{1}{2}}\left( 
\frac{r-r_{-}}{\ell b^{\frac{1}{3}}}\right) ^{-\frac{C}{B}\frac{{r_{-}}^{2}}{%
2{r_{+}}^{2}}}F\left( r\right) \exp \left( \frac{A}{B}\frac{r_{+}-r_{-}}{2{%
r_{+}}^{2}}\frac{t}{\Xi \ell }\right)  \label{eq:4117}
\end{align}
The following definitions have been introduced in order to facilitate the
notation, 
\begin{eqnarray}
A &\equiv &\left( {r_{+}}^{2}+{r_{-}}^{2}\right) ^{2}+2\left(
r_{+}+r_{-}\right) ^{4},  \label{eq:4118} \\
B &\equiv &\ell \left[ \left( r_{+}+r_{-}\right) ^{2}+2{r_{-}}^{2}\right] ,
\\
C &\equiv &\ell \left[ \left( r_{+}+r_{-}\right) ^{2}+2{r_{+}}^{2}\right] ,
\\
D &\equiv &\frac{\ell }{2}\left( r_{+}+r_{-}\right) ^{3}, \\
E &\equiv &\ell {\frac{{\ \left( {r_{+}}^{2}+{r_{-}}^{2}\right) ^{2}+2\left(
r_{+}+r_{-}\right) ^{2}\left( {r_{+}}^{2}+{r_{-}}^{2}+r_{+}r_{-}\right) }}{{%
\ \sqrt{\left( r_{+}+r_{-}\right) ^{2}+2\left( {r_{+}}^{2}+{r_{-}}%
^{2}\right) }}}},
\end{eqnarray}
and finally, 
\begin{eqnarray}
&F\left( r\right) \equiv \left( \frac{1}{\ell ^{2}b^{\frac{2}{3}}}\left[
r^{2}+\left( r_{+}+r_{-}\right) r+\left( {r_{+}}^{2}+{r_{-}}%
^{2}+r_{+}r_{-}\right) \right] \right) ^{-\frac{D}{B}\frac{r_{+}-r_{-}}{2{%
r_{+}}^{2}}}&  \notag \\
&\exp \left( \frac{E}{B}\frac{r_{+}-r_{-}}{2{r_{+}}^{2}}\arctan \frac{%
2r+\left( r_{+}+r_{-}\right) }{\sqrt{\left( r_{+}+r_{-}\right) ^{2}+2\left( {%
r_{+}}^{2}+{r_{-}}^{2}\right) }}\right) .&  \label{eq:4123}
\end{eqnarray}
In this first coordinate patch, $r_{-}<r\leq \infty $, the metric can be
written as, 
\begin{eqnarray}
ds^{2} &=&-\frac{\ell ^{2}b^{\frac{2}{3}}\left( \frac{r-r_{-}}{\ell b^{\frac{%
1}{3}}}\right) ^{1+\frac{C}{B}\frac{{r_{-}}^{2}}{{r_{+}}^{2}}}}{{k_{+}}%
^{2}r^{2}}G_{+}\left( r\right) dUdV  \notag \\
&&+\frac{\alpha \ell }{\sqrt{1-\frac{\alpha ^{2}}{2\ell ^{2}}}}\frac{b^{%
\frac{2}{3}}\left( \frac{r-r_{-}}{\ell b^{\frac{1}{3}}}\right) ^{1+\frac{C}{B%
}\frac{{r_{-}}^{2}}{{r_{+}}^{2}}}}{k_{+}r^{2}}G_{+}\left( r\right) \left(
VdU-UdV\right) d\overline{\varphi }  \notag \\
&&+\left( r^{2}-f(r)\frac{\alpha ^{2}}{1-\frac{\alpha ^{2}}{2\ell ^{2}}}%
\right) d\overline{\varphi }^{2}+\frac{r^{2}}{\ell ^{2}}dz^{2},
\label{eq:4124}
\end{eqnarray}
where, 
\begin{equation}
G_{+}(r)\equiv \frac{r^{2}+\left( r_{+}+r_{-}\right) r+\left( {r_{+}}^{2}+{%
r_{-}}^{2}+r_{+}r_{-}\right) }{F^{2}(r)},  \label{eq:4125}
\end{equation}
and 
\begin{equation}
k_{+}=\frac{A}{B}\frac{r_{+}-r_{-}}{2{r_{+}}^{2}}.  \label{eq:4126}
\end{equation}
We see that the metric given in (\ref{eq:4124}) is regular in this patch,
and in particular is regular at $r_{+}$. It is however singular at $r_{-}$.
To have a metric non-singular at $r_{-}$\ one has to define new Kruskal
coordinates for the patch $0<r<r_{+}$.

For $0<r\leq r_{-}$\ and $r_{-}\leq r<r_{+}$ we have 
\begin{align}
& U=-\left( \frac{r_{+}-r}{\ell b^{\frac{1}{3}}}\right) ^{-\frac{B}{C}\frac{{%
r_{+}}^{2}}{2{r_{-}}^{2}}}\left( \frac{r_{-}-r}{\ell b^{\frac{1}{3}}}\right)
^{\frac{1}{2}}H\left( r\right) \exp \left( \frac{A}{C}\frac{r_{+}-r_{-}}{2{%
r_{-}}^{2}}\frac{t}{\Xi \ell }\right) ,  \notag \\
& V=\left( \frac{r_{+}-r}{\ell b^{\frac{1}{3}}}\right) ^{-\frac{B}{C}\frac{{%
r_{+}}^{2}}{2{r_{-}}^{2}}}\left( \frac{r_{-}-r}{\ell b^{\frac{1}{3}}}\right)
^{\frac{1}{2}}H\left( r\right) \exp \left( -\frac{A}{C}\frac{r_{+}-r_{-}}{2{%
r_{-}}^{2}}\frac{t}{\Xi \ell }\right) ,  \label{eq:4127}
\end{align}
and 
\begin{align}
& U=\left( \frac{r_{+}-r}{\ell b^{\frac{1}{3}}}\right) ^{-\frac{B}{C}\frac{{%
r_{+}}^{2}}{2{r_{-}}^{2}}}\left( \frac{r-r_{-}}{\ell b^{\frac{1}{3}}}\right)
^{\frac{1}{2}}H\left( r\right) \exp \left( \frac{A}{C}\frac{r_{+}-r_{-}}{2{%
r_{-}}^{2}}\frac{t}{\Xi \ell }\right) ,  \notag \\
& V=\left( \frac{r_{+}-r}{\ell b^{\frac{1}{3}}}\right) ^{-\frac{B}{C}\frac{{%
r_{+}}^{2}}{2{r_{-}}^{2}}}\left( \frac{r-r_{-}}{\ell b^{\frac{1}{3}}}\right)
^{\frac{1}{2}}H\left( r\right) \exp \left( -\frac{A}{C}\frac{r_{+}-r_{-}}{2{%
r_{-}}^{2}}\frac{t}{\Xi \ell }\right) ,  \label{eq:4128}
\end{align}
where, 
\begin{eqnarray}
H(r) &=&\left( \frac{1}{\ell ^{2}b^{\frac{2}{3}}}\left[ r^{2}+\left(
r_{+}+r_{-}\right) r+\left( {r_{+}}^{2}+{r_{-}}^{2}+r_{+}r_{-}\right) \right]
\right) ^{\frac{D}{C}\frac{r_{+}-r_{-}}{2{r_{-}}^{2}}}  \notag \\
&&\exp \left( -\frac{E}{C}\frac{r_{+}-r_{-}}{2{r_{-}}^{2}}\arctan \frac{%
2r+\left( r_{+}+r_{-}\right) }{\sqrt{\left( r_{+}+r_{-}\right) ^{2}+2\left( {%
r_{+}}^{2}+{r_{-}}^{2}\right) }}\right) .
\end{eqnarray}
The metric for this second patch can be written as 
\begin{eqnarray}
ds^{2} &=&-\frac{\ell ^{2}b^{\frac{2}{3}}\left( \frac{r_{+}-r}{\ell b^{\frac{%
1}{3}}}\right) ^{1+\frac{B}{C}\frac{{r_{+}}^{2}}{{r_{-}}^{2}}}}{{k_{-}}%
^{2}r^{2}}G_{-}\left( r\right) dUdV  \notag \\
&&-\frac{\alpha \ell }{\sqrt{1-\frac{\alpha ^{2}}{2\ell }}}\frac{b^{\frac{2}{%
3}}\left( \frac{r_{+}-r}{\ell b^{\frac{1}{3}}}\right) ^{1+\frac{B}{C}\frac{{%
r_{+}}^{2}}{{r_{-}}^{2}}}}{k_{-}r^{2}}G_{-}\left( r\right) \left(
VdU-UdV\right) d\overline{\varphi }  \notag \\
&&+\left( r^{2}-f(r)\frac{\alpha ^{2}}{1-\frac{\alpha ^{2}}{2\ell ^{2}}}%
\right) d\overline{\varphi }^{2}+\frac{r^{2}}{\ell ^{2}}dz^{2},
\end{eqnarray}
where, 
\begin{equation}
G_{-}(r)\equiv \frac{r^{2}+\left( r_{+}+r_{-}\right) r+\left( {r_{+}}^{2}+{%
r_{-}}^{2}+r_{+}r_{-}\right) }{H^{2}(r)}  \label{eq:43131}
\end{equation}
and 
\begin{equation}
k_{-}=\frac{A}{B}\frac{r_{+}-r_{-}}{2{r_{-}}^{2}}.  \label{eq:4132}
\end{equation}
The metric is regular at $r_{-}$\ and is singular at $r=0$. To construct the
Penrose diagram we have to define the Penrose coordinates, $\psi $, $\xi $\
by the usual arctangent functions of $U$\ and $V$, 
\begin{equation}
U=\tan \frac{1}{2}(\psi -\xi )\quad \text{\textrm{and}}\quad V=\tan \frac{1}{%
2}(\psi +\xi )  \label{eq:4133}
\end{equation}
From (\ref{eq:4133}), (\ref{eq:4116}) and (\ref{eq:4117}) we see that: (i)
the line $r=\infty $\ is mapped into two symmetrical curved timelike lines,
and (ii) the line $r=r_{+}$\ is mapped into two mutual perpendicular
straight lines at $45^{0}$. From (\ref{eq:4127}) and (\ref{eq:4128}) we see
that: (i) $r=0$\ is mapped into a curved timelike line and (ii) $r=r_{-}$\
is mapped into two mutual perpendicular straight null lines at $45^{0}$. One
has to join these two different patches (see \cite{chandra,bhtz}) and then
repeat them over in the vertical. The result is the Penrose diagram shown in
figure (\ref{fig4.2}). The lines $r=0$\ and $r=\infty $\ are drawn as
vertical lines, although in the coordinates $\psi $\ and $\xi $\ they should
be curved outwards, bulged. It is always possible to change coordinates so
that the lines are indeed vertical. \FRAME{ftbpFU}{1.7331in}{3.1704in}{0pt}{%
\Qcb{\textsl{The Penrose diagram representing the nonextreme charged
rotating black string. }}}{\Qlb{fig4.2}}{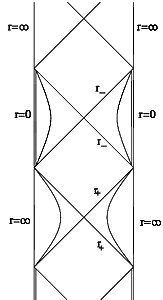}{\raisebox{-3.1704in}{\includegraphics[height=3.1704in]{fig42.jpg}}}

\subsection{(II) Extreme Case:}

The extreme case is given when $Q$\ is connected to $M$\ and $\alpha $\
through the relation, 
\begin{equation}
Q^{6}=\frac{27}{64}M^{2}{\left( 1-\frac{3}{2}\frac{\alpha ^{2}}{\ell ^{2}}%
\right) \left( 1-\frac{1}{2}\frac{\alpha ^{2}}{\ell ^{2}}\right) ^{3}},
\label{eq:4201}
\end{equation}
which can also be put in the form $\alpha ^{2}/\ell
^{2}=2/3-(128/81)Q^{6}[M^{4}(1-\alpha ^{2}/2\ell ^{2})^{3}]^{-1}$\ as above.
In figure (\ref{fig4.1}) we have drawn the line which gives the values of $Q$%
\ and $\alpha $\ (in suitable $M$\ units) compatible with this case. The
event and Cauchy horizons join together in one single horizon $r_{+}$\ given
by 
\begin{equation}
r_{+}=\frac{4Q^{2}\ell }{3M(1-\frac{\alpha ^{2}}{2\ell ^{2}})}
\label{eq:4202}
\end{equation}
The function $f(r)$\ is now, 
\begin{equation}
f(r)=\frac{(r-{r_{+}})^{2}(r^{2}+2r_{+}r+3{r_{+}}^{2})}{r^{2}\ell ^{2}}
\label{eq:4203}
\end{equation}
so the metric (\ref{eq:4101}) turns to 
\begin{eqnarray}
ds^{2} &=&-\frac{(r-{r_{+}})^{2}(r^{2}+2r_{+}r+3{r_{+}}^{2})}{r^{2}\ell ^{2}}%
(\Xi dt-ad\varphi )^{2}  \notag \\
&&+\frac{r^{2}\ell ^{2}dr^{2}}{(r-{r_{+}})^{2}(r^{2}+2r_{+}r+3{r_{+}}^{2})}%
+r^{2}(\Xi dt-ad\varphi )^{2}+\frac{r^{2}}{\ell ^{2}}dz^{2}.  \label{eq:4204}
\end{eqnarray}
There are no Kruskal coordinates. To draw the Penrose diagram we resort
first to the double null coordinates $u$\ and $v$, 
\begin{equation}
u=\frac{1}{\ell }(\Xi t-r_{\ast })\quad \mathrm{and}\quad v=\frac{1}{\ell }%
(\Xi t+r_{\ast })  \label{eq:4205}
\end{equation}
where $r_{\ast }$\ is the tortoise coordinate given by 
\begin{eqnarray}
r_{\ast } &=&\frac{2\ell }{9r_{+}}\ln \left( \frac{r-r_{+}}{\ell b^{\frac{1}{%
3}}}\right) -\frac{\ell }{6(r-r_{+})}  \notag \\
&&-\frac{\ell }{9r_{+}}\ln \left[ \frac{(r^{2}+2r_{+}r+3{r_{+}}^{2})}{\ell
^{2}b^{\frac{2}{3}}}\right] +\frac{7\ell }{18\sqrt{2}r_{+}}\arctan \frac{%
r+r_{+}}{\sqrt{2}r_{+}}.
\end{eqnarray}
Defining the new angular coordinate as before $\overline{\varphi }=\Xi
\varphi -(a/\ell ^{2})t$, the metric (\ref{eq:4204}) is now 
\begin{eqnarray}
ds^{2} &=&-\frac{(r-{r_{+}})^{2}(r^{2}+2r_{+}r+3{r_{+}}^{2})}{\ell ^{2}r^{2}}%
\frac{dt^{2}}{\Xi ^{2}}+\frac{\ell ^{2}r^{2}dr^{2}}{(r-{r_{+}}%
)^{2}(r^{2}+2r_{+}r+3{r_{+}}^{2})}  \notag \\
&&+\frac{af(r)}{\Xi ^{2}}2dtd\overline{\varphi }+(r^{2}-f(r)\frac{a^{2}}{\Xi
^{2}})d\overline{\varphi }^{2}+\frac{r^{2}}{\ell ^{2}}dz^{2}.
\label{eq:4207}
\end{eqnarray}
Now defining the Penrose coordinates \cite{bhtz,carter2} $\psi $\ and $\xi $%
\ via the relations 
\begin{equation}
u=\tan \frac{1}{2}(\psi -\xi )\quad \text{\textrm{and}}\quad v=\tan \frac{1}{%
2}(\psi +\xi ),  \label{eq:4208}
\end{equation}
one can write the metric (\ref{eq:4207}) as: 
\begin{eqnarray}
ds^{2} &=&-\frac{(r-{r_{+}})^{2}(r^{2}+2r_{+}r+3{r_{+}}^{2})}{r^{2}}\frac{%
d\psi ^{2}-d\xi ^{2}}{(\cos \psi +\cos \xi )^{2}}  \notag \\
&&+\frac{af(r)}{\Xi ^{2}}2dtd\overline{\varphi }+(r^{2}-f(r)\frac{a^{2}}{\Xi
^{2}})d\overline{\varphi }^{2}+\frac{r^{2}}{\ell ^{2}}dz^{2},
\label{eq:4209}
\end{eqnarray}
where $t$\ is given implicitly in terms of $\psi $\ and $\xi $. From the
defining Eqs. (\ref{eq:4205}) and (\ref{eq:4208}) we have 
\begin{equation}
\frac{\sin \xi }{\cos \psi +\cos \xi }=\frac{r_{\ast }}{\ell }
\label{eq:4210}
\end{equation}
Then, one can draw the Penrose diagram (see figure (\ref{fig4.3})). \FRAME{%
ftbpFU}{1.4624in}{3.3797in}{0pt}{\Qcb{\textsl{The Penrose diagram for the
extremal charged rotating black string.}}}{\Qlb{fig4.3}}{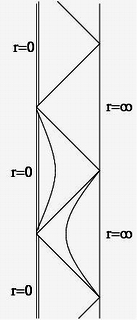}
{\raisebox{-3.3797in}{\includegraphics[height=3.3797in]{fig43.jpg}}
}\ The lines $r=r_{+}$\ are given by the equation $\psi =\pm \xi
+n\pi $\ with $n$\ any integer, and therefore are lines at $45^{0}$. The
lines $r=0$\ and $r=\infty $\ are timelike lines given by an equation of the
form $\sin \xi /(\cos \psi +\cos \xi )=const.$, where the constant is easily
found from $r_{\ast }$. These are not straight vertical lines. However by a
further coordinate transformation it is possible to straighten them out as
it is shown in Fig. (\ref{fig4.3}). The metric (\ref{eq:4209}) is regular at 
$r=r_{+},$\ because the zeros of the denominator and numerator cancel each
other. \vskip0.3cm

\subsection{(III) Naked Singularity}

In this case there are no roots for $f(r)$\ as defined in (\ref{eq:4102}).
Therefore there are no horizons. The singularity is timelike and naked.
Infinity is also timelike. There is an infinite redshift surface if the
following inequality is satisfied 
\begin{equation}
{Q}^{6}\leq \frac{27}{64}(1-\frac{1}{2}\frac{\alpha ^{2}}{\ell ^{2}}%
)^{4}M^{4}.  \label{eq:4301}
\end{equation}
The Penrose diagram is sketched in Fig. (\ref{fig4.4}).\FRAME{ftbpFU}{%
1.9337in}{2.9603in}{0pt}{\Qcb{\textsl{The Penrose diagram for the charged
rotating nacked singularity.}}}{\Qlb{fig4.4}}{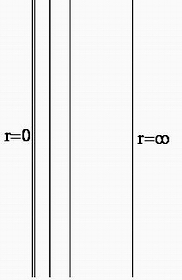}{\raisebox{-2.9603in}{\includegraphics[height=2.9603in]{fig44.jpg}}}

In this chapter, first, we made the black string metric in $4-$dimensions
and then the properties of this metric studied. At the next chapter we
generalize this metric to higher dimensions and its\ thermodynamic
properties will be studied.\ Also; In\ chapter \ref{Higgs}, The `no-hair'
theorem for this metric will be discussed.

\chapter{Thermodynamics of $(n+1)$-Dimensional Charged Rotating Black Branes
\label{Khodam}\protect\footnote{%
We published the paper related to this subject in Phys. Rev. D (see Ref. 
\cite{DehKhod1})}\protect\bigskip}

The theory of AdS/CFT correspondence was given in Chapter \ref{Ads/CFT}.
There, we studied all divergent terms and obtained the counterterms to
finite the Gravitational action. In Chapter \ref{thermodynamics}, we studied
the theory of thermodynamics of black holes and in chapter \ref{Lemosch}, we
gave the metric of a 4-dimensional charged rotating black string that was
obtained by Lemos \cite{Lemos}. In this chapter, we study the thermodynamics
of the $(n+1)$-dimensional charged rotating black brane introduced by Awad 
\cite{Awad2}, and consider its stability.

\section{The Action and Thermodynamic Quantities of Asymptotically AdS
(AAds) Charged Rotating Black Brane\label{Metr}}

The gravitational action for Einstein-Maxwell theory in $(n+1)$ dimensions
for AAdS spacetimes is 
\begin{equation}
I_{G}=-\frac{1}{16\pi }\int_{\mathcal{M}}d^{n+1}x\sqrt{_{-}g}\left(
R-2\Lambda -F^{\mu \nu }F_{\mu \nu }\right) +\frac{1}{8\pi }\int_{\partial 
\mathcal{M}}d^{n}x\sqrt{_{-}\gamma }\Theta (\gamma ),  \label{Actg}
\end{equation}
where $F_{\mu \nu }=\partial _{\mu }A_{\nu }-\partial _{\nu }A_{\mu }$ is
the electromagnetic tensor field and$\ A_{\mu }$ is the vector potential.
The first term is the Einstein-Hilbert volume term with negative
cosmological constant $\Lambda =-n(n-1)/2\ell ^{2}$ and the second term is
the Gibbons Hawking boundary term which is chosen such that the variational
principle is well-defined. The manifold $\mathcal{M}$ has metric $g_{\mu \nu
}$ and covariant derivative $\nabla _{\mu }$. $\Theta $ is the trace of the
extrinsic curvature $\Theta ^{\mu \nu }$ of any boundary(ies) $\partial 
\mathcal{M}$ (that is $^{3}B$ and/or $\Sigma $) of the manifold $\mathcal{M}$%
, with induced metric(s) $\gamma _{i,j}$ (see Appendix \ref{boundary}).

The counterterm for asymptotically AdS spacetimes up to seven dimensions as
was obtained in chapter \ref{Ads/CFT} is 
\begin{eqnarray}
I_{ct} &=&\frac{1}{8\pi }\int_{\partial \mathcal{M}_{\infty }}d^{n}x\sqrt{%
-\gamma }\{\frac{n-1}{\ell }-\frac{\ell \Upsilon (n-3)}{2(n-2)}R  \notag \\
&&-\frac{\ell ^{3}\Upsilon (n-5)}{2(n-4)(n-2)^{2}}\left( R_{ab}R^{ab}-\frac{n%
}{4(n-1)}R^{2}\right) +...\},  \label{Actct1}
\end{eqnarray}
where $R$, $R_{abcd}$, and $R_{ab}$ are the Ricci scalar, Riemann and Ricci
tensors of the boundary metric $\gamma _{ab}$.

The metric of $(n+1)$-dimensional AAdS charged rotating black brane with $k$
rotation parameters is \cite{Awad2} 
\begin{eqnarray}
ds^{2} &=&-f(r)\left( \Xi dt-{{\sum_{i=1}^{k}}}a_{i}d\phi _{i}\right) ^{2}+%
\frac{r^{2}}{\ell ^{4}}{{\sum_{i=1}^{k}}}\left( a_{i}dt-\Xi \ell ^{2}d\phi
_{i}\right) ^{2}  \notag \\
&&\ \text{ }+\frac{dr^{2}}{f(r)}-\frac{r^{2}}{\ell ^{2}}{\sum_{i<j}^{k}}%
(a_{i}d\phi _{j}-a_{j}d\phi _{i})^{2}+r^{2}d\Omega ^{2},  \label{met2}
\end{eqnarray}
where $\Xi =\sqrt{1+\sum_{i}^{k}a_{i}^{2}/\ell ^{2}}$ and $d\Omega ^{2}$ is
the Euclidean metric on the $\left( n-1-k\right) $-dimensional submanifold.
with volume $V_{n-1}$. The maximum number of rotation parameters in $(n+1)$
dimensions is $[(n+1)/2]$, where $[x]$ denotes the integer part of $x$. In
Eq. (\ref{met2}) $f(r)$ is 
\begin{equation}
f(r)=\frac{r^{2}}{\ell ^{2}}-\frac{m}{r^{n-2}}+\frac{q^{2}}{r^{2n-4}},
\label{Fg}
\end{equation}
and the gauge potential is given by 
\begin{equation}
A_{\mu }=-\sqrt{\frac{n-1}{2n-4}}\frac{q}{r^{n-2}}\left( \Xi \delta _{\mu
}^{0}-\delta _{\mu }^{i}a_{i}\right) ,\hspace{1cm}\text{(no sum on }i\text{).%
}  \label{Vecp}
\end{equation}
In $4-$dimensions with $n=3,$ simply, one can see that this metric exchange
to the previous metric of Eq. (\ref{eq:27}). The Einstein equation for this
spacetime can be written as: 
\begin{equation}
R_{\mu \nu }-\frac{1}{2}g_{\mu \nu }R-\frac{n(n-1)}{2\ell ^{2}}g_{\mu \nu
}=8\pi T_{(\mathrm{em})\mu \nu },
\end{equation}
where $R$ is the Ricci scalar. The stress energy-momentum tensor $T_{(%
\mathrm{em})\mu \nu }$ is 
\begin{equation}
T_{(\mathrm{em})\mu \nu }=\frac{1}{8\pi }(2F_{\hspace{0.3161pc}\mu
}^{\lambda }F_{\lambda \nu }-\frac{1}{2}F^{\sigma \lambda }F_{\sigma \lambda
}g_{\mu \nu }).
\end{equation}

Of course, as expressed in chapter \ref{Ads/CFT}, there are also logarithmic
divergences due to the Weyl anomaly and matter field. To obtain the total
action, we first calculate the logarithmic divergences due to the Weyl
anomaly and matter field given in Eqs. (\ref{Ano4}), (\ref{Ano6}), (\ref%
{logem4}), and (\ref{logem6}). The leading metric $\tilde{\gamma}_{ij}^{0}$
in Eq.(\ref{asym}) can be obtained as 
\begin{equation}
\tilde{\gamma}_{ij}^{0}dx^{i}dx^{j}=-\frac{1}{\ell ^{2}}dt^{2}+d\phi
^{2}+d\Omega ^{2}.  \label{gamma0}
\end{equation}
Therefore the curvature scalar $R^{0}(\tilde{\gamma}^{0})$ and Ricci tensor $%
R_{ij}^{0}(\tilde{\gamma}^{0})$ are zero. Also it is easy to show that $%
F_{ij}^{0}$ in Eqs. (\ref{logem4}) and (\ref{logem6}) vanishes. Thus, all
the logarithmic divergences for the $(n+1)$-dimensional charged rotating
black brane are zero. It is also a matter of calculation to show that the
counterterm action due to the electromagnetic field in Eq. (\ref{Actct2}) is
zero. Thus, the total renormalized action is 
\begin{equation}
I=I_{G}+I_{\mathrm{ct}}.  \label{Acttot}
\end{equation}
In order to obtain the Einstein-Maxwell equations by the variation of the
volume integral with respect to the fields, one should impose the boundary
condition $\delta A_{\mu }=0$ on $\partial \mathcal{M}$. Thus the action (%
\ref{Acttot}) is appropriate to study the grand-canonical ensemble with
fixed electric potential \cite{Cal}. To study the canonical ensemble with
fixed electric charge one should impose the boundary condition $\delta
(n^{a}F_{ab})=0$, and therefore the total action is \cite{Haw1} 
\begin{equation}
\tilde{I}=I-\frac{1}{4\pi }\int_{\partial \mathcal{M}_{\infty }}d^{n}x\sqrt{%
_{-}\gamma }n_{a}F^{ab}A_{b}.  \label{Actcan}
\end{equation}
The divergence free stress-energy tensor for $n\leq 6$ was given by Eq. (\ref%
{finitestress}).

As in the case of rotating black hole solutions of Einstein's gravity, the
above metric given by Eqs. (\ref{met2})-(\ref{Vecp}) has two types of
Killing and event horizons. It was proved that a stationary black hole event
horizon should be a Killing horizon in the four-dimensional Einstein gravity 
\cite{Haw1}. The Killing vector of this $(n+1)$-dimensional metric is 
\begin{equation}
\chi =\partial _{t}+{{{\sum_{i=1}^{k}}}}\Omega _{i}\partial _{\phi _{i}},
\label{Kil}
\end{equation}
which is the null generator of the event horizon. The metric of Eqs. (\ref%
{met2})-(\ref{Vecp}) has two inner and outer event horizons located at $%
r_{-} $ and $r_{+}$, if the metric parameters $m$ and $q$ are chosen to be
suitable \cite{Awad2}. The two horizons $r_{-}$ and $r_{+}$\ are the real
roots of $f(r)=0.$ For later use in the thermodynamics of the black brane,
it is better to present an expression for the critical value of the charge
and mass in term of the radius of the event horizon $r_{+}$. It is easy to
show that the metric has two inner and outer horizons provided the charge
parameter, $q$ is less than $q_{\mathrm{crit}}$ given as 
\begin{equation}
q_{\mathrm{crit}}=\sqrt{\frac{n}{n-2}}\frac{r_{+}^{n-1}}{\ell },
\label{qcrit}
\end{equation}
or $m$ is greater than $m_{\mathrm{crit}},$ that is 
\begin{equation}
m_{\mathrm{crit}}=\left( \frac{n-1}{n-2}\right) \frac{2r^{n}}{\ell ^{2}}.
\label{mcrit}
\end{equation}
In the case that $q=q_{\mathrm{crit}}$ or $m=m_{\mathrm{crit}},$ we will
have an extreme black brane.

As mentioned in chapter \ref{Ads/CFT} and \ref{Lemosch}, mass and angular
momentum are conserved charges which may calculated for black holes
(string). By calculating the finite or divergence free stress
energy-momentum tensor and using Eqs. (\ref{consmass}) and (\ref{consang}),
we obtain the total mass as well as the total angular momentum as: 
\begin{eqnarray}
M &=&\frac{V_{n-1}\left( 2\pi \right) ^{k}}{16\pi }m\left[ n\Xi ^{2}-1\right]
,  \label{Mass} \\
J_{i} &=&\frac{V_{n-1}\left( 2\pi \right) ^{k}}{16\pi }n\Xi ma_{i}.
\label{Angmom}
\end{eqnarray}
The Hawking temperature and the angular velocities of the $r_{+}$ can be
calculated as expressed in chapter \ref{thermodynamics}. Hence we\ have 
\begin{eqnarray}
T &=&\frac{1}{\beta _{+}}=\left( \frac{4\pi \Xi }{f^{\prime }(r_{+})}\right)
^{-1}  \notag \\
&=&\frac{nr_{+}^{(2n-2)}-(n-2)q^{2}\ell ^{2}}{4\pi \ell ^{2}\Xi
r_{+}^{(2n-3)}},  \label{Temp} \\
\Omega _{j} &=&\frac{a_{j}}{\Xi \ell ^{2}},  \label{Om}
\end{eqnarray}
where $\beta _{+}$ is the inverse Hawking temperature. These quantities also
was obtained by Awad in Ref. \cite{Awad2}. Equation (\ref{qcrit}) shows that
the temperature $T$ in Eq. (\ref{Temp}) is positive for the allowed values
of the metric parameters and vanishes for the extremal solution.

By using Eqs. (\ref{Actg}), (\ref{Actct1}), (\ref{Acttot}), and (\ref{Actcan}%
), the Euclidean actions in the grand-canonical and the canonical ensemble
can be calculated as 
\begin{eqnarray}
I &=&-\frac{\beta _{+}V_{n-1}\left( 2\pi \right) ^{k}}{16\pi }\frac{%
r_{+}^{(2n-2)}+q^{2}\ell ^{2}}{r_{+}^{(n-2)}\ell ^{2}},  \label{Igcm} \\
\tilde{I} &=&-\frac{\beta _{+}V_{n-1}\left( 2\pi \right) ^{k}}{16\pi }\frac{%
r_{+}^{(2n-2)}-(2n-3)q^{2}\ell ^{2}}{r_{+}^{(n-2)}\ell ^{2}}.  \label{Icm}
\end{eqnarray}
The electric charge $Q$, can be found by calculating the flux of the
electromagnetic field at infinity, yielding 
\begin{equation}
Q=\frac{\Xi V_{n-1}\left( 2\pi \right) ^{k}}{4\pi }\sqrt{\frac{(n-1)(n-2)}{2}%
}q.  \label{Charg}
\end{equation}
The electric potential $\Phi $, measured at infinity with respect to the
horizon, is defined by \cite{Cal} 
\begin{equation*}
\Phi =A_{\mu }\chi ^{\mu }\left| _{r\rightarrow \infty }-A_{\mu }\chi ^{\mu
}\right| _{r=r_{+}},
\end{equation*}
where $\chi $ is the null generators of the event horizon given by Eq. (\ref%
{Kil}). One obtains 
\begin{equation}
\Phi =\sqrt{\frac{(n-1)}{2(n-2)}}\frac{q}{\Xi r_{+}^{(n-2)}}.  \label{Pot}
\end{equation}
The area law of the entropy is universal as was seen in Eq. (\ref{entropy}),
and applies to all kinds of black holes/branes \cite{Hawking99}. At first,
we should calculate the area of event horizon `$A$'$.$ This can be done by
calculating $\int \sqrt{\sigma }d^{n-1}x$. We obtain 
\begin{equation}
A=V_{n-1}\Xi r_{+}^{(n-1)},  \label{area}
\end{equation}
and therefore 
\begin{equation}
S=\frac{\Xi V_{n-1}}{4}r_{+}^{(n-1)}.  \label{Entropy}
\end{equation}
For $n=3$, these quantities given in Eqs. (\ref{Mass})-(\ref{Entropy})
reduce to those calculated in Ref. \cite{Dehghani1}.

\section{Thermodynamics of black brane\label{Therm}}

\subsection{Generalization of Smarr Formula}

At first we obtain the mass as a function of the extensive quantities $S$, $%
J $, and $Q$. Using the expression for the entropy, the mass, the angular
momenta, and the charge given in Eqs. (\ref{Mass}), (\ref{Angmom}), (\ref%
{Charg}), (\ref{Entropy}), and the fact that $f(r_{+})=0$, one can obtain a
Smarr-type formula as 
\begin{equation}
M(S,J,Q)=\frac{(nZ-1)\sqrt{\sum_{i}^{k}J_{i}^{2}}}{n\ell \sqrt{Z(Z-1)}},
\label{Smar}
\end{equation}
where $Z=\Xi ^{2}$ is the positive real root of the following equation:

\begin{equation}
\left( Z-1\right) ^{(n-1)}-\frac{Z}{16S^{2}}\left\{ \frac{4\pi
(n-1)(n-2)\ell SJ}{n[(n-1)(n-2)S^{2}+2\pi ^{2}Q^{2}\ell ^{2}]}\right\}
^{(2n-2)}=0.  \label{Zsmar}
\end{equation}
One may then regard the parameters $S$, $J$, and $Q$ as a complete set of
extensive parameters for the mass $M(S,J,Q)$ and define the intensive
parameters conjugate to $S$, $J$ and $Q$. These quantities are the
temperature $T$, the angular velocities $\Omega _{i}$, and the electric
potential $\Phi ,$ 
\begin{equation}
T=\left( \frac{\partial M}{\partial S}\right) _{J,Q},\ \ \Omega _{i}=\left( 
\frac{\partial M}{\partial J_{i}}\right) _{S,Q},\ \ \Phi =\left( \frac{%
\partial M}{\partial Q}\right) _{S,J}.  \label{Dsmar}
\end{equation}
It is a matter of straightforward calculation to show that the intensive
quantities calculated by Eq. (\ref{Dsmar}) coincide with Eqs. (\ref{Temp}), (%
\ref{Om}), and (\ref{Pot}) found in Sec. (\ref{Metr}). Thus, the
thermodynamic quantities calculated in Sec. (\ref{Metr}) satisfy the first
law of thermodynamics which was expressed in Eq. (\ref{firstlaw}) as:

\begin{equation}
dM=TdS+{{{\sum_{i=1}^k}}}\Omega _idJ_i+\Phi dQ.  \label{Flth}
\end{equation}

\subsection{Thermodynamic potentials}

Now we obtain the thermodynamic potential in the grand-canonical and
canonical ensembles. Using the definition of the Gibbs potential $G(T,\Omega
,\Phi )=I/\beta $ , we obtain

\begin{equation}
G=-\frac{V_{n-1}}{16\pi }\left( \frac{2}{n^{2}(n-1)(1-\sum_{i}^{k}\ell
^{2}\Omega _{i}^{2})}\right) ^{n/2}\left( \gamma ^{2}+n^{2}(n-2)\Phi
^{2}\right) (\gamma \ell )^{(n-2)},  \label{Gibbs}
\end{equation}
where 
\begin{equation}
\gamma =\sqrt{2n-2}T\pi \ell +\sqrt{2(n-1)\pi ^{2}T^{2}\ell
^{2}+n(n-2)^{2}\Phi ^{2}}.  \label{Ggibbs}
\end{equation}
Using the expressions (\ref{Temp}), (\ref{Om}), and (\ref{Pot}) for the
inverse Hawking temperature, the angular velocities and the electric
potential, one obtains

\begin{equation}
G(T,\Omega ,\Phi )=M-TS-\sum_i^k\Omega _iJ_i-\Phi Q ,  \label{Lgibbs}
\end{equation}
which means that $G(T,\Omega ,\Phi )$ is, indeed, the Legendre
transformation of the $M(S,J_i,Q)$ with respect to $S,$ $J_i$, and $Q$. It
is a matter of straightforward calculation to show that the extensive
quantities

\begin{equation}
J_i=-\left( \frac{\partial G}{\partial \Omega _i}\right) _{T,\Phi },\ \
Q=-\left( \frac{\partial G}{\partial \Phi }\right) _{T,\Omega }, \ \
S=-\left( \frac{\partial G}{\partial T}\right) _{\Omega ,\Phi },
\label{Dgibbs}
\end{equation}
turn out to coincide precisely with the expressions (\ref{Angmom}), (\ref%
{Charg}), and (\ref{Entropy}).

For the canonical ensemble, the Helmholtz free energy $F(T,J,Q)$ is defined
as

\begin{equation}
F(T,J,Q)=\frac{\tilde{I}}{\beta }+\sum_{i}^{k}\Omega _{i}J_{i},  \label{Helm}
\end{equation}
where $\overset{\sim }{I}$ is given by Eq. (\ref{Icm}). One can verify that
the conjugate quantities

\begin{equation}
\Omega _{i}=\left( \frac{\partial F}{\partial J_{i}}\right) _{T,Q},\ \ \Phi
=\left( \frac{\partial F}{\partial Q}\right) _{T,J},\ \ S=-\left( \frac{%
\partial F}{\partial T}\right) _{J,Q},  \label{Dhelm}
\end{equation}
agree with expressions (\ref{Om}), (\ref{Pot}), and (\ref{Entropy}). Also it
is worthwhile to mention that $F(T,J,Q)$ is the Legendre transformation of
the $M(S,J_{i},Q)$ with respect to $S$, i.e.

\begin{equation}
F(T,J,Q)=M-TS.  \label{Lhelm}
\end{equation}

\subsection{Stability in the canonical and the grand-canonical ensemble}

The stability of a thermodynamic system with respect to the small variations
of the thermodynamic coordinates, is usually performed by analyzing the
behavior of the entropy $S(M,J,Q)$ around the equilibrium. The local
stability in any ensemble requires that $S(M,J,Q)$ be a convex function of
their extensive variables or its Legendre transformation must be a concave
function of their intensive variables. Thus, the local stability can in
principle be carried out by finding the determinant of the Hessian matrix of 
$S$ with respect to its extensive variables $X_{i}$, $\mathbf{H}%
_{X_{i}X_{j}}^{S}=[\partial ^{2}S/\partial X_{i}\partial X_{j}]$, or the
determinant of the Hessian of the Gibbs function with respect to its
intensive variables $Y_{i}$, $\mathbf{H}_{Y_{i}Y_{j}}^{G}=[\partial
^{2}G/\partial Y_{i}\partial Y_{j}]$ \cite{Cvet,Cal}. Also, one can perform
the stability analysis through the use of the Hessian matrix of the mass
with respect to its extensive parameters \cite{Gub}. In our case the entropy 
$S$ is a function of the mass, angular momenta, and the charge. In the
canonical ensemble, the charge and the angular momenta are fixed parameters,
and therefore the positivity of the thermal capacity $C_{J,Q}=T(\partial
S/\partial T)_{J,Q}$ is sufficient to assure the local stability. The
thermal capacity $C_{J,Q},$ at constant charge and angular momenta is

\begin{eqnarray}
C_{J,Q} &=&\frac{\Xi V_{n-1}}{4}r^{(n-1)}[nr^{(2n-2)}-(n-2)q^{2}\ell
^{2}][r^{(2n-2)}+q^{2}\ell ^{2}]  \notag \\
&&\text{ }\times \ [(n-2)\Xi ^{2}+1]\{(n-2)q^{4}\ell ^{4}[(3n-6)\Xi
^{2}-(n-3)]  \label{Cap} \\
&&\ -2q^{2}\ell ^{2}r^{(2n-2)}[(3n-6)\Xi ^{2}-n^{2}+3]+nr^{(4n-4)}[(n+2)\Xi
^{2}-(n+1)]\}^{-1}.  \notag
\end{eqnarray}
Figure (\ref{fig5.1}) shows the behavior of the heat capacity as a function
of the charge parameter. It shows that $C_{J,Q}$ is positive in various
dimensions and goes to zero as $q$ approaches its critical value (extreme
black brane). Thus, the $(n+1)$-dimensional AAdS charged rotating black
brane is locally stable in the canonical ensemble.\FRAME{ftbpFU}{3.7455in}{%
2.9395in}{0pt}{\Qcb{$C_{J,Q}$\textsl{\ versus }$q$\textsl{\ for }$l=1$%
\textsl{, }$r_{+}=0.8$\textsl{, }$n=4$\textsl{\ (solid), }$n=5$\textsl{\
(dotted), and }$n=6$\textsl{\ (dashed).}}}{\Qlb{fig5.1}}{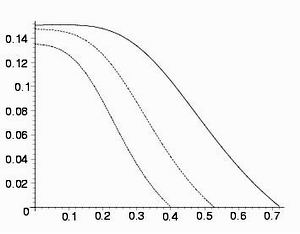}
{\raisebox{-2.9395in}{\includegraphics[height=2.9395in]{fig51.jpg}}}

In the grand-canonical ensemble, we find it more convenient to work with the
Gibbs potential $G(T,\Omega _{i},\Phi )$. Here the thermodynamic variables
are the temperature, the angular velocities, and the electric potential.
After some algebraic manipulation, we obtain

\begin{equation}
\left| \mathbf{H}_{T,\Omega _{i},\Phi }^{G}\right| =\left( \frac{n-3}{n}%
\right) \left[ \frac{nm}{16\pi }\right] ^{k}\left[ \frac{(n-2)\Xi ^{2}+1}{%
r_{+}^{n}+(\frac{n-2}{2})m\ell ^{2}}\right] (\ell \Xi )^{2k+2}\Xi
^{2}r^{3n-4}.  \label{dHes}
\end{equation}
As one can see from Eq. (\ref{dHes}), $\left| \mathbf{H}_{T,\Omega _{i},\Phi
}^{G}\right| $ is positive for all the phase space, and therefore the $(n+1)$%
-dimensional AAdS charged rotating black brane is locally stable in the
grand-canonical ensemble. As mentioned in the previous paragraph, one can
also perform the local stability analysis through the use of the determinant
of the Hessian matrix of $M$ with respect to $S$, $J$ and $Q$, which has the
same result, since $\left| \mathbf{H}_{S,J_{i},Q}^{M}\right| =\left| \mathbf{%
H}_{T,\Omega _{i},\Phi }^{G}\right| ^{-1}$.

\subsection{ Logarithmic correction to the Bekenstein-Hawking entropy}

In recent years, there are several works in literature suggesting that for a
large class of black holes, the area law of the entropy receives additive
logarithmic corrections due to thermal fluctuation of the object around its
equilibrium \cite{Maj}. Typically, the corrected formula has the form 
\begin{equation}
S=S_0-K\ln (S_0)+...,  \label{Logc1}
\end{equation}
where $S_0$ is the standard Bekenstein-Hawking term and $K$ is a number. In
Ref. \cite{Das} an expression has been found for the leading-order
correction of a generic thermodynamic system in terms of the heat capacity $%
C $ as \cite{Das} 
\begin{equation}
S=S_0-K\ln (CT^2).  \label{Logc2}
\end{equation}

Equation (\ref{Logc2}) has been considered by many authors for
Schwarzschild-AdS, Reissner-Nordstrom-AdS, BTZ, and slowly Kerr-AdS
spacetimes \cite{Od2}. Thus, it is worthwhile to investigate its application
for the charged rotating black brane considered in this chapter. Using Eqs. (%
\ref{Temp}), (\ref{Entropy}), and (\ref{Cap}) with $q=0$, one obtains

\begin{equation}
S=S_{0}-\frac{n+1}{2(n-1)}\ln (S_{0})-\Gamma _{n}(\Xi ),  \label{Logq0}
\end{equation}
where $\Gamma _{n}(\Xi )$ is a positive constant depending on $\Xi $ and $n$%
. Equation (\ref{Logq0}) shows that the correction of the entropy is
proportional to the logarithm of the area of the horizon. For small values
of $q$ the logarithmic correction in Eq. (\ref{Logc2}) can be expanded in
terms of the power of $S_{0}$ as 
\begin{equation}
S=S_{0}-\frac{n+1}{2(n-1)}\ln (S_{0})-\Gamma _{n}(\Xi )+\frac{\ell ^{2}\Xi
^{2}[(n-4)\Xi ^{2}+2]}{16[(n+2)\Xi ^{2}-n-1]}\frac{q^{2}}{S_{0}^{2}}+...
\label{Logq}
\end{equation}
Again the leading term is a logarithmic term of the area.

\part{\protect\bigskip \protect\bigskip\ HAIRY BLACK HOLE\label{part2}}

\chapter{The Theory of Abelian Higgs Hair for Black Holes\label{Higgstheory}}

Black hole `\emph{hair}' is defined to be any field(s) associated with a
stationary black hole configuration which can be detected by asymptotic
observers but which cannot be identified with the electromagnetic or
gravitational degrees of freedom. Back in the heyday of black hole physics a
number of results were proven \cite{Israel,Wald,Carter} which seemed to
imply that black holes `have \emph{no-hair}'. These results implied that
given certain assumptions the only information about a black hole which an
observer far from the hole can determine experimentally is summarized by the
electric charge, magnetic charge, angular momentum and mass of the hole.
Such uniqueness results are referred to as `\emph{no-hair}' theorems. These
important results would seem to imply that a black hole horizon can support
only these limited gauge charges; for a long time people thought that other
matter fields simply could not be associated with a black hole. Thus, for
example, lepton or baryon number were not good quantum numbers for black
holes, despite being defined for a neutron star. However, this idea was to
some extent discredited when various authors \cite{Bizon1}, using numerical
techniques, discovered black hole solutions of the Einstein- Yang-Mills
equations that support Yang-Mills fields which can be detected by asymptotic
observers \cite{Bartnik}; one therefore says that these black holes are `%
\emph{coloured}'. Once the solutions of \cite{Bartnik} were discovered, it
wasn't long before other people were finding similar solutions in
Einstein-non-Abelian gauge systems \cite{Smoller}.

However, these exotic solutions do not violate the original no-hair results
since all such solutions are known to be unstable (see e.g. \cite{Bizon2}).
Since the original no-hair theorems assumed a stationary picture they simply
do not apply to coloured holes. On the other hand, coloured holes do still
exist and so they are said to `evade' the usual no-hair results. These
results teach us that we have to tread carefully when we start talking about
black hole hair. There are other amusing tricks which allow one to evade
no-hair theorems. We will stick with our definition of hair as any property
which can be measured by asymptotic observers. Furthermore, we shall follow 
\cite{Achucarro} and use the term `dressing' for the question of whether or
not fields actually live on the horizon.

With all of this in mind, we want to analyze the extent to which hair is
present in situations where we allow the topology of some field
configurations to be non-trivial; in particular, an interesting question is
whether or not topological defects, such as domain walls, strings, or
textures \cite{Vilenkin}, can act as `hair' for a black hole. In \cite%
{Achucarro} evidence was presented that a Nielsen-Olesen ($U(1)$) vortex can
act as `long' hair for a Schwarzschild black hole.

Recently it has been shown that these ideas can be extended to the case of
anti-de Sitter (AdS) and de Sitter (dS) spacetimes. For asymptotically AdS
spacetimes, it has been shown that conformally coupled scalar field can be
painted as hair \cite{Eli2}. Another asymptotically AdS hairy black hole
solution has been investigated in Ref. \cite{Torii2}. Also it was shown that
there exist a solution to the $SU(2)$ Einstein-Yang-Mills equations which
describes a stable Yang-Mills hairy black hole, that is asymptotically AdS 
\cite{Eli1}. Although the idea of Nielson-Olesen vortices has been first
introduced in flat spacetimes \cite{NO} , but recently it has been extended
to (A)dS spacetimes \cite{DehHigg1,Ghez1}. The existence of long range
Nielson-Olesen vortex as hair for asymptotically AdS black holes has been
investigated in Refs. \cite{DehHiggs2,Dehjal} for Schwarzschild-AdS black
hole and charged black string. The explicit calculations which can
investigate the existence of a long range Nielson-Olesen vortex solution as
a stable hair for a stationary black hole solution is escorted with much
more difficulties due to the rotation parameter \cite{Ghez2}. In the next
chapter, we study the Abelian Higgs hair for a four dimensional rotating
charged black string that is a stationary model for Einstein-Maxwell
equation with cylindrical symmetry.

\section{Abelian Higgs Field as a Source of the Topological Defects}

On a cold day, ice forms quickly on the surface of a pond. But it does not
grow as a smooth, featureless covering. Instead, the water begins to freeze
in many places independently, and the growing plates of ice join up in
random fashion, leaving zig--zag boundaries between them. These irregular
margins are an example of what physicists call `\emph{topological defects}'
-- \textsl{defects} because they are places where the crystal structure of
the ice is disrupted, and \textsl{topological} because an accurate
description of them involves ideas of symmetry embodied in topology, the
branch of mathematics that focuses on the study of continuous surfaces.

Current theories of particle physics likewise predict that a variety of
topological defects would almost certainly have formed during the early
evolution of the universe. Just as water turns to ice (a phase transition)
when the temperature drops, so the interactions between elementary particles
run through distinct phases as the typical energy of those particles falls
with the expansion of the universe. When conditions favor the appearance of
a new phase, it generally crops up in many places at the same time, and when
separate regions of the new phase run into each other, topological defects
are the result. The detection of such structures in the modern universe
would provide precious information on events in the earliest instants after
the big bang.

A central concept of particle physics theories attempting to unify all the
fundamental interactions is the concept of symmetry breaking. As the
universe expanded and cooled, first the gravitational interaction, and
subsequently all other known forces would have begun adopting their own
identities. In the context of the standard hot big bang theory the
spontaneous breaking of fundamental symmetries is realized as a phase
transition in the early universe. Such phase transitions have several
exciting cosmological consequences and thus provide an important link
between particle physics and cosmology.

There are several symmetries which are expected to break down in the course
of time. In each of these transitions the spacetime gets `oriented' by the
presence of a hypothetical force field called the `Higgs field', named for
Peter Higgs. This field orientation signals the transition from a state of
higher symmetry to a final state where the system under consideration obeys
a smaller group of symmetry rules. As a simple analogy we may consider the
transition from liquid water to ice; the formation of the crystal structure
ice (where water molecules are arranged in a well defined lattice), breaks
the symmetry possessed when the system was in the higher temperature liquid
phase, when every direction in the system was equivalent. In the same way,
it is precisely the orientation in the Higgs field which breaks the highly
symmetric state between particles and forces. Kibble \cite{Kibble76} first
saw the possibility of defect formation when he realized that in a cooling
universe phase transitions proceed by the formation of uncorrelated domains
that subsequently coalesce, leaving behind relics in the form of defects. In
the expanding universe, widely separated regions in space have not had
enough time to `communicate' amongst themselves and are therefore not
correlated, due to a lack of causal contact. It is therefore natural to
suppose that different regions ended up having arbitrary orientations of the
Higgs field and that, when they merged together, it was hard for domains
with very different preferred directions to adjust themselves and fit
smoothly. In the interfaces of these domains, defects form.

Different models for the Higgs field lead to the formation of a whole
variety of topological defects, with very different characteristics and
dimensions. Some of the proposed theories have symmetry breaking patterns
leading to the formation of `domain walls' (mirror reflection discrete
symmetry): incredibly thin planar surfaces trapping enormous concentrations
of mass--energy which separate domains of conflicting field orientations,
similar to two--dimensional sheet--like structures found in ferromagnets.
Within other theories, cosmological fields get distributed in such a way
that the old (symmetric) phase gets confined into a finite region of space
surrounded completely by the new (non--symmetric) phase. This situation
leads to the generation of defects with linear geometry called `cosmic
strings'. Theoretical reasons suggest these strings (vortex lines) do not
have any loose ends in order that the two phases not get mixed up. This
leaves infinite strings and closed loops as the only possible alternatives
for these defects to manifest themselves in the early universe. `Magnetic
monopole' is another possible topological defect. Cosmic strings bounded by
monopoles is yet another possibility in grand unified theories (GUT) phase
transitions.

Cosmic strings are without any doubt the topological defect most thoroughly
studied, both in cosmology and solid--state physics (vortices).

\section{Structure formation from defects}

In this section we will provide just a quick description of the remarkable
cosmological features of cosmic strings. Many of the proposed observational
tests for the existence of cosmic strings are based on their gravitational
interactions. In fact, the gravitational field around a straight static
string is very unusual \cite{Vilenkin81}. A simple computations indicates
that space is flat outside of an infinite straight cosmic string and
therefore test particles in its vicinity should not feel any gravitational
attraction.

In fact, a full general relativistic analysis confirms this and test
particles in the space around the string feel no Newtonian attraction;
however there exists something unusual, a sort of wedge missing from the
space surrounding the string and called the `\emph{deficit angle}', usually
noted $\Delta $, that makes the topology of space around the string that of
a cone (Fig. (\ref{fig6.1})). To see this, consider the metric of a source
with energy--momentum tensor \cite{Vilenkin81}, \cite{Gott85} 
\begin{equation}
T_{\mu }^{\nu }=\delta (x)\delta (y)\mathrm{diag}(\mu ,0,0,T)\ .
\end{equation}
In the case with $T=\mu $ (a rather simple equation of state) this is the
effective energy--momentum tensor of an unperturbed string with string
tension $\mu $ as seen from distances much larger than the thickness of the
string (a Goto--Nambu string).\FRAME{ftbpFU}{4.9644cm}{6.4207cm}{0pt}{\Qcb{%
\textsl{Cosmic strings affect surrounding spacetime by removing a small
angular wedge which is called deficit angle }$\Delta $\textsl{, (}$\Delta
\approx $\textsl{\ }$10^{-5}$ \textsl{radian), creating a conelike geometry. 
}}}{\Qlb{fig6.1}}{fig6.1.jpg}{\includegraphics{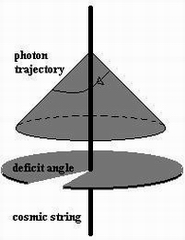}}

The gravitational field around the cosmic string [neglecting terms of order $%
(G\mu)^2$] is found by solving the linearized Einstein equations with the
above $T_\mu ^\nu$. One gets 
\begin{equation}
h_{00} = h_{33} = 4G(\mu -T) \ln(r/r_0 ) ,  \label{h00}
\end{equation}
\begin{equation}
h_{11} = h_{22} = 4G(\mu +T) \ln(r/r_0 ) ,
\end{equation}
where $h_{\mu \nu} = g_{\mu \nu} - \eta _{\mu \nu}$ is the metric
perturbation, the radial distance from the string is $r = (x^2 + y^2 )^{1/2}$%
, and $r_0$ is a constant of integration.

For an ideal, straight, unperturbed string, the tension and mass per unit
length are $T=\mu =\mu _{0}$ and one gets 
\begin{equation}
h_{00}=h_{33}=0,\ \ \ h_{11}=h_{22}=8G\mu _{0}\ln (r/r_{0}).
\end{equation}
By a coordinate transformation one can bring this metric to a locally flat
form 
\begin{equation}
ds^{2}=dt^{2}-dz^{2}-dr^{2}-(1-8G\mu _{0})r^{2}d\phi ^{2},
\end{equation}
which describes a conical and flat (Euclidean) space with a wedge of angular
size $\Delta =8\pi G\mu _{0}$ (the deficit angle) removed from the plane and
with the two faces of the wedge identified.

\section{The Abelian Higgs Vortex}

The Lagrangian of Einstein gravity in the presence of electromagnetic and
abelian Higgs field is 
\begin{equation}
\mathcal{L}=\mathcal{R}-2\Lambda -F_{\mu \nu }F^{\mu \nu }+\mathcal{L}%
_{Higgs}.
\end{equation}
where $F_{\mu \nu }=\partial _{\lbrack \mu }A_{\nu ]},$ with electromagnetic
field $A_{\mu }.$ The $\mathcal{L}_{Higgs}$ is the Lagrangian of Higgs field
which defines as follow 
\begin{equation}
\mathcal{L}_{Higgs}=-{\frac{1}{16\pi }}\mathcal{F}_{\mu \nu }\mathcal{F}%
^{\mu \nu }-{\frac{1}{2}}|D_{\mu }\Phi |^{2}-{\xi }\left( |\Phi |^{2}-{\eta }%
^{2}\right) ^{2}~.  \label{Haction}
\end{equation}
The matter content of the abelian Higgs system consists of the complex Higgs
field, $\Phi ,$ and a $U(1)$ gauge field $B_{\nu }$ with strength, $\mathcal{%
F}_{\mu \nu }=\partial _{\lbrack \mu }B_{\nu ]}.$ Both the Higgs scalar and
the gauge field become massive in the broken symmetry phase. The gauge
covariant derivative is $D_{\mu }=\nabla _{\mu }+ieB_{\mu },$ where $\nabla
_{\mu }$ is the spacetime covariant derivative. The parameter $\eta $ is the
symmetry breaking energy scale and $\xi $ is the Higgs coupling. These can
be related to the Higgs mass by $m_{Higgs}=$ $2\eta \sqrt{\xi }$. There is
another relevant mass scale, i.e., that of the vector field in the broken
phase, $m_{vector}$ $=\sqrt{2}e\eta $. On length scales smaller than $%
m_{vector}^{-1},$ $m_{Higgs}^{-1},$ the vector and Higgs field behave as
essentially massless. It is also convenient to define the Bogomolnyi
parameter $\beta =2\xi /e^{2}=m_{Higgs}^{2}/$ $m_{vector}^{2}.$ This
Lagrangian is invariant under the action of the Abelian group $\mathbf{G}%
=U(1)$. This is the Abelian Higgs model \cite{Itzykson}. The fields in $%
\mathcal{L}_{H}$ will be treated as `test field', i.e., their energy
momentum tensor is supposed to yield a negligible contribution to the source
of the gravitational field. Notice that we have two different gauge fields, $%
F$ and $\mathcal{F}$, and each is treated in a different manner. It is only $%
\mathcal{F}$ that couples to the Higgs scalar field and is therefore subject
to spontaneous symmetry breaking. The other gauge field, $F$, can be thought
of as the free massless Maxwell field which apart from modifying the
background geometry, its dynamic will be of little concern to us here. For
detailed study about the Higgs mechanism, we refer to \cite{Vilenkin}, \cite%
{Hindmarsh} and \cite{Gangui}. In this part, we use units in which $8\pi
G=c=1$.

The action (\ref{Haction}) is invariant under the following transformations, 
\begin{equation}
\Phi \rightarrow \Phi e^{i\Lambda (x)},\text{ \ \ \ \ \ \ \ \ \ }B_{\mu
}\rightarrow B_{\mu }-\nabla _{\mu }\Lambda (x),
\end{equation}
which is spontaneously broken in the ground state, $\Phi =\eta e^{i\Lambda
_{0}}.$ Besides this ground state, another solution, the vortex, is present
when the phase of $\Phi (x)$ is a non-single valued quantity. To better
describe this, define the real fields $X$, $P_{\mu }$, $\omega (x)$ , by 
\begin{equation}
\Phi (x^{\mu })=\eta X(x^{\mu })e^{i\omega (x^{\mu })},
\end{equation}
\begin{equation}
B_{\mu }(x^{\mu })=\frac{1}{e}\left[ P_{\mu }(x^{\mu })-\nabla _{\mu }\omega
(x^{\mu })\right] .  \label{Pfield}
\end{equation}
The flux of Higgs gauge field $B_{\mu }$ is quantized by identify the vortex
line. The flux is given by 
\begin{equation}
\Phi _{H}=\int \mathcal{F}_{\mu \nu }d\sigma ^{\mu \nu }=\oint B_{\mu
}(x^{\mu })dx^{\mu }.  \label{flux}
\end{equation}
Using the Eq. (\ref{Pfield}), then the integration (\ref{flux}) without any $%
P$ field is 
\begin{equation}
\Phi _{H}=-\frac{1}{e}\oint \nabla _{\mu }\omega (x)dx^{\mu }.
\end{equation}
The line integral over the gradient of phase $\Phi $ does not necessarily
vanish. The only requirement on the phase is that $\Phi $ is single valued,
i.e., \ $\oint d\omega =2\pi N$ ($N$ is an integer). In this case a vortex
is present. The integer $N$ is called the winding number of the vortex. If $%
N\neq 0$, and if the spatial topology is trivial, then, by continuity, the
integration loop must encircle a point of unbroken symmetry $(X=0)$, namely,
the vortex core. Therefore the flux of Higgs gauge field can be obtained as: 
\begin{equation}
\Phi _{H}=\mathcal{-}\frac{2\pi N}{e}.
\end{equation}
Thus, the flux of vortex lines is quantized, and $-2\pi /e$ being the
quantum.

The field equations that follow by varying $X$ in the action (\ref{Haction})
and employing a suitable choice of gauge, are 
\begin{equation}
\nabla _{\mu }\nabla ^{\mu }X-XP_{\mu }P^{\mu }-4\xi \eta ^{2}X(X^{2}-1)=0,
\label{Fi1}
\end{equation}
while by varying $B_{\mu }$ one finds 
\begin{equation}
\nabla _{\mu }\tilde{F}^{\mu \nu }-4\pi e^{2}\eta ^{2}P^{\nu }X^{2}=0.
\label{Fi2}
\end{equation}
By varying the action with respect to $g_{\mu \nu },$ one obtains 
\begin{equation}
G_{\mu \nu }-\frac{3}{\ell ^{2}}g_{\mu \nu }=(\mathcal{T}_{\mu \nu }^{em}+%
\mathcal{T}_{\mu \nu }^{Higgs}),  \label{Fi3}
\end{equation}
where $\tilde{F}^{\mu \nu }=\nabla ^{\mu }P^{\nu }-\nabla ^{\nu }P^{\mu }$
is the field strength of the corresponding gauge field $P^{\mu }$, and $%
\mathcal{T}_{\mu \nu }^{em}$ and $\mathcal{T}_{\mu \nu }^{Higgs}$ are the
stress energy tensors of the electromagnetic and Higgs fields given by 
\begin{eqnarray}
&&\mathcal{T}_{\mu \nu }^{em}=2F_{\mu }^{\sigma }F_{\sigma \nu }-\frac{1}{2}%
F^{2}g_{\mu \nu }  \label{Elstr} \\
&&\mathcal{T}_{\mu \nu }^{Higgs}=\eta ^{2}\nabla _{\mu }X\nabla _{\nu
}X+\eta ^{2}X^{2}P_{\mu }P_{\nu }+\frac{1}{4\pi e^{2}}\mathcal{F}_{\mu
\sigma }\mathcal{F}_{v}^{\sigma }+g_{\mu \nu }\mathcal{L}_{H}.  \label{Hstr}
\end{eqnarray}
\ The field $\omega $\ is not dynamical. Some of the static black holes may
be dressed by the vortices of Nielsen-Olesen type \cite{NO} appear as
cylindrically symmetric solutions, 
\begin{equation}
\Phi =X(r_{c})e^{iN\phi },\text{ \ \ \ \ \ \ \ \ \ \ }P_{\phi }=NP(r_{c}),
\end{equation}
where $r_{c}$ is the cylinder radial coordinate, and all other components of 
$P_{\mu }$ are zero.

The Abelian Higgs hair for \emph{rotating} charged black string will be
studied in the next chapter.

\chapter{Abelian Higgs Hair for Charged Rotating black string \label{Higgs}%
\protect\footnote{%
The paper related to this subject has been submitted to Can. J. Phys. (see
Ref. \cite{DehKhod2})}\protect\bigskip}

The metric of the charged rotating black string which had been considered in
Chapter \ref{Lemosch}, can be rewritten as:\ 
\begin{equation}
ds^{2}=-\Gamma \left( \Xi dt-ad\phi \right) ^{2}+\frac{r^{2}}{\ell ^{4}}%
\left( adt-\Xi \ell ^{2}d\phi \right) ^{2}+\frac{dr^{2}}{\Gamma }+\frac{r^{2}%
}{\ell ^{2}}dz^{2},  \label{met1}
\end{equation}
where $\Gamma =$ $f(r)$ and the other\ parameters were defined in chapter %
\ref{Lemosch} (see Eqs.(\ref{eq:4102}) and (\ref{Xi-A}).

We seek a cylindrically symmetric solution for the Higgs field equations (%
\ref{Fi1}) and (\ref{Fi2}) in the background of a charged rotating black
string. Thus, we assume that the fields $X$ and $P_{\mu }$ are functions of $%
r$. As mention in the previous chapter, for the case of vanishing rotation
parameter (static case), one can choose the Nielson-Olesen type gauge field
as $P_{\mu }\left( r\right) =\left( 0,0,Np(r),0\right) $. Indeed, the field
equations (\ref{Fi1}) and (\ref{Fi2}) reduces to two equations for the two
unknown functions $X(r)$ and $P(r)$. Here, for stationary case (non
vanishing rotation parameter), the field equations (\ref{Fi1}) and (\ref{Fi2}%
) reduces to three equations and therefore one may use the following gauge
choice 
\begin{equation}
P_{\mu }\left( r\right) =\left( S(r),0,NP(r),0\right) .  \label{P}
\end{equation}
The field equations (\ref{Fi1}) and (\ref{Fi2}) reduce to 
\begin{eqnarray}
&&r^{2}{\Gamma }\,X^{\prime \prime }+\left( 4\ell ^{-2}\,r^{3}-b\ell \right)
X^{\prime }-4\,r^{2}X\left( X^{2}-1\right) -X\left( aS+\Xi NP\right) ^{2} 
\notag \\
&&-{{r}^{2}\Gamma ^{-1}X\left( \Xi S+Na\ell ^{-2}P\right) ^{2}=}0,
\label{Eqm1} \\
&&N({r}^{3}{\ell }^{2}{\Gamma }\,{\Xi }^{2}-{r}^{5}a^{2}\ell ^{-2})P^{\prime
\prime }+N(2\,{r}^{4}+b{\ell }^{3}r{\Xi }^{2}-2\,{\lambda }^{2}{\ell }^{4}{%
\Xi }^{2})P^{\prime }  \notag \\
&&+{\ell }^{2}a\Xi ({{\lambda }^{2}{\ell }^{2}}-{b\ell {r}})\left(
rS^{\prime \prime }-{S}^{\prime }+{\alpha \,rX^{2}{\Gamma }^{-1}S}\right)
-a\Xi {{\lambda }^{2}{\ell }^{4}S}^{\prime }  \notag \\
&&+N\alpha {r}^{3}X^{2}({\ell }^{2}{\Xi }^{2}-{{r}^{2}a^{2}\ell ^{-2}{\Gamma 
}^{-1}})P=0,  \label{Eqm2} \\
&&{r}^{3}\left( {\Xi }^{2}{r}^{2}-{\Gamma }\,{a}^{2}\right) S^{\prime \prime
}+[2\,{r}^{4}-{a}^{2}({b\ell r}-2\,{{\lambda }^{2}{\ell }^{2})]}S^{\prime
}+\alpha ^{2}{r}^{3}\Gamma ^{-1}\left( {{\Xi }^{2}{r}^{2}}-{\Gamma a}%
^{2}\right) \times  \notag \\
&&\,XS-N\Xi \,a[\left( rP^{\prime \prime }-{P}^{\prime }+{\alpha r\,X^{2}{%
\Gamma }^{-1}P}\right) ({{\lambda }^{2}{\ell }^{2}}-{b\ell {r)}}-{{\lambda }%
^{2}{\ell }^{2}P}^{\prime }]=0,  \label{Eqm3}
\end{eqnarray}
where $\alpha =4\pi e^{2}/\xi $ and the prime denotes a derivative with
respect to $r$. It is worthwhile to mention that even in the pure flat or
(anti-)de Sitter spacetimes no exact analytic solutions are known for
equations (\ref{Fi1}) and (\ref{Fi2}) coupled with (\ref{Fi3}). Even in
situation of no electromagnetic charge or any horizon. Thus we should try to
solve these coupled differential equations approximately. In the first order
approximation, we solve Eqs. (\ref{Fi1}) and (\ref{Fi2}) in the background
of charged rotating black string and then we will carry out numerical
calculations to solve Eq. (\ref{Fi3}). For asymptotically AdS spacetimes, it
had showed that the Abelian Higgs equations of motion in the background of
charged black string spacetime (static case) have vortex solution \cite%
{Dehjal}. Here we want to investigate the influence of rotation parameter on
the vortex solutions.

\section{Numerical solutions\label{Num}}

We pay attention now to the numerical solutions of Eqs. (\ref{Eqm1})-(\ref%
{Eqm3}) outside the black string horizon. First, we must take appropriate
boundary conditions. Since at a large distance from the horizon the metric (%
\ref{met1}) reduces to AdS spacetime, we demand that our solutions go to the
solutions of the vortex equations in AdS spacetime given in \cite{DehHigg1}.
This requires that we demand ($X\rightarrow 1$, $P\rightarrow 0$) as $r$
goes to infinity and ($X=0$, $P=1$) on the horizon. For consistency with the
non-rotating case \cite{Dehjal}, we take $S=0$ on the horizon. Also one may
note that the electric field, $\tilde{F}_{tr}$ which is proportional to $%
S^{\prime }$ should be zero as $r$ goes to infinity. We employ a grid of
points $r_{i}$ with division $dr$, where $r_{i}$ goes from $r_{H}$ to some
large value of $r$ ($r_{\infty }$) which is much greater than $r_{H}$. We
rewrite Eqs. (\ref{Eqm1})-(\ref{Eqm3}) in a finite deference language and
use the successive over relaxation method \cite{Num} to calculate the
numerical solutions of $X(r),$ $P(r)$ and $S(r)$ for different values of the
rotation parameter and winding number. The numerical results of calculations
are shown in Figs. (\ref{fig7.1})-(\ref{fig7.9}). In these figures, first,
we investigate the influence of rotation parameter on the solutions of field
equations (\ref{Eqm1})-(\ref{Eqm3}). We carry out all the calculation for $%
\ell =1$, $\lambda =0.2$ and $b=0.3$ for which the radius of horizon is $%
r_{H}=0.6173$. Figures (\ref{fig7.1}) and (\ref{fig7.2}) show the behavior
of the electric, $E_{Higgs}=\tilde{F}_{tr}$, and the magnetic, $H_{Higgs}=%
\tilde{F}_{\phi r}$, fields associated with the field $P_{\mu }$
respectively for different values of the rotation parameters. We use the
subscript `Higgs' for these electromagnetic field to emphasis that they are
coupled with the Higgs scalar field $\Phi $. As one can see in Fig. (\ref%
{fig7.1}), $E_{Higgs}$ is zero for $a=0$, and becomes larger as $a$
increases. The magnetic field $H_{Higgs}$ is plotted in Fig. (\ref{fig7.2})
for different values of angular momentum. This figure shows that the
variation of $H_{Higgs}$ with respect to the rotation parameter is very
slow. Overall, these figures show that the vortex thickness decreases as the
rotation parameter increases. As we mentioned, in the case of non vanishing
rotation parameter we encounter with the electric field $E_{Higgs}$. One may
compute the source of this electric field through the use of Gauss law
numerically. It is notable that the computation of this electric type charge
for $E_{Higgs}$ shows that it increases as $a$ becomes larger. This is
analogous to the rotating solutions of Einstein-Maxwell equation discussed
in the context of cosmic string theory for which the electric charge of the
string is proportional to the rotation parameter of the string \cite{Lem}.
The effect of rotation on field $X(r)$ also is shown in Figs. (\ref{fig7.3}).

Next, we investigate the influence of the winding number $N$ on the
solutions of field equations (\ref{Eqm1})-(\ref{Eqm3}) for the case of
rotating charged black string. The results for $a=0.5$ are shown in Figs. (%
\ref{fig7.4})-(\ref{fig7.6}) for different values of $N$. As in the case of
asymptotically flat, dS, and AdS spacetimes considered in Refs. \cite%
{Achucarro,Dehjal, DehHiggs2,DehHigg1, Ghez1}, increasing the winding number
yields a greater vortex thickness.

The effects of charge per unit length, $\lambda ,$ or mass per unit length, $%
b,$ parameters on the solutions of field equations (\ref{Eqm1})-(\ref{Eqm3})
for the case of rotating charged black string are shown in Figs. (\ref%
{fig7.7})-(\ref{fig7.9}). As one can see from Fig. (\ref{fig7.7}), the
vortex thickness decreases slowly by increasing the charge per unit length
and from Fig.$\ $(\ref{fig7.8}), the dependence of $H_{Higgs}$ on $\lambda $
is very small (almost negligible) , but $E_{Higgs}$ will become stronger by
increasing the charge per unit length (see Fig. (\ref{fig7.9})).

\section{Asymptotic Behavior of the Solutions of Einstein-Maxwell-Higgs
Equation\label{Self}}

In previous section we found the solutions of Higgs field equation in the
background of charged rotating black string. Here, we want to solve the
coupled Einstein-Maxwell-Higgs differential equation (\ref{Fi1})-(\ref{Fi3}%
). This is a formidable problem even for flat or AdS spacetimes, and no
exact solutions have been found for these spacetimes yet. Indeed, besides
the electromagnetic stress energy tensor, the energy-momentum tensor of the
Higgs field is also a source for Einstein equation (\ref{Fi3}). However,
some physical results can be obtained by making some approximations. First,
we assume that the thickness of the skin covering the black string is much
smaller than all the other relevant length scales. Second, we assume that
the gravitational effects of the Higgs field are weak enough so that the
linearized Einstein-Abelian Higgs differential equations are applicable. We
choose $g_{\mu \nu }\simeq g_{\mu \nu }^{(0)}+\varepsilon g_{\mu \nu }^{(1)}$%
, where $g_{\mu \nu }^{(0)}$ is the rotating charged black string metric in
the absence of the Higgs field and $\ g_{\mu \nu }^{(1)}$ is the first order
correction to the metric. Employing the two assumptions concerning the
thickness of the vortex and its weak gravitational field, the first order
approximation to Einstein equation (\ref{Fi3}) can be written as: 
\begin{equation}
G_{\mu \nu }^{(1)}-\frac{3}{\ell ^{2}}g_{\mu \nu }^{(1)}=\mathcal{T}_{\mu
\nu }^{(0)},  \label{Eineq}
\end{equation}
where $G_{\mu \nu }^{(1)}$ is the first order correction to the Einstein
tensor due to $g_{\mu \nu }^{(1)}$ and $\mathcal{T}_{\mu \nu }^{(0)}$ is the
energy-momentum tensor of the Higgs field in the rotating charged black
string background metric with components: 
\begin{eqnarray}
\mathcal{T}_{t}^{t(0)}(r) &=&\{-\Gamma \ell ^{4}\,\left( r^{2}\Xi ^{2}-{{%
\Gamma }\,{a}^{2}}\right) S^{\prime ^{2}}-\,\alpha r^{2}\ell ^{4}\,{\Gamma }%
^{2}\,X^{\prime ^{2}}-\,{N}^{2}\Gamma \left( \,\Xi ^{2}\ell ^{4}{\Gamma }-{{a%
}^{2}{r}^{2}}\right) P^{\prime ^{2}}  \notag \\
&&-{\alpha \,{N}^{2}X^{2}\ell }^{4}\Gamma {P^{2}}-\alpha a^{2}\,\left( {%
\Gamma }\,{\ell }^{2}-{r^{2}}\right) \left( {{N}^{2}P^{2}}-S^{2}\ell
^{2}\right) X^{2}-{\alpha r}^{2}\ell ^{4}{\,X^{2}S^{2}}  \notag \\
&&-2\alpha r^{2}\ell ^{4}\Gamma \left( X^{2}-1\right) ^{2}\}/2r^{2}\ell
^{4}\Gamma ,  \notag \\
\mathcal{T}_{\phi }^{\phi (0)}(r) &=&\{\,\Gamma \ell ^{4}\left( -{{\Gamma }\,%
{a}^{2}}+{r}^{2}{\Xi }^{2}\right) S^{\prime ^{2}}-\,\alpha r^{2}\ell ^{4}\,{%
\Gamma }^{2}\,X^{\prime ^{2}}+\,{N}^{2}\Gamma \left( {\Gamma }\,\Xi ^{2}{%
\ell }^{4}-{{a}^{2}{r}^{2}}\right) P^{\prime ^{2}}  \notag \\
&&+\,\alpha \ell ^{4}\,X^{2}\left( {{N}^{2}{\Gamma }P^{2}}+{{r}^{2}S^{2}}%
\right) +\,\alpha \,X^{2}{a}^{2}\left( {\Gamma }\,{\ell }^{2}-{{r}^{2}}%
\right) \left( {{N}^{2}P^{2}}-{\ell }^{2}S^{2}\right)  \notag \\
&&-2\alpha \Gamma r^{2}\ell ^{4}\,\left( X^{2}-1\right) ^{2}\}/2r^{2}\ell
^{4}\Gamma ,  \notag \\
\mathcal{T}_{\phi }^{t(0)}(r) &=&\{2{N}^{2}\ell ^{2}\Gamma a\Xi \,\left( {%
\Gamma }\,{\ell }^{2}-{{r}^{2}}\right) P^{\prime ^{2}}+2\Gamma \ell
^{4}N\left( {{\Gamma }\,{a}^{2}}-{r}^{2}{\Xi }^{2}\right) P^{\prime
}S^{\prime }-{r^{2}\ell ^{2}PS}  \notag \\
&&+2\ell ^{2}NX^{2}\alpha \left( P{a}\left( {\Gamma }\,{\ell }^{2}-{{r}^{2}}%
\right) \left( aS+{\Xi \,NP}\right) \right) \}/2r^{2}\ell ^{4}\Gamma , 
\notag \\
\mathcal{T}_{r}^{r(0)}(r) &=&\{\Gamma \ell ^{4}\,\left( {{\Gamma }\,{a}%
^{2}-\Xi }^{2}{r}^{2}\right) S^{\prime ^{2}}+\,\alpha r^{2}\ell ^{4}\,{%
\Gamma }^{2}X^{\prime ^{2}}+\,{N}^{2}\Gamma \left( {\Gamma }\,\Xi ^{2}{\ell }%
^{4}-{{a}^{2}{r}^{2}}\right) P^{\prime ^{2}}  \notag \\
&&+2a\Xi \ell ^{2}\,N\Gamma \left( {{\Gamma \ell }^{2}}-{r}^{2}\right)
P^{\prime }S^{\prime }-2\alpha r^{2}\ell ^{4}\Gamma \,\left( X^{2}-1\right)
^{2}-\,\alpha \,X^{2}{a}^{2}\times  \notag \\
&&\left( {\Gamma }\,{\ell }^{2}-{r}^{2}\right) \left( {{N}^{2}P^{2}}+{\ell }%
^{2}S^{2}\right) -2\alpha \,\ell ^{2}X^{2}Na\Xi \,SP\left( {\Gamma }\,{\ell }%
^{2}-{r}^{2}\right)  \notag \\
&&-\,\alpha \ell ^{4}\,X^{2}\left( {{\Gamma N}^{2}P^{2}}-{{r}^{2}S^{2}}%
\right) \}/2r^{2}\ell ^{4}\Gamma ,  \notag \\
\mathcal{T}_{z}^{z(0)}(r) &=&\{\Gamma \ell ^{4}\left( -{{\Gamma }\,{a}^{2}}+{%
r}^{2}{\Xi }^{2}\right) S^{\prime ^{2}}-\,\alpha \,r^{2}\ell ^{4}{\Gamma }%
^{2}\,X^{\prime ^{2}}-\,{N}^{2}\Gamma \left( {\Gamma }\,{\ell }^{4}\Xi ^{2}-{%
{a}^{2}{r}^{2}}\right) P^{\prime ^{2}}  \notag \\
&&-2a\ell ^{2}\Xi \,N\Gamma \left( {{\Gamma \ell }^{2}}-{r}^{2}\right)
P^{\prime }S^{\prime }-2\alpha r^{2}\ell ^{4}\Gamma \,\left( X^{2}-1\right)
^{2}-\,\alpha \,X^{2}{a}^{2}\times  \notag \\
&&\left( {\Gamma }\,{\ell }^{2}-{{r}^{2}}\right) \left( {{N}^{2}P^{2}}+{\ell 
}^{2}S^{2}\right) -2\alpha \ell ^{2}\,X^{2}Na\Xi \,SP\left( {\Gamma }\,{\ell 
}^{2}-{{r}^{2}}\right)  \notag \\
&&-\alpha \,\ell ^{4}X^{2}\left( {{N}^{2}{\Gamma }P^{2}}-{{r}^{2}S^{2}}%
\right) \}/2r^{2}\ell ^{4}\Gamma ,  \label{Thiggs}
\end{eqnarray}
where $X$, $P$, and $S$ are the solutions of the Abelian Higgs system. The
behavior of $T_{\mu \nu }(r)$ is shown in Fig. (\ref{fig7.10}). As we
mentioned in the last section, the vortex thickness decreases as the
rotation parameter increases. This fact is more clear in Fig. (\ref{fig7.11}%
) which shows $T_{t}^{t(0)}$ for various values of $a$.

For convenience, we use the following form of the metric which has
cylindrical symmetry 
\begin{equation}
ds^{2}=-\widetilde{A}(r)dt^{2}+\widetilde{B}(r)dr^{2}+\widetilde{C}%
(r)dtd\phi +\widetilde{D}(r)d\phi ^{2}+\widetilde{E}(r)dz^{2}.
\label{ABCmetric}
\end{equation}
In order to solve numerically Eq. (\ref{Eineq}), it is better to write the
metric function $\widetilde{A}(r)$ to $\widetilde{E}(r)$ as 
\begin{eqnarray}
\widetilde{A}(r) &=&A_{0}(r)[1+\varepsilon A(r)],  \notag \\
\widetilde{B}(r) &=&B_{0}(r)[1+\varepsilon B(r)],  \notag \\
\widetilde{C}(r) &=&C_{0}(r)[1+\varepsilon C(r)],  \notag \\
\widetilde{D}(r) &=&D_{0}(r)[1+\varepsilon D(r)],  \notag \\
\widetilde{E}(r) &=&E_{0}(r)[1+\varepsilon E(r)],  \label{ABCexpa}
\end{eqnarray}
where $A_{0}(r)=\Gamma \Xi ^{2}-r^{2}a^{2}\ell ^{-4}$, $B_{0}(r)=\Gamma
^{-1} $, $C_{0}=2a\Xi (\Gamma -r^{2}\ell ^{-2})$, $D_{0}(r)=r^{2}\Xi
^{2}-\Gamma a^{2}$, $E_{0}(r)=r^{2}\ell ^{-2}$, yielding the metric of the
stationary rotating charged black string in four dimensions. The Einstein
equations (\ref{Eineq}) in terms of the functions $A(r)$ to $E(r)$ are given
in the Appendix \ref{A-E}. Here we want to obtain the behavior of these
functions for large values of the coordinate $r$. As one can see from Fig. (%
\ref{fig7.10}), the components of the energy-momentum tensor rapidly go to
zero outside the skin, so the situation is the same as what happened in the
static black string spacetime considered in \cite{Dehjal}. One can solve the
linearized Einstein equation for large values of $r$ numerically. The
results which are displayed in Fig. (\ref{fig7.12}) show that $%
A(r)=B(r)=E(r)=0$, and $2C(r)=D(r)=2$. Hence the metric (\ref{ABCmetric})
can be written as 
\begin{eqnarray}
ds^{2} &=&-A_{0}(r)dt^{2}+B_{0}(r)dr^{2}+(1+\varepsilon )C_{0}(r)dtd\phi 
\notag \\
&&+(1+2\varepsilon )D_{0}(r)d\phi ^{2}+E_{0}(r)dz^{2}.  \label{metdef1}
\end{eqnarray}
It is worthwhile to mention that the metric (\ref{metdef1}) is the first
order solution in $\varepsilon $ of the Einstein-Maxwell-Higgs equations far
from the thin string. Of course, one may note that the metric (\ref{metdef1}%
) is the first order approximation of the following metric 
\begin{equation}
ds^{2}=-\Gamma \left( \Xi dt-a\gamma d\phi \right) ^{2}+\frac{r^{2}}{\ell
^{4}}\left( adt-\Xi \ell ^{2}\gamma d\phi \right) ^{2}+\frac{dr^{2}}{\Gamma }%
+\frac{r^{2}}{\ell ^{2}}dz^{2},  \label{Metdef}
\end{equation}
which is the exact solution of Einstein-Maxwell gravity. In Eq. (\ref{Metdef}%
), $\gamma $ is defined as $\gamma =1+\varepsilon $. The above metric
describes a stationary rotating charged black string with a deficit angle $%
\Delta =2\pi \varepsilon $. The size of deficit angle $\Delta $ is
proportional to $2\pi \int rT_{t}^{t(0)}dr$ \cite{Achucarro}. Numerical
computation shows that the absolute values of this integral decreases as the
rotation parameter increases, which can also been from Fig. (\ref{fig7.11}).
So, using a physical Lagrangian based model, we have established that the
presence of the Higgs field induces a deficit angle in the rotating charged
black string metric which decreases as the rotation parameter increases.

\FRAME{ftbpFU}{3.2465in}{2.3817in}{0pt}{\Qcb{$E_{Higgs}\times 10^{3}$\textsl{%
\ versus }$r$\textsl{\ for }$N=1$\textsl{, }$a=0$\textsl{\ (touch the
horizontal axis), }$0.25$\textsl{\ (dotted), }$0.5$\textsl{\ (solid) and }$%
0.7$\textsl{\ (bold).}}}{\Qlb{fig7.1}}{ehiggsa.jpg}{\raisebox{-2.3817in}{\includegraphics[height=2.3817in]{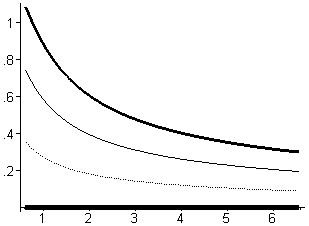}}}\FRAME{ftbpFU}{3.0588in}{2.0487in}{0pt}{\Qcb{$H_{Higgs}$\textsl{\
versus }$r$\textsl{\ for }$N=1$\textsl{, }$a=0$\textsl{\ (solid) and }$0.7$%
\textsl{\ (bold).}}}{\Qlb{fig7.2}}{hhiggsa.jpg}{\raisebox{-2.0487in}{\includegraphics[height=2.0487in]{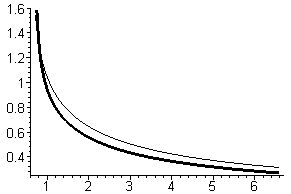}}}\FRAME{ftbpFU}{3.3088in}{2.4336in}{0pt}{\Qcb{$X(r)$\textsl{\
versus }$r$\textsl{\ for various }$a$\textsl{\ (all curves touch each other).%
}}}{\Qlb{fig7.3}}{xhiggsa.jpg}{\raisebox{-2.4336in}{\includegraphics[height=2.4336in]{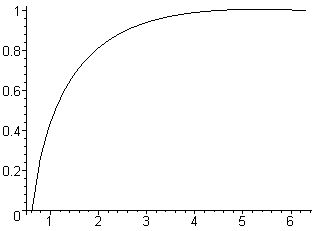}}}\FRAME{ftbpFU}{2.9032in}{2.1638in}{%
0pt}{\Qcb{$E_{Higgs}\times 10^{3}$\textsl{\ versus }$r$\textsl{\ for }$N=1$%
\textsl{\ (dotted), }$3$\textsl{\ (solid), and }$5$\textsl{\ (bold).}}}{\Qlb{%
fig7.4}}{ehiggsn.jpg}{\raisebox{-2.1638in}{\includegraphics[height=2.1638in]{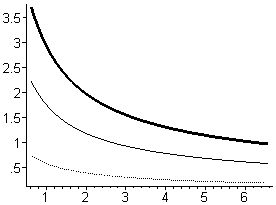}}}\FRAME{ftbpFU}{3.0277in}{2.2883in}{%
0pt}{\Qcb{$H_{Higgs}$\textsl{\ versus }$r$\textsl{\ for }$N=1$\textsl{\
(dotted), }$3$\textsl{\ (solid), and }$5$\textsl{\ (bold).}}}{\Qlb{fig7.5}}{%
hhiggsn.jpg}{\raisebox{-2.2883in}{\includegraphics[height=2.2883in]{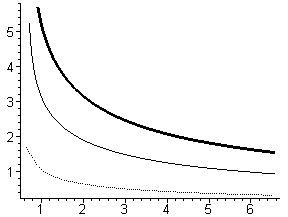}}}\FRAME{ftbpFU}{3.3399in}{2.4448in}{%
0pt}{\Qcb{$X(r)$\textsl{\ versus }$r$\textsl{\ for }$N=1$\textsl{\ (dotted), 
}$3$\textsl{\ (solid), and }$5$\textsl{\ (bold).}}}{\Qlb{fig7.6}}{xhiggsn.jpg%
}{\raisebox{-2.4448in}{\includegraphics[height=2.4448in]{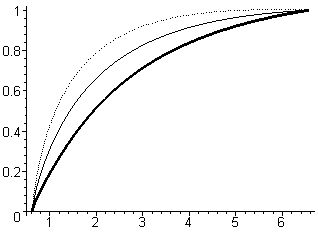}}}\FRAME{ftbpFU}{3.3615in}{2.5892in}{0pt}{\Qcb{$X_{Higgs}$\textsl{\
versus }$r$\textsl{\ for fix }$r_{+}=0.6173,$ $N=1$\textsl{, }$a=0.25$ 
\textsl{and} $\protect\lambda =$\textsl{\ }$0.2$\textsl{\ (dotted), }$0.4$%
\textsl{\ (solid) and }$0.5$\textsl{\ (bold).}}}{\Qlb{fig7.7}}{xhiggsq.jpg}{%
\raisebox{-2.5892in}{\includegraphics[height=2.5892in]{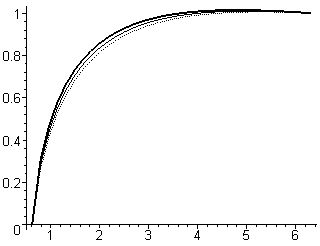}}}\FRAME{ftbpFU}{2.9032in}{1.9969in}{0pt}{\Qcb{$H_{Higgs}$\textsl{\
versus }$r$\textsl{\ for fix }$r_{+}=0.6173,$ $N=1$\textsl{, }$a=0.25$ 
\textsl{and various} $\protect\lambda $ \textsl{(the dependence on }$\protect%
\lambda $ \textsl{is very small).}}}{\Qlb{fig7.8}}{hhiggsq.jpg}
{\raisebox{-1.9969in}{\includegraphics[height=1.9969in]{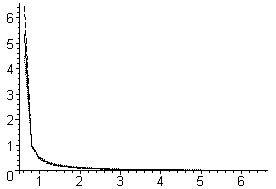}}}
\FRAME{ftbpFU}{8.1934cm}{6.6053cm}{0pt}{\Qcb{$E_{Higgs}\times
10^{3}$\textsl{\ versus }$r$\textsl{\ for fix }$r_{+}=0.6173,$ $N=1$\textsl{%
, }$a=0.25$ \textsl{and} $\protect\lambda =$\textsl{\ }$0.2$\textsl{\
(dotted), }$0.4$\textsl{\ (solid) and }$0.5$\textsl{\ (bold).}}}{\Qlb{fig7.9}%
}{ehiggsq.jpg}{\includegraphics{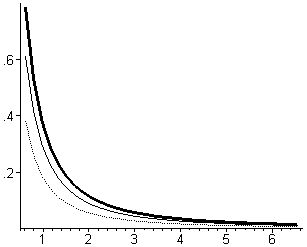}}\FRAME{ftbpFU}{3.4238in}{2.412in}{0pt%
}{\Qcb{$T_{t}^{t(0)}=T_{z}^{z(0)}$\textsl{\ (dotted), }$T_{\protect\varphi %
}^{\protect\varphi (0)}$\textsl{\ (solid), }$T_{\protect\varphi }^{t(0)}$%
\textsl{\ (bold) and }$T_{r}^{r(0)}$\textsl{\ (thick-bold), versus }$r$%
\textsl{\ for }$N=1$\textsl{, }$a=0.5$\textsl{.}}}{\Qlb{fig7.10}}{stress.jpg%
}{\raisebox{-2.412in}{\includegraphics[height=2.412in]{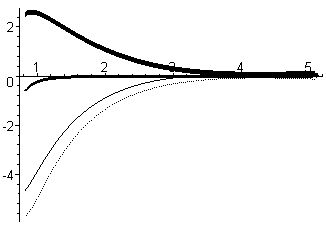}}}\FRAME{ftbpFU}{3.3088in}{2.4543in}{0pt}{\Qcb{$\left|
T_{t}^{t(0)}\right| $\textsl{\ versus }$r$\textsl{\ for }$a=0$\textsl{\
(solid) and }$a=0.7$\textsl{\ (bold).}}}{\Qlb{fig7.11}}{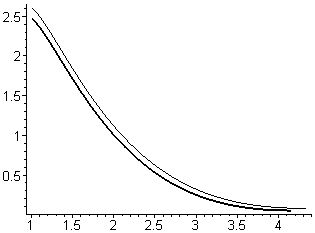}{%
\raisebox{-2.4543in}{\includegraphics[height=2.4543in]{stresscomp.jpg}}}\FRAME{ftbpFU}{3.2984in}{2.444in}{0pt}{\Qcb{$A(r)$\textsl{, }$B(r)$%
\textsl{\ and }$E(r)$\textsl{\ touch the horizontal axis, }$C(r)$\textsl{\
(solid) and }$D(r)$\textsl{\ (bold).}}}{\Qlb{fig7.12}}{abcd.jpg}
{\raisebox{-2.444in}{\includegraphics[height=2.444in]{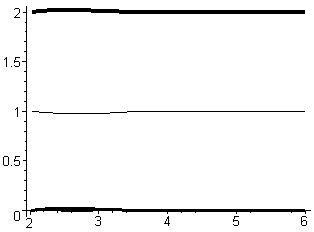}}}

\chapter{Conclusion\label{conclusion}}

In this thesis, two aspects of asymptotically charged rotating black branes
in various dimensions have been studied. In part \ref{part1}, the
thermodynamics of these spacetimes has been investigated, while in the
second part the no-hair theorem for the four-dimensional case of these
spacetimes has been considered. In chapter 1 of this thesis we give a brief
review of the conjecture of AdS/CFT correspondence, which presents an
equivalence between a gravitational theory in an ($n+1$)-dimensional AAdS
spacetime and\ a conformal field theory on the $n$-dimensional boundary of
the bulk. This conjecture furnished a means for calculating the action and
thermodynamic quantities intrinsically without reliance on any reference
spacetime. Chapter 2 devoted to a brief review of the classical laws of
thermodynamics of black holes in Einstein gravity, and the analogy between
thermodynamic of black holes and thermodynamics in thermal physics. Also,
the quantum aspect of black holes was discussed, and it was described that
the laws of thermodynamics of black holes is not only a mathematical
analogy, but have also physical interpretation. In chapter 3, we focus on
the asymptotically charged rotating black string with zero curvature
horizon, and the maximal analytical extension of this solution was studied.
In chapter 4, the thermodynamics of asymptotically charged rotating black
branes in various dimensions, which first was introduced by Awad \cite{Awad2}%
, was studied. We calculated the conserved quantities and the Euclidean
actions of charged rotating black branes both in canonical and
grand-canonical ensemble through the use of counterterms renormalization
procedure. Also we obtained the charge and electric potential of the black
brane in an arbitrary dimension. We found that the logarithmic divergencies
associated to the Weyl anomalies and matter field are zero. We obtained a
Smarr-type formula for the mass as a function of the extensive parameters $S$%
, $J$\ and $Q$, calculated the temperature, the angular velocity, and the
electric potential, and showed that these quantities satisfy the first law
of thermodynamics. Using the conserved quantities and the Euclidean actions,
the thermodynamics potentials of the system in the canonical and
grand-canonical ensemble were calculated. We found that the Helmholtz free
energy, $F(T,J,Q)$, is a Legendre transformation of the mass with respect to 
$S$ and the Gibbs potential is a Legendre transformation of the mass with
respect to $S,$\ $J$\ and $Q$\ in the grand-canonical ensemble.

Also, we studied the phase behavior of the charged rotating black branes in $%
(n+1)$ dimensions and showed that there is no Hawking-Page phase transition
in spite of the angular momentum of the branes. Indeed, we calculated the
heat capacity and the determinant of the Hessian matrix of the Gibbs
potential with respect to $S$, $J$ and $Q$ of the black brane and found that
they are positive for all the phase space, which means that the brane is
stable for all the allowed values of the metric parameters. This analysis
has also be done through the use of the determinant of the Hessian matrix of 
$M(S,J,Q)$ with respect its extensive variables and we got the same phase
behavior. This phase behavior is in commensurable with the fact that there
is no Hawking-Page transition for black object whose horizon is
diffeomorphic to $\mathbb{R}^{p}$ and therefore the system is always in the
high temperature phase \cite{Wit}.

Finally, we obtained the logarithmic correction of the entropy due to the
thermal fluctuation around the thermal equilibrium. For the case of
uncharged rotation black brane, we found that only a term which is
proportional to $\ln (\mathrm{area})$ will appear. But we found that for the
charged rotating black brane, the correction contains other powers of the '$%
area$' including the logarithmic term.

In part \ref{part2} of this thesis, we investigated the no-hair theorem for
the four-dimensional AAdS charged rotating black string. In chapter 6, we
studied the cosmological defects and the theory of the Abelian Higgs field
for an arbitrary spacetime. We obtained the field equations for
Einstein-Maxwell-Higgs system in the background of a stationary rotating
charged black string. Since there is no analytic solutions for
Einstein-Maxwell-Higgs system, even for the flat spacetimes, we attempted to
solve them numerically. We obtained the numerical solutions for various
values of rotation parameter and found that for a fixed horizon radius, by
increasing the rotation parameter the vortex thickness decrease very slowly.
Also the numerical solutions for various values of winding number and charge
per unit length $\lambda ,$\ were obtained. These solutions shows that the
vortex thickness increases as the winding number increases and is unchanged
for various $\lambda $ with constant $r_{+}$.

The main difference between the case of Abelian Higgs field in the
background of static black string considered in \cite{Dehjal}, and this work
is that the time component of the gauge field coupled to the Higgs scalar
field is not zero for non zero rotation parameter. Indeed, we found that for
the case of rotating black string, there exist an electric field coupled to
the Higgs scalar field. This electric field increases as the rotation
parameter becomes larger. Numerical calculations show that the electric
charge which creates this electric field grows up as the rotation parameter
increases. This is analogous to the results that Dias and Lemos have found
recently for the magnetic rotating string \cite{Lem}. They showed that the
charge per unit length of a rotating string in the Einstein-Maxwell gravity
increases as the rotation parameter becomes larger, and we found that the
electric charge of the field $F_{\mu \nu }^{\prime }$ coupled to the Higgs
field has the same feature.

Also, the effect of a thin vortex on pure AdS spacetime was studied. By
including the self-gravity of a thin vortex in the rotating charged black
string background in the first order approximation, we found out that the
effect of a thin vortex on the stationary charged rotating black string is
to create a deficit angle in the metric, as in the case of pure AdS \cite%
{Deh1}, Schwarzschild-AdS \cite{Deh2}, Kerr-AdS, and Reissner-Nordstr\"{o}%
m-AdS \cite{Ghez2} spacetimes. We found that the deficit angle decreases as
the rotation parameter increases.

\bigskip \appendix

\chapter{The Symmetries of Anti-de Sitter Spacetimes and The Conformal Field
Theory \label{Symmetry}}

Information on the geometrical properties of AdS spaces can be found in most
advanced text books on general relativity (see for example Ref. \cite%
{Weinberg}). Field theories on AdS spaces have been considered first in
Refs. \cite{Fronsdal,Avis}.

On the other hand, conformal field theories in dimensions $n>2$\ have
enjoyed a growing attention after great successes of the case $n=2.$\
Earlier studies can be found in \cite{Ferrara}, and more recent relevant
references are \cite{Erdmenger}.

Here, we give a brief review of AdS space which focuses on two points.
First, an AdS space is explicitly constructed. Secondly, the AdS symmetries
are found, and the symmetry algebra is represented in a form which reveals
its isomorphism to the conformal algebra.

The review of the basics of CFTs first recalls the definition of conformal
transformations and the explicit expressions for conformal transformations
of Euclidean space. Secondly, the expressions for the symmetry algebra
operators acting on quasi-primary conformal fields are given. Quasi-primary
conformal fields are important, because they form the basic field content of
any CFT.

\section{The Geometry of AdS$_{n+1}$}

As is well known, an AdS space is a maximally symmetric space which can be
represented as a hyperboloid embedded into a higher dimensional Minkowski
space. The following considerations apply to the AdS spaces with Euclidean
signature, but many results notably those regarding the symmetry algebra,
can be straightforwardly carried over to the general case. Let the
dimensions of the AdS and the embedding Minkowski spaces be $n+1$\ and $n+2,$%
\ respectively, and let the embedding be defined by 
\begin{equation}
y^{A}y^{B}\eta _{AB}=-\ell ^{2},\hspace{1.0421pc}y^{-1}>0,  \label{2.1}
\end{equation}
where $\ell $\ is the `radius' of the hyperboloid, and $A,B=-1,0,...,n.$\
The Minkowski metric tensor is given by 
\begin{equation}
\eta _{-1-1}=-1,\hspace{1.0343pc}\eta _{\mu \nu }=\delta _{\mu \nu },\hspace{%
1.0343pc}and\hspace{1.0343pc}\eta _{-1\mu }=0
\end{equation}
(with $\mu =0,1,...n$). The metric 
\begin{equation}
ds^{2}=dy^{A}dy^{B}\eta _{AB}  \label{2.3}
\end{equation}
readily represents the AdS metric, if one takes the coordinates $y^{\mu }$\
as AdS coordinates and define $y^{-1}$\ via equation (\ref{2.1}).

It is useful to introduce a new set of AdS coordinates by 
\begin{equation}
x^{0}=\frac{\ell ^{2}}{y^{0}+y^{-1}}\text{ \ \ and}\hspace{1.019pc}x^{i}=%
\frac{x^{0}y^{i}}{l}.  \label{2.4}
\end{equation}

The domain of the new variables is given by $0<x^{0}<\infty ,$\ $x^{i}\in
R(i=1,...,n).$\ More over the AdS metric (\ref{2.3}) take the form 
\begin{equation}
ds^{2}=\frac{l^{2}}{(x^{0})^{2}}\delta _{\mu \nu }dx^{\mu }dx^{\nu }.
\label{2.5}
\end{equation}
Obviously, $AdS_{n+1}$\ is an open space, i.e. it does not possess a
boundary. However, it is useful to consider the boundary of the coordinate
patch, namely the conformally compactified Euclidean space given by $x^{0}=0$%
\ plus the single point $x_{0}=\infty $, as a pseudo boundary, which will be
called the AdS horizon.

For completeness the expressions for the affine connections, the curvature
tensor, Ricci tensor and curvature scalar are provided here:

\begin{eqnarray}
\Gamma _{\hspace{0.409pc}\nu \lambda }^{\mu } &=&\frac{1}{x_{0}}\left(
\delta _{0}^{\mu }\delta _{\nu \lambda }-\delta _{\nu }^{\mu }\delta
_{0\lambda }-\delta _{\lambda }^{\mu }\delta _{0\nu }\right) , \\
R_{\hspace{0.409pc}\nu \lambda \rho }^{\mu } &=&\frac{1}{x_{0}^{2}}\left(
\delta _{\rho }^{\mu }\delta _{\nu \lambda }-\delta _{\lambda }^{\mu }\delta
_{\nu \rho }\right) , \\
R_{\mu \nu } &=&-\frac{n}{x_{0}^{2}}\delta _{\mu \nu }, \\
R &=&-n(n+1)\ell ^{-2}.  \label{2.9}
\end{eqnarray}
\medskip

\subsection{The Symmetry Group:}

The definition (\ref{2.1}) of anti--de Sitter space is invariant under
transformations of the embedding Minkowski space of the form $(y^{\prime
})^{A}=R_{\hspace{0.4135pc}B}^{A}y^{B},$\ where the $(n+2)\times (n+2)$\
matrix $\ R$\ satisfies $R^{T}\eta R=\eta $\ and \ $R_{\hspace{0.4135pc}%
-1}^{-1}>0$. The group of such matrices consists of two subsets, one being
the Lie group $SO(n+1,1),$\ whereas the other can be represented by $%
\mathcal{I}\times SO(n+1,1)$\ with an inversion $\mathcal{I}$. In\ the first
symmetry, One can introduce a conformal basis of $SO(n+1,1)$\ and show that
the conformal algebra and $SO(n+1,1)$\ algebra are isomorphic \cite{Muck}.

In the second set of symmetries, one encounters the inversion $\mathcal{I}$,
whose action on the coordinates $y^{A}$\ of the embedding Minkowski space
can be defined by the matrix 
\begin{equation}
\mathcal{I}_{\hspace{0.4167pc}B}^{A}=\delta _{B}^{A}-2\delta _{0}^{A}\delta
_{B}^{0}.
\end{equation}
Obviously, 
\begin{equation}
\mathcal{II}=1,
\end{equation}
which must hold in every representation. Using Eq. (\ref{2.4}), the
transformation induced on the $x^{\mu }$\ coordinates is found to be 
\begin{equation}
x^{\prime \mu }=l^{2}\frac{x^{\mu }}{\mathbf{x}^{2}}.  \label{2.21}
\end{equation}
\medskip

\section{Basics of Conformal Field Theory in $n$-Dimensions}

Let $g_{ij}$\ be the metric tensor of some $n$--dimensional manifold with
respect to some coordinates $x^{i}$. A transformation $\mathbf{x}\rightarrow 
\mathbf{x}^{\prime }(\mathbf{x})$, under which the metric tensor changes as 
\begin{equation}
g_{ij}^{\prime }(\mathbf{x})=[\lambda (\mathbf{x})]^{2}g_{ij}(\mathbf{x})
\label{2.29}
\end{equation}
is called a conformal transformation. Conformal transformations are a
generalization of ordinary symmetry transformations, in which every symmetry
transformation satisfies Eq. (\ref{2.29}) with $\lambda (\mathbf{x})=1$.

Given such a transformation $\mathbf{x}\rightarrow \mathbf{x}^{\prime }$,
one can define the transformation matrix 
\begin{equation}
\mathcal{R}_{\hspace{0.409pc}\nu }^{\mu }(\mathbf{x})=\lambda (x)\frac{%
\partial x^{\prime \mu }}{\partial x^{\nu }},
\end{equation}
which satisfies 
\begin{equation}
g_{\mu \rho }^{\prime }(\mathbf{x}^{\prime })\mathcal{R}_{\hspace{0.4105pc}%
\nu }^{\mu }(\mathbf{x})\mathcal{R}_{\hspace{0.409pc}\lambda }^{\rho }(%
\mathbf{x})=g_{\nu \lambda }^{\prime }(\mathbf{x})
\end{equation}
Obviously, the conformal transformations form a group.

\subsection{\ Conformal Symmetry}

A very important case, of conformal symmetry which has been studied, is the
case of conformal symmetry on a flat manifold, which will be assumed to be
Euclidean here. Let $g_{ij}=\delta _{ij}$. Then, for $n>2$, the
transformations satisfying equation (\ref{2.29}) are \cite{Francesco}:

\ \ 
\begin{equation}
\frame{$%
\begin{array}[t]{l}
\ \ \text{dilatations: \ \ \ }x^{\prime i}=cx^{i} \\ 
\text{translations:}\ \ \ \ x^{\prime i}=x^{i}+b^{i} \\ 
\ \ \ \text{\ rotations:}\ \text{\ \ \ }x^{\prime i}=R_{\hspace{0.4135pc}%
j}^{i}x^{j} \\ 
\ \ \ \ \text{inversion: \ \ \ }x^{\prime i}=\frac{x^{i}}{\mathbf{x}^{2}},%
\end{array}
$}  \label{2.32}
\end{equation}
as well as any combination of the above. It is assumed that $c>0$\ and $R\in
O(n)$.

It has been shown by a direct calculations that the infinitesimal versions
of the transformations (\ref{2.32}) are identical with the symmetry
transformations of $AdS_{n+1}$\ restricted to its horizon, $x_{0}=0$. Hence,
the same is true for the finite transformations connected to the identity.
Moreover, the inversion is identical to the AdS inversion formula (\ref{2.21}%
) restricted to $x_{0}=0$. Thus, the AdS symmetry transformations directly
correspond to conformal transformations of the AdS horizon.

\chapter{Boundaries of the Spacetime\label{boundary}}

Let spacetime be an ($n+1$)-dimensional Lorentzian manifold equipped with a
metric $g_{\mu \nu }$\ . We consider a region of this manifold, $\mathcal{M}$%
, and we define various tensors on the boundary. We require that the region $%
\mathcal{M}$\ \ have the topology of the direct product of a spacelike
hypersurface, $\Sigma $\ , with a real (timelike) interval. This requirement
allows me to foliate the manifold into leaves $\Sigma _{t}$\ of constant
foliation parameter $t.$\ The vector $u^{\mu }$\ is the future-directed
timelike vector to $\Sigma _{t}.$\ The boundary of the region $\mathcal{M}$\
is the union of \ `initial' and `final' hypersurfaces, $\Sigma _{i}$\ and $%
\Sigma _{f}$, and the timelike hypersurface $^{3}B$\ . This timelike
hypersurface can also be foliated into the spacelike quasilocal surfaces $%
B_{t}$. The (spacelike) outwards normal vector to the boundary element $%
^{3}B $\ \ is $n^{\mu }$\ while the bi-normal to the quasilocal surface $B$\
is $n^{\mu \nu }=2u^{[\mu }n^{\nu ]}.$\ For simplicity, we enforce the
condition $g_{\mu \nu }u^{\mu }n^{\nu }=0$\ and $g_{\mu \nu }n^{\mu }n^{\nu
}=1.$\ In Fig (\ref{fig2.1}) you can see these definitions.\FRAME{ftbpFU}{%
2.1223in}{2.1223in}{0pt}{\Qcb{\textsl{The manifold, its boundaries and the
normal vectors to these boundaries.}}}{\Qlb{fig2.1}}{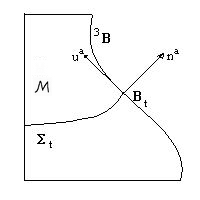}
{\raisebox{-2.1223in}{\includegraphics[height=2.1223in]{fig21.jpg}}}
\medskip

\section{The Timelike Boundary\label{Spacelike}}

Now we examine the geometry on the timelike boundary $^{3}B$\ . The two
fundamental forms on $^{3}B$\ are the induced metric $\gamma _{\mu \nu }$\ = 
$g_{\mu \nu }-n_{\mu }n_{\nu }$, and the extrinsic curvature $\Theta _{\mu
\nu }=-\frac{1}{2}\pounds _{n}\gamma _{\mu \nu }.$\ The restriction of the
induced metric to the boundary $^{3}B$\ can be viewed as the physical metric
on the $n$-dimensional manifold $^{3}B$\ . Alternately, one can view the
operator $\gamma _{\hspace{0.4151pc}\nu }^{\mu }$\ as a projection operator
that will take a vector on the tangent space of $\mathcal{M}$\ to a vector
on the tangent space of $^{3}B$\ . The derivative operator compatible with
the metric $\gamma _{\mu \nu }$\ is $D$\ .

The second fundamental form can be thought of as the failure of the vectors $%
n^{\mu }$\ to coincide when parallel-transported along the boundary $^{3}B$\
. The definition of extrinsic curvature yields 
\begin{eqnarray}
\Theta _{\mu \nu } &=&-\frac{1}{2}\pounds _{n}\gamma _{\mu \nu }  \notag \\
&=&-\frac{1}{2}\pounds _{n}(g_{\mu \nu }-n_{\mu }n_{\nu }) \\
&=&-(\nabla _{\mu }n_{\nu }-n^{\lambda }n_{\mu }\nabla _{\lambda }n_{\nu }) 
\notag \\
&=&-\gamma _{\mu }^{\alpha }\nabla _{\alpha }n_{\nu },  \notag
\end{eqnarray}
which is the difference between the normal vector and the parallel transport
(along $^{3}B$\ ) of a nearby normal. In the third line we have used the
hypersurface orthogonality of the unit vector, the definition of the Lie
derivative, and the compatibility of the metric with the derivative operator 
$\nabla $. From this viewpoint, the second fundamental form describes how
curved the surface $^{3}B$\ \ is.

\section{The Spacelike Boundary}

On the spacelike boundary $\Sigma $, the induced metric is $h_{ab}$\ = $%
g_{ab}$\ + $u_{a}u_{b}$\ while the extrinsic curvature of $\Sigma $\
embedded in $\mathcal{M}$\ is $Kab=-\frac{1}{2}\pounds _{u}h_{ab}=-h_{%
\hspace{0.4105pc}a}^{c}\nabla _{c}u_{b}.$\ Here, the operator $h_{\hspace{%
0.412pc}a}^{c}$\ is a projection operator onto the tangent space of\ $\Sigma
.$\ Define the lapse function as the normalization of the unit normal $u^{a}$%
\ relative to the vector $t^{a}$\ : $N=-t^{a}u_{a}=(u^{a}\nabla _{a}t)^{-1}$%
. The shift vector $V^{a}=h_{\hspace{0.4105pc}b}^{a}t^{b}$\ is the
projection of the vector $t^{a}$\ onto the surface $\Sigma $. Then the
vector $t^{a}$ can be decomposed into a portion normal to $\Sigma $\ and a
portion tangent to $\Sigma $\ : $t^{a}=Nu^{a}+V^{a}$.

Let $\mathcal{D}$\ be the derivative operator compatible with the metric $%
h_{ab}$. It is straightforward to show that $\mathcal{D}_{a}T_{\hspace{0.5cm}%
d...e}^{b...c}=h_{\hspace{0.5289pc}f}^{b}...h_{e}^{\hspace{0.5289pc}j}h_{a}^{%
\hspace{0.5289pc}k}\nabla _{k}T_{\hspace{0.5cm}i...j}^{f...g}$. By the
Gauss-Codacci relations which are covariant expressions of the bulk Einstein
tensor $G_{\mu \nu }=R_{\mu \nu }-\frac{1}{2}g_{\mu \nu }R$\ in terms of the
boundary Einstein tensor $G_{ab}(h)$\ (which only depends on the induced
metric $h_{ab}$\ ) and the extrinsic curvature $K_{ab}$\cite{Waldbook1}, we
can write 
\begin{equation}
R_{abcd}[h]=h_{a}^{\hspace{0.4231pc}j}h_{b}^{\hspace{0.4247pc}k}h_{c}^{%
\hspace{0.4231pc}l}h_{d}^{\hspace{0.4247pc}m}R_{jklm}[g]+2K_{d[a}K_{b]c}
\label{Gauss1}
\end{equation}
and 
\begin{equation}
2\mathcal{D}_{[b}K_{\hspace{0.4231pc}a]}^{b}=n^{c}h_{\hspace{0.5309pc}%
a}^{d}R_{cd}[g].  \label{Gauss2}
\end{equation}
An immediate consequence of (\ref{Gauss1}) is 
\begin{eqnarray}
2u^{a}u^{b}G_{ab}[g] &=&h^{ac}h^{bd}R_{abcd}[g]  \label{Gauss3} \\
&=&R[h]+K^{2}-K^{ab}K_{ab}.  \notag
\end{eqnarray}
\medskip

\section{The Quasilocal Surface}

Because $u^{a}$\ and $n^{a}$\ are taken to be orthogonal, the quasilocal
surface $B$\ can be viewed either as the boundary of $\Sigma $\ or as a leaf
in the foliation of $^{3}B$\ . The induced metric on $B$\ is 
\begin{equation}
\sigma _{ab}=g_{ab}+u_{a}u_{b}-n_{a}n_{b}.  \label{sigma}
\end{equation}

The extrinsic curvature of $B$\ embedded in $\Sigma $\ is $k_{ab}=\sigma _{%
\hspace{0.4135pc}c}^{a}D_{c}n_{b}.$\ A straightforward analysis yields the
following relationship between the various extrinsic curvatures \cite{BrownY}%
: 
\begin{equation}
\Theta _{ab}=k_{ab}+u_{a}u_{b}n_{c}u^{d}\nabla _{d}u^{c}+2\sigma _{\hspace{%
0.4135pc}(a}^{c}u_{b)}n^{d}K_{cd}  \label{Extr1}
\end{equation}
Thus, the projection of $\Theta _{ab}$\ onto $B$\ is the extrinsic curvature 
$k_{ab}$. The quantities $\Theta _{ab}$\ and $K_{ab}$\ are related by $%
\sigma _{\hspace{0.412pc}c}^{a}u^{b}\Theta _{ab}=-\sigma _{\hspace{0.4135pc}%
c}^{a}n^{b}K_{ab}$. Finally, the trace of (\ref{Extr1}) yields 
\begin{equation}
\Theta =k-n_{c}a^{c}
\end{equation}
where $\Theta $\ and $k$\ are the traces of the extrinsic curvatures $\Theta
_{ab}$\ and $k_{ab}$\ respectively.

Variations of the induced metric, $\gamma _{ab}$\ , on the timelike boundary 
$^{3}B$\ can be decomposed into pieces that are normal-normal,
normal-tangential, and tangential-tangential to the quasilocal surface $B$: 
\begin{equation}
\delta \gamma _{ab}=\sigma _{\hspace{0.4199pc}c}^{a}\sigma _{\hspace{0.4215pc%
}b}^{d}\delta \sigma _{cd}-\frac{2}{N}u_{a}u_{b}\delta N-\frac{2}{N}%
u_{(a}\sigma _{b)c}\delta N^{c}.
\end{equation}
The variation of the metric on the quasilocal surface can also be decomposed
into a variation of the square-root of the determinant, $\sqrt{\sigma }$,
plus a variation of the conformally invariant part of the metric $\zeta
_{ab} $: 
\begin{equation}
\delta \sigma _{ab}=\frac{2}{n-2}(\frac{\sigma _{ab}}{\sqrt{\sigma }})\delta 
\sqrt{\sigma }+(\sqrt{\sigma })^{\frac{2}{n-2}}\delta \zeta _{ab},
\label{Extr2}
\end{equation}
where $\zeta _{ab}=(\sqrt{\sigma })^{-2/(n-2)}\sigma _{ab}.$ The second term
in equation (\ref{Extr2}) represents changes in the `shape' of the
quasilocal surface that preserve the determinant whereas the changes in the
determinant (given by the first term) reflect a change in the `size' of the
quasilocal surface while maintaining the same shape.

\chapter{The Einstein Equations in Terms of the Functions $A,...,E$ \label%
{A-E}}

The Einstein equation mentioned in Sec. (\ref{Self}) in terms of the metric
functions $A(r)$ to $E(r)$ are 
\begin{align*}
& {\Xi }^{2}\digamma _{1}(A+D-2\,C-4{\lambda }^{2}{\ell }^{4}a^{2}\Gamma
^{2}({\Gamma }\,{\Xi }^{2}{\ell }^{4}-{r}^{2}a^{2})D+4{\lambda }^{2}{\ell }%
^{8}{\Xi }^{2}\Gamma ^{2}\times \\
& \left( {\Gamma }\,a^{2}-{r}^{2}{\Xi }^{2}\right) A-{r}^{4}\ell ^{2}\Gamma
^{2}\left\{ 4r\ell ^{4}\,{\Xi }^{2}{\Gamma }+\left( b\ell ^{3}\,-4{r}%
^{3}a^{2}\right) \right\} B^{\prime }-{\Xi }^{2}a^{2}{\ell }^{8}r\Gamma
^{2}\times \\
& \{2\,{b}^{2}r^{2}+{\lambda }^{2}(4{\lambda }^{2}{\ell }^{2}-6\,b\ell
r)\}A^{\prime }-12{\ell }^{4}\,{r}^{6}\Gamma ^{2}{B}-\,{\Xi ^{2}r}^{2}\ell
^{2}\Gamma \digamma _{{2}}\left( A^{\prime }+D^{\prime }-2\,C^{\prime
}\right) \\
& +\ell ^{2}\Gamma ^{2}{r}^{4}\{4r\ell ^{4}{\Gamma }\,{\Xi }^{2}+\left(
4r^{3}-b\ell ^{3}\right) \left( a^{2}-\ell ^{2}\right) \}\left( E^{\prime
}+D^{\prime }\right) +2\,{\Gamma }^{3}r^{6}\ell ^{6}\,\left( E^{\prime
\prime }+D^{\prime \prime }\right) \\
& +\Gamma ^{2}\ell ^{4}a^{2}\{2{\Xi }^{2}{\ell }^{4}[\,{b}^{2}r+{\lambda }%
^{2}(2{\lambda }^{2}{\ell }^{2}r^{-1}-\,3b\ell )]+r^{3}(4{\lambda }^{2}{\ell 
}^{2}-3b\ell r)\}D^{\prime } \\
& -2\,{\Xi }^{2}a^{2}{\lambda }^{2}{\ell }^{6}r\Gamma ^{2}({\Gamma r\ell }%
^{2}-r^{3}-{b\ell ^{3})}\left( D^{\prime \prime }-C^{\prime \prime }\right)
=4r^{6}\ell ^{6}\Gamma ^{2}\mathcal{T}_{t}^{t(0)},
\end{align*}
\begin{eqnarray*}
&&8\Gamma ^{2}(\,{\Gamma \ell }^{2}-{{r}^{2}}){\lambda }^{2}\Xi \,{\ell }%
^{6}a^{2}D+8\,{\lambda }^{2}\Xi \,{\ell }^{6}\Gamma ^{2}a\{\ell ^{2}r^{2}-({%
\Gamma \ell }^{2}-{{r}^{2}})a^{2}\}C \\
&&+a{{\Xi }^{3}\digamma _{{3}}\left( A+D-2\,C\right) +}r^{4}\ell ^{4}a\Xi
\Gamma ^{2}(4r\ell ^{2}{\Gamma }-4{{r}^{3}}+\,{b\ell }^{3})\left( B^{\prime
}-E^{\prime }-2D^{\prime }\right) \\
&&-{r}^{2}{\ell }^{2}\,{a\Xi \,\Gamma {F}_{{5}}(r,b,\lambda )\left(
A^{\prime }-D^{\prime }\right) }-2{r}^{2}{\ell }^{2}{\Xi a\Gamma \digamma _{{%
4}}\left( C^{\prime }-D^{\prime }\right) }+2{\ell }^{4}\Gamma ^{2}\Xi
\,a\,\times \\
&&\{{\ell }^{5}{r}^{2}\left( {\Xi }^{2}-1\right) {\Gamma }\,\left( {{\lambda 
}^{2}\ell }-br\right) +{\Xi }^{2}{r}^{6}({r}^{2}-{\Gamma \ell }^{2})\}\left(
D^{\prime \prime }-C^{\prime \prime }\right) =4r^{6}\ell ^{6}\Gamma ^{2}%
\mathcal{T}_{\varphi }^{t(0)},
\end{eqnarray*}
\begin{eqnarray*}
&&\digamma {_{{6}}\left( A+D-2\,C\right) }+8\,a^{2}{\lambda }^{2}{\ell }%
^{6}\Gamma ^{2}\{{\Gamma \ell }^{2}{\Xi }^{2}-{{r}^{2}\left( -1+{\Xi }%
^{2}\right) \}}\left( C-A\right) +4r^{2}\ell ^{6}\times \\
&&{{\lambda }^{2}\Gamma }^{2}{\left( \,{\Xi }^{2}\ell ^{2}+a^{2}\right) A+}%
r^{2}\Gamma \ell ^{-4}{{\Xi }^{2}}\,\digamma {_{{7}}\left( A^{\prime
}+D^{\prime }-2\,C^{\prime }\right) }-4{\Xi }^{2}a^{2}{\ell }^{6}r\Gamma
^{2}\times \\
&&\{{b}^{2}r\ell ^{2}+{\lambda }^{2}(2\,{\Gamma r\ell }^{2}-{b\ell }^{3}-2{r}%
^{3})\}C^{\prime }+\,\ell ^{4}r^{2}\Gamma ^{2}{\digamma _{{8}}A^{\prime }}%
+r^{4}\ell ^{4}\Gamma ^{2}({4r}^{3}-b\ell ^{3})E^{\prime } \\
&&+r^{4}\ell ^{4}\Gamma ^{2}\{4r\,a^{2}{\Gamma }-{\Xi }^{2}({4r}^{3}-b\ell
^{3})\}\left( B^{\prime }-E^{\prime }\right) +2r^{6}\ell ^{6}\,{\Gamma }%
^{3}\,\left( E^{\prime \prime }+A^{\prime \prime }\right) \\
&&-2\,{\Xi }^{2}a^{2}{\lambda }^{2}{\ell }^{6}r\Gamma ^{2}({\Gamma \ell }%
^{2}r-{{r}^{3}}-{b\ell }^{3})\left( A^{\prime \prime }-C^{\prime \prime
}\right) -12{\ell }^{4}\,{r}^{6}\Gamma ^{2}{B}=4r^{6}\ell ^{6}\Gamma ^{2}%
\mathcal{T}_{\varphi }^{\varphi (0)},
\end{eqnarray*}
\begin{eqnarray*}
&&{{\Xi }^{2}\digamma _{{9}}\left( A+D-2\,C\right) }-8\,r^{2}{{\lambda }^{2}{%
\Xi }^{2}{\ell }^{6}\Gamma C}+4r^{2}{\lambda }^{2}{\ell }^{4}\Gamma {\left(
\,{\Xi }^{2}\ell ^{2}-a^{2}\right) D} \\
&&+r^{4}\ell ^{2}\Gamma \left( {4r}^{3}-b\ell ^{3}\right) E^{\prime
}+r^{4}\ell ^{2}\Gamma \{4r\,a^{2}{\Gamma }-{\Xi }^{2}({4r}^{3}-b\ell
^{3})\}D^{\prime } \\
&&+r^{4}\Gamma \{4r\ell ^{4}\,{\Xi }^{2}{\Gamma }-a^{2}({4r}^{3}-b\ell
^{3})\}A^{\prime }+\,{\Xi }^{2}a^{2}b{\ell }^{3}r^{2}\Gamma ({\Gamma \ell }%
^{2}-{{r}^{2})\times } \\
&&\left( A^{\prime }+D^{\prime }-2C^{\prime }\right) -{12\,{r}^{6}{\ell }%
^{2}\Gamma {B}=4r^{6}\ell ^{4}\Gamma }\mathcal{T}_{r}^{r(0)},
\end{eqnarray*}
\begin{eqnarray*}
&&{{\Xi }^{2}\digamma _{{10}}\left( A+D-2\,C\right) }-16r^{2}{\lambda }^{2}{%
\ell }^{6}\Gamma ^{2}{\left( \,{\Xi }^{2}\ell ^{2}-a^{2}\right) D}+(32\,{{%
\lambda }^{2}{\Xi }^{2}{\ell }^{8}r}^{2}\Gamma ^{2}){C} \\
&&-4{\ell }^{4}{r}^{4}{\Gamma }^{2}({4r}^{3}-b\ell ^{3}){B}^{\prime }+4{\ell 
}^{4}{r}^{2}{\Gamma }^{2}\{8r^{3}\ell ^{2}\,{\Gamma }+(4{{\lambda }^{2}{\ell 
}^{2}}-\,{3b\ell r)}\left( {a}^{2}-2\ell ^{2}\right) \}D^{\prime } \\
&&+4{\ell }^{2}{r}^{2}{\Gamma }\,{{\Xi }^{2}\digamma _{{11}}\left( A^{\prime
}+D^{\prime }-2C^{\prime }\right) }+\{8\,{r}^{4}+\,\left( 3\,{\Xi }%
^{2}-2\right) b\ell ^{2}r-4{{\lambda }^{2}{\ell }^{4}{\Xi }^{2}\}\times } \\
&&4{\ell }^{4}{r}^{3}{\Gamma }^{2}A^{\prime }-8\,{\Xi }^{2}a^{2}{\lambda }%
^{2}{\ell }^{6}r\Gamma ^{2}\left( r\ell ^{2}{\Gamma }-{r}^{3}-b\ell
^{3}\right) \left( A^{\prime \prime }+D^{\prime \prime }-2C^{\prime \prime
}\right) +8\,r^{6}\ell ^{6}{\Gamma }^{3} \\
&&\,\times \left( A^{\prime \prime }+D^{\prime \prime }\right) -48{\ell }%
^{4}r^{6}\Gamma ^{2}{B}=16r^{6}{\ell }^{6}\Gamma ^{2}\mathcal{T}_{z}^{z(0)},
\end{eqnarray*}
where $\mathcal{T}_{\mu }^{\nu (0)}$ is the energy momentum tensor of the
Higgs field given in Eqs. (\ref{Thiggs}) and the functions $\digamma _{i}$'s
are: 
\begin{eqnarray*}
\digamma _{1} &=&{\Xi }^{4}(32\,{\Gamma }^{3}{\ell }^{6}{r}^{4}-{\Gamma }^{2}%
{\ell }^{10}{b}^{2}+{r}^{4}{b}^{2}{\ell }^{6}-32{\Gamma }\,{\ell }^{2}{r}%
^{8}-8\,{\Gamma }^{3}{\ell }^{9}rb+8{\Gamma }^{2}{\ell }^{7}{r}^{3}b \\
&&+16\,{r}^{10}-16\,{\Gamma }^{4}{\ell }^{8}{r}^{2}-8{r}^{7}b{\ell }^{3}+8\,{%
\Gamma }\,{\ell }^{5}{r}^{5}b)+{\Xi }^{2}({\Gamma }^{2}{\ell }^{10}{b}%
^{2}+12\,{\Gamma }^{4}{\ell }^{8}{r}^{2} \\
&&-28\,{\Gamma }^{3}{\ell }^{6}{r}^{4}-12\,{\Gamma }^{2}{\ell }^{4}{r}%
^{6}+60\,{\Gamma }\,{\ell }^{2}{r}^{8}-32\,{r}^{10}+16\,{r}^{7}b{\ell }%
^{3}-12{\Gamma }\,{\ell }^{5}{r}^{5}b \\
&&+4\,{\Gamma }^{3}{\ell }^{9}rb-8\,{\Gamma }^{2}{\ell }^{7}{r}^{3}b-2\,{r}%
^{4}{b}^{2}{\ell }^{6})+4\,{\Gamma }^{4}{\ell }^{8}{r}^{2}+{r}^{4}{b}^{2}{%
\ell }^{6}-8\,{r}^{7}b{\ell }^{3} \\
&&+4\,{\Gamma }^{3}{\ell }^{9}rb+16\,{r}^{10}+4\,{\Gamma }\,{\ell }^{5}{r}%
^{5}b+12\,{\Gamma }^{2}{\ell }^{4}{r}^{6}-28\,{\Gamma }\,{\ell }^{2}{r}%
^{8}-4\,{\Gamma }^{3}{\ell }^{6}{r}^{4}, \\
\digamma _{2} &=&\,{\Xi }^{4}(4{r}^{7}-4\,{\Gamma }^{3}{\ell }^{6}r-b{\ell }%
^{3}{r}^{4}-b{\ell }^{7}{\Gamma }^{2}+2\,b{\ell }^{5}{\Gamma }\,{r}^{2}+12{%
\Gamma }^{2}{\ell }^{4}{r}^{3}-12\,{\Gamma }\,{\ell }^{2}{r}^{5}) \\
&&+{\Xi }^{2}(2\,{r}^{4}b{\ell }^{3}-8\,{r}^{7}-b{\ell }^{7}{\Gamma }^{2}+16{%
r}^{5}{\Gamma }\,{\ell }^{2}-8\,{r}^{3}{\Gamma }^{2}{\ell }^{4}-{r}^{2}b{%
\ell }^{5}{\Gamma )}-{r}^{4}b{\ell }^{3} \\
&&+4\,{r}^{7}+2\,{\Gamma }^{2}{\ell }^{7}b-{\Gamma }\,{\ell }^{5}{r}^{2}b+4\,%
{\Gamma }^{3}{\ell }^{6}r-4\,{\Gamma }^{2}{\ell }^{4}{r}^{3}-4\,{\Gamma }\,{%
\ell }^{2}{r}^{5}, \\
\digamma _{3} &=&\,{\Xi }^{2}(32{\Gamma }\,{\ell }^{2}{r}^{8}-32{\Gamma }^{3}%
{\ell }^{6}{r}^{4}+8\,{r}^{7}b{\ell }^{3}-{r}^{4}{b}^{2}{\ell }^{6}-8\,{%
\Gamma }^{2}{\ell }^{7}{r}^{3}b-8{\Gamma }\,{\ell }^{5}{r}^{5}b \\
&&+8\,{\Gamma }^{3}{\ell }^{9}rb+{\Gamma }^{2}{\ell }^{10}{b}^{2}-16\,{r}%
^{10}+16\,{\Gamma }^{4}{\ell }^{8}{r}^{2})+8\,{\Gamma }\,{\ell }^{5}{r}%
^{5}b+16\,{r}^{10} \\
&&+32\,{\Gamma }^{3}{\ell }^{6}{r}^{4}+8\,{\Gamma }^{2}{\ell }^{7}{r}%
^{3}b-8\,{\Gamma }^{3}{\ell }^{9}rb-16\,{\Gamma }^{4}{\ell }^{8}{r}^{2}-32\,{%
\Gamma }\,{\ell }^{2}{r}^{8} \\
&&+{r}^{4}{b}^{2}{\ell }^{6}-8\,{r}^{7}b{\ell }^{3}-{b}^{2}{\ell }^{10}{%
\Gamma }^{2}, \\
\digamma _{4} &=&\,{\Xi }^{4}(4{r}^{7}-b{\ell }^{7}{\Gamma }^{2}-b{\ell }^{3}%
{r}^{4}+12\,{\Gamma }^{2}{\ell }^{4}{r}^{3}-12\,{\Gamma }\,{\ell }^{2}{r}%
^{5}-4\,{\Gamma }^{3}{\ell }^{6}r+2\,b{\ell }^{5}{\Gamma }\,{r}^{2}) \\
&&+{\Xi }^{2}(8\,{r}^{5}{\Gamma }\,{\ell }^{2}-b{\ell }^{7}{\Gamma }^{2}-4\,{%
r}^{7}-4{r}^{3}{\Gamma }^{2}{\ell }^{4}+{r}^{4}b{\ell }^{3}) \\
&&+2\,{\Gamma }^{2}{\ell }^{7}b+4\,{\Gamma }^{3}{\ell }^{6}r-4\,{\Gamma }^{2}%
{\ell }^{4}{r}^{3}, \\
\digamma _{5} &=&\,{\Xi }^{4}(4{\Gamma }^{3}{\ell }^{6}r-4{r}^{7}+b{\ell }%
^{7}{\Gamma }^{2}-2\,b{\ell }^{5}{\Gamma }\,{r}^{2}-12\,{\Gamma }^{2}{\ell }%
^{4}{r}^{3}+12\,{\Gamma }\,{\ell }^{2}{r}^{5}+b{\ell }^{3}{r}^{4}) \\
&&+\,{\Xi }^{2}(2{r}^{2}b{\ell }^{5}{\Gamma }-b{\ell }^{7}{\Gamma }^{2}+4{r}%
^{7}+12\,{r}^{3}{\Gamma }^{2}{\ell }^{4}-12\,{r}^{5}{\Gamma }\,{\ell }%
^{2}-4\,{\Gamma }^{3}{\ell }^{6}r-{r}^{4}b{\ell }^{3}) \\
&&-{\Gamma }\,{\ell }^{5}{r}^{2}b-4\,{\Gamma }^{2}{\ell }^{4}{r}^{3}+4\,{%
\Gamma }\,{\ell }^{2}{r}^{5}, \\
\digamma _{6} &=&{\Xi }^{6}({b}^{2}{\ell }^{10}{\Gamma }^{2}+16{\Gamma }^{4}{%
\ell }^{8}{r}^{2}-32{r}^{4}{\Gamma }^{3}{\ell }^{6}+32\,{r}^{8}{\Gamma }\,{%
\ell }^{2}+8{r}^{7}b{\ell }^{3}+8\,{\Gamma }^{3}{\ell }^{9}rb \\
&&-{b}^{2}{\ell }^{6}{r}^{4}-8\,{r}^{5}{\Gamma }\,{\ell }^{5}b-16\,{r}^{10}-8%
{r}^{3}{\Gamma }^{2}{\ell }^{7}b)+{\Xi }^{4}{r}^{4}{b}^{2}{\ell }^{6}+2\,{%
\Xi }^{4}(30{r}^{4}{\Gamma }^{3}{\ell }^{6} \\
&&+4b{\ell }^{3}{r}^{7}+6b{\ell }^{7}{\Gamma }^{2}{r}^{3}-8\,b{\ell }^{9}{%
\Gamma }^{3}r-18\,{r}^{8}{\Gamma }\,{\ell }^{2}+6\,b{\ell }^{5}{\Gamma }\,{r}%
^{5}-4{r}^{6}{\Gamma }^{2}{\ell }^{4}+8{r}^{10} \\
&&-{b}^{2}{\ell }^{10}{\Gamma }^{2}+16{\Gamma }^{4}{\ell }^{8}{r}^{2})+4\,{%
\Xi }^{2}({r}^{6}{\Gamma }^{2}{\ell }^{4}-6{r}^{4}{\Gamma }^{3}{\ell }^{6}-{r%
}^{5}{\Gamma \ell }^{5}b+2{\Gamma }^{3}{\ell }^{9}rb \\
&&+\,{r}^{8}{\Gamma }\,{\ell }^{2}+4{\Gamma }^{4}{\ell }^{8}{r}^{2}+{\Gamma }%
^{2}{\ell }^{10}{b}^{2}/4)+4(\,{\Gamma }^{2}{\ell }^{4}{r}^{6}-\,{\Gamma }%
^{3}{\ell }^{6}{r}^{4}-\,{\Gamma }^{2}{\ell }^{7}{r}^{3}b),
\end{eqnarray*}
\begin{eqnarray*}
\digamma _{7} &=&{\Xi }^{4}(11\,rb{\ell }^{5}{\lambda }^{4}-10\,{r}^{2}{b}%
^{2}{\ell }^{4}{\lambda }^{2}-4\,{\lambda }^{6}{\ell }^{6}+3\,{r}^{3}{b}^{3}{%
\ell }^{3})+{\Xi }^{2}(3\,{r}^{6}{b}^{2}-7\,{r}^{5}b\ell {\lambda }^{2} \\
&&-6\,{r}^{3}{b}^{3}{\ell }^{3}-22\,rb{\ell }^{5}{\lambda }^{4}+8\,{\lambda }%
^{6}{\ell }^{6}+4\,{r}^{4}{\lambda }^{4}{\ell }^{2}+20\,{r}^{2}{b}^{2}{\ell }%
^{4}{\lambda }^{2})-3\,{r}^{6}{b}^{2} \\
&&+7\,{r}^{5}b\ell {\lambda }^{2}-4\,{r}^{4}{\lambda }^{4}{\ell }^{2}+11\,rb{%
\ell }^{5}{\lambda }^{4}+3\,{r}^{3}{b}^{3}{\ell }^{3}-4\,{\lambda }^{6}{\ell 
}^{6}-10\,{r}^{2}{b}^{2}{\ell }^{4}{\lambda }^{2}, \\
\digamma _{8} &=&4\,{\Gamma }\,{\ell }^{2}{r}^{3}+2\,{\ell }^{3}{r}^{2}{\Xi }%
^{2}b+8\,{r}^{3}{\Xi }^{2}{\Gamma }\,{\ell }^{2}-4\,{\ell }^{3}{\Xi }^{4}b{r}%
^{2}+8\,{\Xi }^{4}{r}^{5}+4\,{\ell }^{5}{\Xi }^{4}b{\Gamma } \\
&&-4\,{\ell }^{5}{\Xi }^{2}b{\Gamma }+8\,{\Xi }^{4}{\Gamma }^{2}{\ell }%
^{4}r-16\,{\Xi }^{4}{\Gamma }\,{\ell }^{2}{r}^{3}-8\,{\Xi }^{2}{\Gamma }^{2}{%
\ell }^{4}r+4\,{r}^{5}-{r}^{2}b{\ell }^{3}, \\
\digamma _{9} &=&{\Xi }^{2}(12\,{r}^{4}{\Gamma }^{2}{\ell }^{4}-24\,{r}^{6}{%
\Gamma }\,{\ell }^{2}-4\,{\ell }^{7}b{\Gamma }^{2}r-{\ell }^{8}{b}^{2}{%
\Gamma }+4\,{r}^{3}b{\ell }^{5}{\Gamma }+12\,{r}^{8}) \\
&&16\,{r}^{4}{\Gamma }^{2}{\ell }^{4}+28\,{r}^{6}{\Gamma }\,{\ell }^{2}+{b}%
^{2}{\ell }^{8}{\Gamma }-8\,{r}^{3}{\Gamma }\,{\ell }^{5}b+4\,b{\ell }^{7}{%
\Gamma }^{2}r-12\,{r}^{8}, \\
\digamma _{10} &=&24\,{\Gamma }^{3}{\ell }^{9}rb-16\,{\Gamma }^{2}{\ell }^{7}%
{r}^{3}b+{\Xi }^{2}(76\,{\Gamma }^{3}{\ell }^{6}{r}^{4}-40\,{\Gamma }^{4}{%
\ell }^{8}{r}^{2}+20\,{\Gamma }^{2}{\ell }^{7}{r}^{3}b \\
&&-48\,{\Gamma }^{2}{\ell }^{4}{r}^{6}+28\,{\Gamma }\,{\ell }^{2}{r}^{8}-16\,%
{r}^{10}-24\,{\Gamma }^{3}{\ell }^{9}rb-2\,{\Gamma }^{2}{\ell }^{10}{b}%
^{2}-4\,{\Gamma }\,{\ell }^{5}{r}^{5}b \\
&&-{r}^{4}{b}^{2}{\ell }^{6}+8\,{r}^{7}b{\ell }^{3})+40\,{\Gamma }^{4}{\ell }%
^{8}{r}^{2}-72\,{\Gamma }^{3}{\ell }^{6}{r}^{4}+44\,{\Gamma }^{2}{\ell }^{4}{%
r}^{6}+16\,{r}^{10} \\
&&+4\,{\Gamma }\,{\ell }^{5}{r}^{5}b-28\,{\Gamma }\,{\ell }^{2}{r}^{8}+{r}%
^{4}{b}^{2}{\ell }^{6}-8\,{r}^{7}b{\ell }^{3}+2\,{b}^{2}{\ell }^{10}{\Gamma }%
^{2}, \\
\digamma _{11} &=&{\Xi }^{2}(4\,{r}^{7}-{r}^{4}b{\ell }^{3}-28\,{r}^{3}{%
\Gamma }^{2}{\ell }^{4}-5\,{r}^{2}b{\ell }^{5}{\Gamma }+6\,b{\ell }^{7}{%
\Gamma }^{2}+8\,{r}^{5}{\Gamma }\,{\ell }^{2}+16\,{\Gamma }^{3}{\ell }^{6}r)
\\
&&+{r}^{4}b{\ell }^{3}-8\,{\Gamma }\,{\ell }^{2}{r}^{5}+5{\Gamma \ell }^{5}{r%
}^{2}b-4\,{r}^{7}-6\,{\Gamma }^{2}{\ell }^{7}b+28\,{\Gamma }^{2}{\ell }^{4}{r%
}^{3}-16\,{\Gamma }^{3}{\ell }^{6}r.
\end{eqnarray*}

\bigskip

\end{document}